\documentclass[a4paper,fleqn]{article}
\usepackage{amsmath}
\usepackage{amssymb}
\usepackage{axodraw}
\usepackage[pdfborder={0 0 0},breaklinks=true]{hyperref}
\usepackage{breakurl}
\usepackage{array}
\usepackage{calc}
\usepackage{longtable}
\usepackage{fullpage}
\usepackage{pstricks}
\usepackage{pst-text}
\usepackage{graphicx}
\usepackage{xspace}
\usepackage{ifthen}
\usepackage{booktabs}
\usepackage{cite}
\usepackage{mcite}
\newcommand{\tops}{\texorpdfstring}
\newcommand{\Sherpa}{S\protect\scalebox{0.8}{HERPA}\xspace}
\newcommand{\Comix}{C\protect\scalebox{0.8}{OMIX}\xspace}
\newcommand{\Apacic}{A\protect\scalebox{0.8}{PACIC++}\xspace}
\newcommand{\Amegic}{A\protect\scalebox{0.8}{MEGIC++}\xspace}

\newcommand{\Amisic}{A\protect\scalebox{0.8}{MISIC++}\xspace}
\newcommand{\Ahadic}{A\protect\scalebox{0.8}{HADIC}\xspace}
\newcommand{\Hadrons}{H\protect\scalebox{0.8}{ADRONS}\xspace}
\newcommand{\Photons}{P\protect\scalebox{0.8}{HOTONS}\xspace}
\newcommand{\Photonspp}{P\scalebox{0.8}{HOTONS++}\xspace}  

\newcommand{\Phasic}{P\protect\scalebox{0.8}{HASIC}\xspace}
\newcommand{\Model}{M\protect\scalebox{0.8}{ODEL}\xspace}
\newcommand{\NLOJET}{N\protect\scalebox{0.8}{LOJET++}\xspace}
\newcommand{\MCFM}{M\protect\scalebox{0.8}{CFM}\xspace}
\newcommand{\MCatNLO}{M\protect\scalebox{0.8}{C}@N\protect\scalebox{0.8}{LO}\xspace}
\newcommand{\Softsusy}{S\protect\scalebox{0.8}{OFT}S\protect\scalebox{0.8}{USY}\xspace}
\newcommand{\FeynRules}{F\protect\scalebox{0.8}{EYN}R\protect\scalebox{0.8}{ULES}\xspace}

\newcommand{\Madgraph}{M\protect\scalebox{0.8}{AD}G\protect\scalebox{0.8}{RAPH}\xspace}
\newcommand{\Whizard}{W\protect\scalebox{0.8}{HIZARD}\xspace}

\newcommand{\Pythia}{P\protect\scalebox{0.8}{YTHIA}\xspace}

\newcommand{\Alpgen}{A\protect\scalebox{0.8}{LPGEN}\xspace}
\newcommand{\Herwig}{H\protect\scalebox{0.8}{ERWIG}\xspace}
\newcommand{\Herwigpp}{H\scalebox{0.8}{ERWIG++}\xspace} 
  
\newcommand{\Helac}{H\protect\scalebox{0.8}{ELAC}\xspace}
\newcommand{\Phegas}{P\protect\scalebox{0.8}{HEGAS}\xspace}
\newcommand{\Haag}{H\protect\scalebox{0.8}{AAG}\xspace}
\newcommand{\Vegas}{V\protect\scalebox{0.8}{EGAS}\xspace}
\newcommand{\Rambo}{R\protect\scalebox{0.8}{AMBO}\xspace}

\newcommand{\Photos}{P\protect\scalebox{0.8}{HOTOS}\xspace}
\newcommand{\Comphep}{CompH\protect\scalebox{0.8}{EP}\xspace}
\newcommand{\Calchep}{CalcH\protect\scalebox{0.8}{EP}\xspace}
\newcommand{\LHAPDF}{L\protect\scalebox{0.8}{HAPDF}\xspace}
\newcommand{\SLHA}{S\protect\scalebox{0.8}{LHA}\xspace}
\newcommand{\HepMC}{H\protect\scalebox{0.8}{EPMC}\xspace}
\newcommand{\HepEvt}{H\protect\scalebox{0.8}{EPEVT}\xspace}
\newrgbcolor{mred}{.8 .1 .1}
\newrgbcolor{brick}{.6 .1 .1}
\newrgbcolor{olive}{.1 .6 .1}
\newrgbcolor{dgreen}{.1 .4 .1}
\newrgbcolor{ogray}{.85 .85 .85}
\newrgbcolor{lgray}{.6 .6 .6}
\newrgbcolor{dgray}{.4 .4 .4}
\newrgbcolor{dyellow}{.8 .7 .01}
\newrgbcolor{dblue}{.05 .05 .3}
\newrgbcolor{lblue}{.1 .1 .6}
\long\def\symbolfootnote[#1]#2{\begingroup%
\def\thefootnote{\fnsymbol{footnote}}\footnote[#1]{#2}\endgroup}

\newcommand{\rbr}[1]{\left( #1\right)}

\newcommand{\sbr}[1]{\left[ #1\right]}
\newcommand{\done}{{\rm d}}

\newcommand{\dthree}{{\rm d}^3}

\newcommand{\order}{\mathcal{O}}
\newcommand{\QED}{{\mbox{\scalebox{0.6}{QED}}}}
\newcommand{\mc}[1]{\mathcal{#1}}

\newcommand{\sst}{\scriptstyle}

\newcolumntype{a}[2]{>{\raggedleft}p{#1}@{}p{#2}}
\newcolumntype{B}[1]{>{\centering}p{#1}}

\newcommand{\myfigcaption}[2]{
  \refstepcounter{figure}
  \small{\bf Fig. \thefigure\hspace*{2ex}}
  \parbox[t]{#1-\widthof{\bf Fig. \thefigure\hspace*{4ex}}}{#2}
}
\newcommand{\myfigure}[3]{
  \begin{figure}[#1]
    \begin{center}
      #2\\\myfigcaption{\textwidth}{#3}
    \end{center}
  \end{figure}
}
\newcommand{\mywidefigure}[3]{
  \begin{figure*}[#1]
    \begin{center}
      #2\\\myfigcaption{\widthof{#2}}{#3}
    \end{center}
  \end{figure*}
}
\newcommand{\mytabcaption}[2]{
  \refstepcounter{table}
  \small{\bf Tab. \thetable\hspace*{2ex}}
  \parbox[t]{#1-\widthof{\bf Tab. \thetable\hspace*{4ex}}}{#2}
}
\newcommand{\mytable}[3]{
  \begin{table}[#1]
    \begin{center}
      #2\\\mytabcaption{\widthof{#2}}{#3}
    \end{center}
  \end{table}
}
\newcommand{\bt}{\begin{tabular}}
\newcommand{\et}{\end{tabular}}
\def\be{\begin{equation}}
\def\ee{\end{equation}}
\def\bc{\begin{center}}
\def\ec{\end{center}}
\newcommand{\nnb}{\nonumber}
\newcommand{\bea}{\begin{eqnarray}}
\newcommand{\eea}{\end{eqnarray}}
\newcommand{\bit}{\begin{itemize}}
\newcommand{\eit}{\end{itemize}}
\newcommand{\Miss}[1]{\slash\hspace*{-1.5ex}#1}

\newcommand{\neua}{\tilde{\chi}^0_1}
\newcommand{\neub}{\tilde{\chi}^0_2}
\newcommand{\neuc}{\tilde{\chi}^0_3}
\newcommand{\neud}{\tilde{\chi}^0_4}
\newcommand{\chap}{\tilde{\chi}_1^+}
\newcommand{\cham}{\tilde{\chi}_1^-}
\newcommand{\chbp}{\tilde{\chi}_2^+}
\newcommand{\chbm}{\tilde{\chi}_2^-}
\newcommand{\pone}{\phantom{0}}
\newcommand{\pten}{\phantom{00}}

\newcommand{\pth}{\phantom{0000}}
\newcommand{\tabc}[7]{$ #1 $ & #2 & #3 & #4 & #5 & #6 & #7 \\}
\newcommand{\mcent}{\multicolumn{1}{c}{---}}

\makeatletter
\newif\if@preliminary
\@preliminaryfalse
\def\preliminary{\@preliminarytrue}
\def\preprintno#1{\def\@preprintno{#1}}
\def\address#1{\def\@address{#1}}

\def\abstract#1{\def\@abstract{#1}}
\renewcommand\abstractname{ABSTRACT}
\newlength\preprintnoskip
\newlength\abstractwidth
\renewcommand\maketitle{\begin{titlepage}%
  \let\footnotesize\small
  \hfill\parbox{\preprintnoskip}{%
  \begin{flushright}\@preprintno\end{flushright}}\hspace*{1cm}
  \vskip 60\p@
  \begin{center}%
    {\Large\bf\boldmath \@title \par}\vskip 1cm%
    {\sc\@author \par}\vskip 3mm%
    {\@address \par}%
    \if@preliminary
      \vskip 2cm {\large\sf PRELIMINARY DRAFT \par \@date}%
    \fi
  \end{center}\par
  \@thanks
  \vfill
  \begin{center}%
    \parbox{\abstractwidth}{\centerline{\abstractname}%
    \vskip 3mm%
    \@abstract}
  \end{center}
  \end{titlepage}%
  \setcounter{footnote}{0}%
  \let\thanks\relax\let\maketitle\relax
  \gdef\@thanks{}\gdef\@author{}\gdef\@address{}%
  \gdef\@title{}\gdef\@abstract{}\gdef\@preprintno{}
}%
\makeatother

\begin{document}
\title{Event generation with \Sherpa~1.1}
\preprintno{SLAC-PUB-13420\\ZU-TH 17/08\\DCPT/08/138\\IPPP/08/69\\Edinburgh 2008/30\\FERMILAB-PUB-08-477-T\\MCNET/08/14}
\author{
 T.~Gleisberg$^{1}$,
 S.~H\"oche$^{2}$,
 F.~Krauss$^{3}$,
 M.~Sch\"onherr$^{4}$,
 S.~Schumann$^{5}$,
 F.~Siegert$^{3}$,
 J.~Winter$^{6}$
}
\address{\it%
$^{1}$ Stanford Linear Accelerator Center, Stanford University, Stanford, CA 94309, USA\\
$^{2}$ Institut f{\"u}r Theoretische Physik, Universit{\"a}t Z{\"u}rich, CH-8057 Z{\"u}rich, Switzerland\\
$^{3}$ Institute for Particle Physics Phenomenology, Durham University, Durham DH1 3LE, UK\\
$^{4}$ Institut f\"ur Kern- und Teilchenphysik, TU Dresden, D--01062 Dresden, Germany\\
$^{5}$ School of Physics and Astronomy, The University of Edinburgh, Edinburgh EH9 3JZ, UK\\
(Future address: Institut f\"ur Theoretische Physik, Universit\"at Heidelberg, D-69120, Heidelberg, Germany)\\
$^{6}$ Fermi National Accelerator Laboratory, Batavia, IL 60510, USA
}

\abstract{
In this paper the current release of the Monte Carlo event generator Sherpa, 
version 1.1, is presented. Sherpa is a general-purpose tool for the simulation of 
particle collisions at high-energy colliders. It contains a very flexible
tree-level matrix-element generator for the calculation of hard scattering 
processes within the Standard Model and various new physics models. The 
emission of additional QCD partons off the initial and final states is 
described through a parton-shower model. To consistently combine 
multi-parton matrix elements with the QCD parton cascades the approach of 
Catani, Krauss, Kuhn and Webber is employed. A simple model of multiple 
interactions is used to account for underlying events in hadron--hadron 
collisions. The fragmentation of partons into primary hadrons is described 
using a phenomenological cluster-hadronisation model. A comprehensive 
library for simulating tau-lepton and hadron decays is provided. Where 
available form-factor models and matrix elements are used, allowing for 
the inclusion of spin correlations; effects of virtual and real 
QED corrections are included using the approach of Yennie, Frautschi 
and Suura.}

% \PACS{
% {12.38.-t}{} \and {13.20.-v}{} \and {13.25.-k}{} \and {13.35.Dx}{} \and {13.66.-a}{} \and {13.85.-t}{} \and {13.87.-a}{} 
% }

\maketitle
\addtocounter{page}{1}
\tableofcontents
\section*{Introduction}

The LHC will pose new challenges to both the experimental and the theoretical community. 
It will operate at the highest centre-of-mass energies ever reached in a collider experiment
and provide an enormous luminosity, leading to tremendously large event rates.  On the experimental 
side, the huge phase space in conjunction with the exciting and intricate physics programme
of the LHC, ranging from high-precision flavour physics at comparably low scales to the
discovery of new particles in the TeV range necessitated the development and refinement
of triggers and analysis techniques.  In addition, from a more technological point of view,
data acquisition, storage and processing therefore required the creation of a world-wide 
network satisfying the greatly increased computing needs.  On the theoretical side, on the
other hand, demands for higher precision to correctly model signals of new physics and their
backgrounds led to the rethinking of calculation and simulation paradigms and to the 
construction of a new generation of modern tools.  Maybe the most prominent manifestations 
of these paradigm shifts are the new, full-fledged event generators, which certainly will 
prove to be indispensable tools for data analysis.  Currently, these
new simulation programs holding many new features are replacing the
well-established traditional ones.  In many cases, they 
allow a wider range of applications; for example, the typically more
modular frameworks have alleviated the incorporation of new physics 
models. Often the new tools also offer higher precision in the
simulation, because better and more accurate techniques at various
stages of the event generation have become available over the past years.
Ultimately, this led to a drastic improvement e.g.\ in the description of Standard 
Model backgrounds to signals for new physics.  Thus it is no surprise that there is a 
widespread belief that these new tools will have a sizable impact on
the understanding of LHC physics.  The construction, maintenance,
validation and extension of event generators is therefore one of the
principal tasks of particle-physics phenomenology today.

\subsection*{The inner working of event generators}
\myfigure{t}{
  \includegraphics[width=0.49\textwidth]{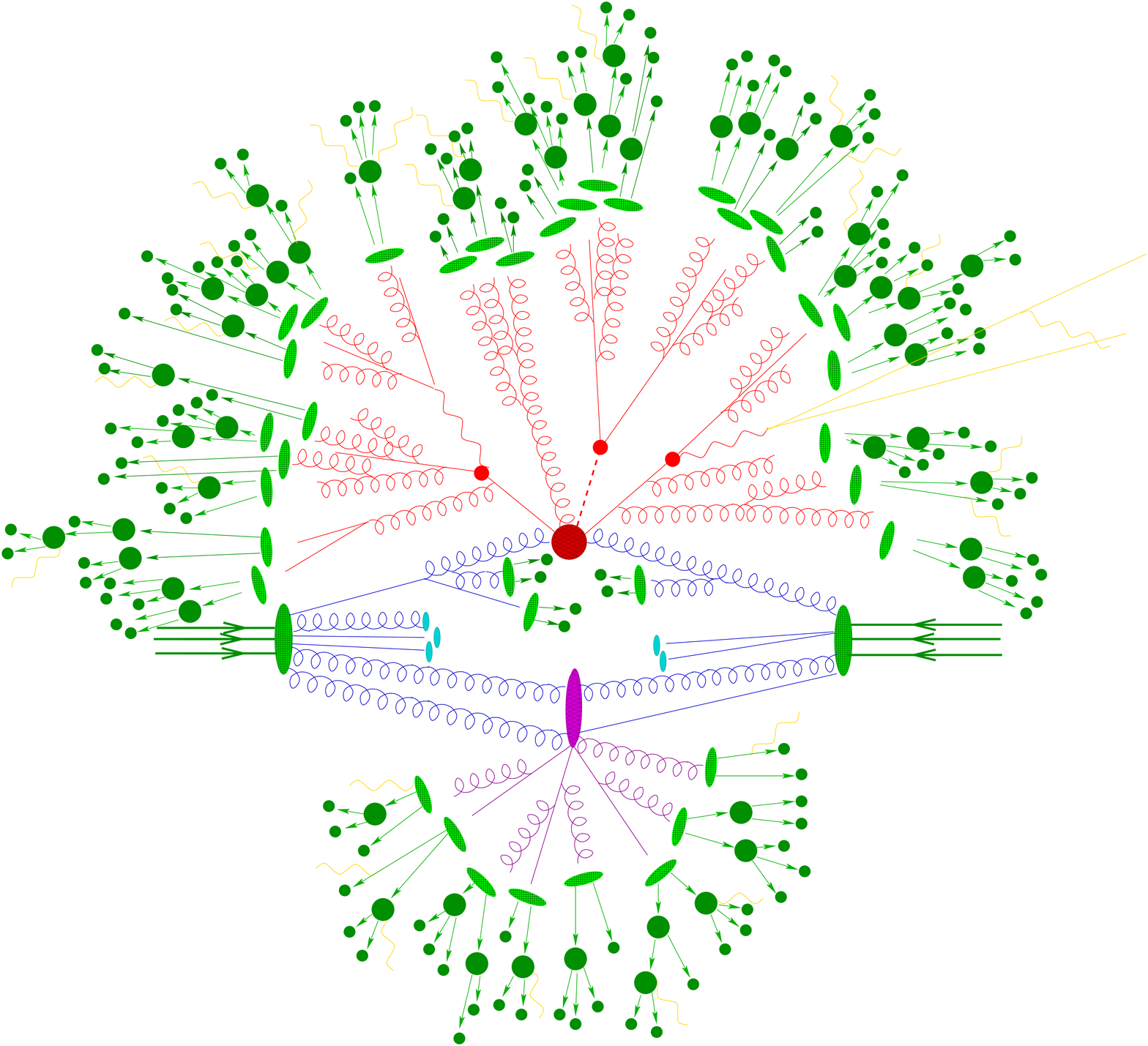}}{
  Pictorial representation of a $t\bar{t}h$ event as produced by 
  an event generator. The hard interaction (big red blob) is followed
  by the decay of both top quarks and the Higgs boson (small red blobs).
  Additional hard QCD radiation is produced (red) and a secondary
  interaction takes place (purple blob) before the final-state partons
  hadronise (light green blobs) and hadrons decay (dark green blobs).
  Photon radiation occurs at any stage (yellow).
  \label{fig:event}}
Fig.~\ref{fig:event} pictorially represents a hadron-collider event, where a $t\bar{t}h$
final state is produced and evolves by including effects of QCD bremsstrahlung in the initial and final
state, the underlying event, hadronisation and, finally, the decays of unstable hadrons 
into stable ones.  Event generators usually rely on the factorisation of such events into 
different well-defined phases, corresponding to different kinematic regimes.  In the description
of each of these phases different approximations are employed.  In general the central
piece of the event simulation is provided by the hard process (the
dark red blob in the figure), which can be calculated in fixed order
perturbation theory in the coupling constants owing to the
correspondingly high scales.  This part of the simulation is handled
by computations based on matrix elements,
which are either hard-coded or provided by special programs called parton-level or 
matrix-element (ME) generators.  The QCD evolution described by parton showers then connects
the hard scale of coloured parton creation with the hadronisation scale where the transition 
to the colourless hadrons occurs.  The parton showers model multiple QCD bremsstrahlung 
in an approximation to exact perturbation theory, which is accurate to leading logarithmic 
order.  At the hadronisation scale, which is of the order of a few $\Lambda_{QCD}$, QCD partons are transformed into primary hadrons 
(light green blobs) by applying purely phenomenological fragmentation models 
having typically around ten parameters to be fitted to data.  The
primary hadrons finally are decayed into particles that can be
observed in detectors.  In most cases effective theories or simple
symmetry arguments are invoked to describe these decays.  Another
important feature associated with the decays is QED bremsstrahlung,
which is simulated by techniques that are accurate at leading
logarithmic order and, eventually, supplemented with exact first-order
results.  A 
particularly difficult scenario arises in hadronic collisions, where remnants of the 
incoming hadrons may experience secondary hard or semi-hard interactions.  This underlying event
is pictorially represented by the purple blob in Fig.~\ref{fig:event}.  Such effects
are beyond QCD factorisation theorems and therefore no complete first-principles theory is 
available.  Instead, phenomenological models are employed again, with
more parameters to be adjusted by using comparisons with data.

\subsection*{Modern event generators}
The most prominent examples of event generators are the highly successful, well-established
programs \Pythia~\cite{Sjostrand:2006za} and \Herwig~\cite{Corcella:2002jc}.  They have been 
constructed over the past decades alongside with experimental discoveries and most of the 
features visible in past and present experiments can be described by them.  However, 
the need for higher precision to meet the challenges of new energy
scales occuring at the LHC, the 
complexity of final states at those scales, the necessity of maintenance and the wish to 
easily implement new physics models have demanded those codes to be rewritten in a modern 
programming language providing a higher level of modularity.  Object-oriented frameworks
meet the latter requirements, and owing to the community's preference
for {\tt C++}, the new
generation of event generators is constructed in this programming language.  This led to improved re-implementations in form of the programs \Pythia~8~\cite{Sjostrand:2007gs} 
and \Herwigpp~\cite{Bahr:2008pv} -- the successors of the
{\tt Fortran} versions mentioned above -- and to the construction of
the \Sherpa event generator ~\cite{Gleisberg:2003xi}.

In conjunction, in the past decade codes for next-to-leading order
calculations have been made available to the public; prominent
examples include \MCFM~\cite{MCFM}
and \NLOJET~\cite{Nagy:2003tz}.  Furthermore, methods have been proposed for the consistent 
matching of next-to-leading order corrections with parton-shower algorithms~\cite{Frixione:2002ik,*Frixione:2003ei,Frixione:2007vw}. 
Corresponding methods are implemented for example in \MCatNLO~\cite{Frixione:2006gn}, which is 
based on the {\tt Fortran} version of \Herwig, in 
\Herwig++~\cite{LatundeDada:2006gx,*Hamilton:2008pd,*LatundeDada:2008bv} and in some more 
specialised programs~\cite{Frixione:2007nu,*Alioli:2008gx}.  However, the full next-to-leading 
order calculations underlying these new techniques are very complex
and challenging and until today only processes with up to five external legs are under control.  On the other hand, many important experimental signatures rely on final states with 
higher multiplicities.  This has triggered substantial activity in
perfecting techniques and tools at tree-level accuracy, such that by now several codes are available 
that can compute corresponding cross sections and generate events in a fully automated way.  
The most prominent examples include \Alpgen~\cite{Mangano:2002ea}, 
\Comphep/\Calchep~\cite{Boos:2004kh,*Pukhov:2004ca}, 
\Helac-\Phegas~\cite{Kanaki:2000ey,*Papadopoulos:2000tt,*Cafarella:2007pc}, 
\Madgraph~\cite{Stelzer:1994ta,*Alwall:2007st}, \Whizard~\cite{Kilian:2007gr} and 
\Amegic~\cite{Krauss:2001iv}.  Currently only \Amegic is part of, and
integrated in, a full-fledged event generator, namely the \Sherpa
framework.  In order to translate the multi-particle parton-level
events, which are provided by these tools at leading order, into
hadron-level events, several algorithms have been developed, all aiming at preserving the logarithmic accuracy
of the parton shower and supplementing it with the exact perturbative leading order result
for given jet multiplicities.  In \cite{Catani:2001cc} an algorithm achieving this goal
in $e^+e^-$ annihilations into hadrons has been presented and it has been extended to
hadronic collisions in \cite{Krauss:2002up}.  A similar algorithm for the dipole shower
has been discussed in \cite{Lonnblad:2001iq,*Lavesson:2005xu}, whereas
a more different one has been published in \cite{Mangano:2001xp,*Mangano:2006rw}.  All these approaches have been 
compared in \cite{Hoche:2006ph,Alwall:2007fs} and a good agreement has been established.

\section{\protect\Sherpa's event generation framework}
\label{sec:framework}

\Sherpa~\cite{Gleisberg:2003xi} is an acronym for ``Simulation of High Energy Reactions of 
PArticles''.  The program is a complete event generation framework that has been constructed 
from scratch and entirely written in the modern, object oriented programming language 
{\tt C++}.  

\subsection*{Construction paradigm}

The construction of \Sherpa has been pursued in a way largely defined by the following three
paradigms:
\begin{itemize}
\item Modularity.  Different physics aspects are implemented in almost independent modules,
relying on a small number of framework and support modules, like, e.g., the event record etc..
Modularity allows, for example, to have more than one matrix-element generator or parton shower 
in parallel, with the user be in charge of making a choice.  The
central module, \Sherpa, steers the interplay of all other parts and the actual generation procedure.
\item Bottom-to-Top.  Physics modules are typically developed in their own right, being 
tested and validated before they are incorporated into the full event generation framework.  
This in turn results in a quite flexible, minimal structure underlying the organisation
of event generation.
\item Separation of interface and implementation.  In order to facilitate the two requirements
above, \Sherpa relies on a structure where the (nearly independent) physics modules are 
accessible only through physics-specific handlers.  These handlers assist \Sherpa in
generating the event at different stages, each of which is steered through a specific
implementation of {\tt Event\_\-Phase\_\-Handler}, such as {\tt Signal\_\-Process} or 
{\tt Jet\_\-Evolution}.  An example for such an interplay of event phase and physics handler 
is the {\tt Matrix\_\-Element\_\-Handler}, enabling the generation of parton-level events 
either by the built-in hard-coded matrix elements or by the matrix-element generator 
\Amegic.  This is relevant for two event stages, the generation of the signal 
process and owing to the multijet merging procedure the evolution of
the jets.
\end{itemize}
This overall structure fully reflects the paradigm of Monte Carlo event generation by factorising 
the simulation into well-defined, almost independent phases. Accordingly, each 
{\tt Event\_\-Phase\_\-Handler} encapsulates in an abstract way a different aspect of event 
generation for high-energy particle reactions.  This abstraction is
then replaced by real physics using handlers, which ensure that the
overall event generation framework can be blind to the finer details
of the underlying physics and its implementation in form of a physics
module.

\subsection*{Physics modules}
In the following the main modules currently distributed with \Sherpa will be listed
and briefly described.  For a more in-depth discussion, the reader
will be referred to either the corresponding section in this paper or
to the original literature.
\begin{itemize}
\item \Amegic~\cite{Krauss:2001iv}.\\[1mm]
  This is \Sherpa's default matrix-element generator based on Feynman 
  diagrams, which are translated to helicity amplitudes using the methods 
  of~\cite{Kleiss:1985yh,*Ballestrero:1992dv}. It will be described in detail
  in Sec.~\ref{sec:mes}. \Amegic has been thoroughly tested for multiparticle 
  production in the Standard Model~\cite{Gleisberg:2003bi}. 
  Its MSSM implementation has been validated in \cite{Hagiwara:2005wg}. 
  \Amegic employs the Monte Carlo phase-space integration library \Phasic. 
  For the evaluation of initial-state (laser backscattering, initial-state radiation)
  and final-state integrals, the adaptive multi-channel method
  of \cite{Kleiss:1994qy,*Berends:1994pv} is used by default together
  with a Vegas optimisation \cite{Lepage:1980dq} of the single
  channels. In addition, final-state integration accomplished 
  by \Rambo \cite{Kleiss:1985gy}
  and \Haag \cite{vanHameren:2002tc} is supported.\\
\item \Apacic~\cite{Kuhn:2000dk,Krauss:2005re}.\\[1mm]
  \Apacic generates initial- and final-state parton showering.
  The shower evolution is governed by the DGLAP equations and is ordered 
  in parton virtualities. Coherence effects are accounted for by explicit
  ordering of the opening angles in subsequent branchings.
  All features needed for a consistent merging with matrix elements
  \cite{Catani:2001cc,Krauss:2002up} are included, however, the main
  part of the merging procedure is implemented in the \Sherpa module itself.\\
\item \Amisic~\cite{Alekhin:2005dx}.\\[1mm]
  This module simulates multiple parton interactions according to 
  \cite{Sjostrand:1987su}. In \Sherpa the treatment of multiple interactions 
  has been extended by allowing the simultaneous evolution of an independent 
  parton shower in each of the subsequent collisions.
  This shower evolution is handled by \Apacic.\\
\item \Ahadic.\\[1mm]
  \Ahadic is \Sherpa's hadronisation package for translating
  the partons (quarks and gluons) into primordial hadrons.  The algorithm 
  is based on the cluster-fragmentation ideas presented in
  \cite{Gottschalk:1982yt,*Gottschalk:1983fm,*Webber:1983if,*Gottschalk:1986bv}, 
  which are also implemented in the \Herwig event generators.  It should be
  noted that \Ahadic, essentially based on \cite{Winter:2003tt}, indeed differs 
  from the original versions, cf.\ Sec.~\ref{sec:ahadic}.\\
\item \Hadrons~\cite{hadrons}.\\[1mm]
  \Hadrons is the module for simulating hadron and $\tau$-lepton decays.
  The resulting decay products respect full spin correlations (if desired).
  Several matrix elements and form-factor models have been implemented,
  such as the K\"uhn-Santamar\'ia model or form-factor parametrisations
  from Resonance Chiral Theory for the $\tau$-leptons and form factors from
  heavy	quark effective theory or light-cone sum rules for hadron decays.
  For further details, see Sec.~\ref{sec:decays}.\\
\item \Photons~\cite{Schoenherr:2008av}.\\[1mm]
  The \Photons module holds routines to add QED radiation to
  hadron and $\tau$-lepton decays based on the YFS algorithm 
  \cite{Yennie:1961ad}. The structure of \Photons is designed such that 
  the formalism can be extended to scattering processes and to a systematic 
  improvement to higher orders in perturbation theory, cf.\ \cite{Schoenherr:2008av}.  
  The application of \Photons therefore fully accounts for corrections that 
  are usually added by the application of \Photos \cite{Barberio:1993qi}.
\end{itemize}
\Sherpa itself is the steering module that initialises, controls and evaluates 
the different phases during the process of event generation.
Furthermore, all routines for the combination of parton showers and 
matrix elements, which are independent of the specific parton shower
are found in this module. For details concerning the implementation
of these routines, see Sec.~\ref{sec:ckkw}. 

In addition to the main modules of \Sherpa, there is a set of tools
providing basic routines for event generation, general methods for the 
evaluation of helicity amplitudes, some generic matrix elements, etc..
Interfaces to commonly used structures like the \LHAPDF package 
\cite{Whalley:2005nh}, or \SLHA inputs \cite{Skands:2003cj} exist
as well as interfaces to most frequently used output formats like
\HepEvt and \HepMC.

In the following the different phases of event generation and the respective 
realisation within the \Sherpa program are discussed in detail.
Some examples are presented to illustrate the capabilities of the code and
highlight its special features.

\section{Parton distributions and beam spectra}
\label{sec:beampdf}
Within \Sherpa the incoming beam of particles is resolved in two steps.
\begin{itemize}
\item Beamstrahlung, which leads to interacting photons
  in the hard process or photon generation through laser 
  backscattering is described through appropriate spectra.
\item The parton content of photons and hadrons is parametrised
  by appropriate parton distributions.
\end{itemize}

\subsection{Beam spectra}
In the context of leptonic incoming beams, e.g.\ at a 
future linear collider, photon beams can be prepared through 
laser backscattering off the (potentially polarised) electrons at
very high luminosity \cite{Badelek:2001xb}. The respective scenario
can be simulated in \Sherpa through a parametrisation that depends 
on the energy and polarisation of the incoming lepton beams as well 
as on the concrete laser specifications \cite{Zarnecki:2002qr}.

When considering collisions of hadronic beams, e.g.\ at the Fermilab Tevatron 
or the CERN LHC, there exists a two-photon component of the cross section,
which can reasonably be described in the framework of the equivalent-photon 
approximation (EPA). The EPA relates the ha\-dronic cross section to the 
interaction cross section of real photons through a two-photon luminosity 
function \cite{Budnev:1974de}. The corresponding implementation in \Sherpa 
has been described in some detail in \cite{Archibald:2008aa}.

\subsection{Parton distributions}
\Sherpa provides a variety of interfaces to standard PDF sets,  
used to parametrise the parton content of incoming particles in 
hadron collisions. The most commonly employed LHAPDF package 
\cite{Whalley:2005nh} is interfaced such that the full wealth of
PDF sets contained in this package is made available throughout 
the code.

If photons generated through laser backscattering or radiated off 
hadrons or nuclei are themselves resolved, their quark and gluon content 
may be parametrised by a photon PDF, see for example \cite{Gluck:1991ee,
  *Gluck:1991jc}.  In this context the event generation is essentially 
equivalent to the procedure for hadronic initial states, although with 
varying centre-of-mass energy.  This can easily be understood, since on the 
perturbative side of the simulation it only matters that there is a PDF. 
No information about the incoming beam being a photon is needed.

\section{Hard matrix elements\index{Matrix Elements} and phase-space integration}
\label{sec:mes}

Matrix elements for the hard scattering process can be provided either
by a fast internal library of hard-coded $2\to2$ processes or in an
automated way by using the matrix element generator \Amegic.

\Amegic~\cite{Krauss:2001iv}, acronym for ``A Matrix Element Generator
in C++'', is a multi-purpose parton-level generator. It is a convenient
tool for calculating cross sections of nearly arbitrary scattering
processes at tree level based on e.g.\ the following models:
\begin{itemize}
  \item Standard Model,
  \item Extension of the SM by a general set of anomalous triple and
    quartic gauge couplings \cite{Appelquist:1980vg,*Appelquist:1993ka},
  \item Extension of the SM by a single complex scalar
	\cite{Dedes:2008bf},
  \item Extension of the SM by a fourth generation,
  \item Extension of the SM by an axigluon
	\cite{Pati:1975ze,*Hall:1985wz,*Frampton:1987dn,*Frampton:1987ut,*Bagger:1987fz},
  \item Two-Higgs-Doublet Model,
  \item Minimal Supersymmetric Standard Model, 
  \item ADD model of large extra dimensions
    \cite{ArkaniHamed:1998rs,*Antoniadis:1998ig}.
\end{itemize}
Besides calculating production
and decay rates \Amegic is used to generate parton-level events within 
the event simulation framework of \Sherpa. Therefore, these partonic
events can easily be supplemented by parton showers and linked to the
hadronisation, which in turn yields realistic hadronic final states.

\begin{figure}[!t]
\begin{minipage}{0.48\textwidth}
 \begin{center}
  {
   \scalebox{0.8}{\begin{picture}(300,100)
   \Vertex(80,50){3}
   \Line(40,50)(80,50)
   \Line(80,50)(110,20)
   \Line(80,50)(110,80)
   \ArrowLine(65,55)(66,55)\Line(55,55)(65,55)
   \ArrowLine(94,71)(95,72)\Line(87,64)(94,71)
   \ArrowLine(94,29)(95,28)\Line(87,36)(94,29)
   \Text(25,50)[l]{$1$}
   \Text(120,90)[l]{$2$}
   \Text(120,10)[l]{$3$}
   \Text(124,50)[l]{$=(c_LP_L+c_RP_R)\bf{T}_{123}\bf{L}_{123}$}
   \end{picture}}
  }\\
  \myfigcaption{\textwidth}{Generic form of a
  three-point interaction vertex.
  The interaction of particles $1,\,2$ and $3$, with $1$ incoming while
  $2$ and $3$ are outgoing, is defined by the left- ($c_L$) and 
  right-handed ($c_R$) coupling, the $SU(3)$ colour operator $\bf{T}$, 
  and the Lorentz structure of the interaction $\bf{L}$.\label{fig:vertex}}
 \end{center}
\end{minipage}
\begin{minipage}{0.48\textwidth}
  \bc
  \scalebox{0.8}{
    \begin{tabular}{cc}
      \includegraphics[width=3.6cm]{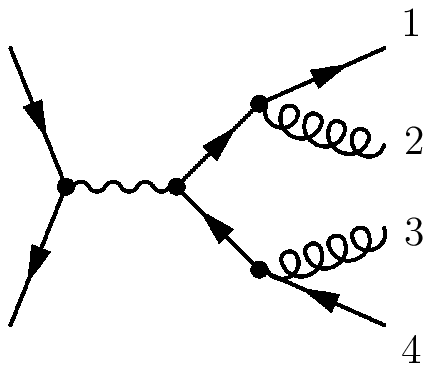} &
      \begin{minipage}[t]{4.5cm}
        \vspace*{-2cm}
        $\begin{array}{l}
          D_s(12,34) \times P_0(12)\times P_0(34)\\
          \times D_a(1,2)\times D_a(4,3)
        \end{array}
        $
      \end{minipage}
    \end{tabular}
  }\\
  \myfigcaption{\textwidth}{
    Translation of a Feynman diagram into a phase-space parametrisation.
    $D_{s,a}$ denote symmetric or asymmetric decays -- the latter ones 
    reproduce the typical feature of collinear emissions of particles
    notorious for gauge theories with massless spin-one bosons. The
    propagator terms for massless particles, $P_0$, peak at the
    minimally allowed invariant mass.\label{fig:exchan}}
  \ec
\end{minipage}
\end{figure}

\medskip
Given a physics model and a certain process \Amegic automatically
generates the corresponding set of tree-level Feynman diagrams from
the complete set of interaction vertices possessed by the 
model.\footnote{
  Note that unitary gauge is considered for the vertices;
  moreover \Amegic is currently limited to three- and four-point
  interactions.} A vertex is thereby defined by the incoming and
outgoing particles, the left- and right-handed coupling, the
explicit $SU(3)$ colour structure of the interaction, and the related
Lorentz structure. For a schematic representation of a generic
three-point function, see Fig.~\ref{fig:vertex}. Four-point
interactions have one more outgoing particle, otherwise are set up
completely equivalently.

The available colour operators $\bf{T}$ are
$1$, $\delta_{ij}$, $\delta_{ab}$, $T^a_{ij}$, $f_{abc}$ and products 
of those, e.g.\ $f_{abe}f_{cde}$. Some vertices are 
decomposed into several colour or Lorentz structures, for example the
$4$-gluon vertex is represented by three products of colour 
structures with a respective Lorentz operator. Because of this quite
general and explicit treatment of colour new physics models can be
implemented in a straightforward manner even though they might imply
interactions with colour structures absent in the SM.

The Feynman diagrams constructed from the vertices are translated into 
helicity amplitudes relying on a formalism similar to the one described 
in \cite{Kleiss:1985yh,*Ballestrero:1992dv} and extended to include 
also spin-two particles in \cite{Gleisberg:2003ue}. During this
translation process the chained Lorentz structure of each single
diagram is mapped onto suitable products and sums of
helicity-amplitude building blocks. The diagrams are then grouped into
sets of amplitudes with a common colour structure. Based on them, the
exact matrix of colour factors between amplitudes is explicitly
calculated using $SU(3)$ algebra relations. A number of refinements of
the helicity method have been implemented in the code to eventually
speed up the matrix-element evaluation and cope with issues that arise
when dealing with extensions of the SM. Concerning the latter, cf.\
Sec.~\ref{sec:mssm}.

For external massive or massless gauge bosons, explicit polarisations 
are enabled. This allows to calculate polarised cross sections.
Furthermore, the numerators of spin-one and spin-two propagators can
be replaced by sums over suitably defined polarisations for off-shell
particles, thereby disentangling nested Lorentz structures emerging
for amplitudes with many internal bosons. As a result, the generic
$3$- and $4$-point Lorentz structures of the considered physics model
are the only basic helicity-amplitude building blocks, which are
needed by \Amegic to construct arbitrary processes. The complete sets
of vertex structures appearing in the SM, the MSSM and the ADD model,
plus a comprehensive list of generalised triple and quartic
gauge-boson interactions have been implemented. Fortunately, these
sets also form the basis for many other extensions of the Standard
Model and, hence, such BSM models can be incorporated without the
need to provide new helicity-amplitude building blocks.

\begin{figure}[t!]
\bc
\includegraphics[width=0.8\linewidth]{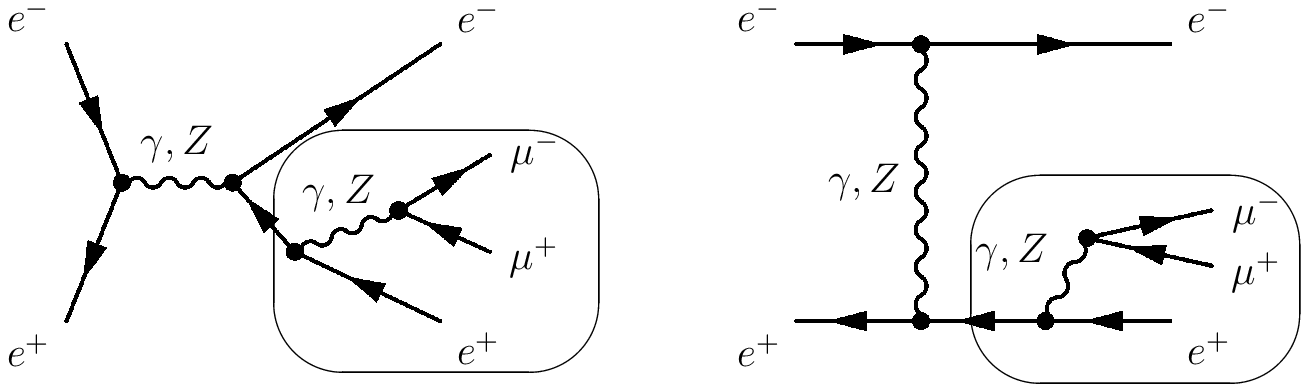}\\
\myfigcaption{\textwidth}{
  Factoring out common pieces of amplitudes with identical colour 
  structure. The subgraphs within the boxes are equal, hence, can be
  factored out such that the two amplitudes can be added.\label{fig:superamp}}
\ec
\end{figure}

Employing the helicity formalism allows to accelerate the 
matrix-element evaluation by making use of symmetries among different 
Feynman graphs. One example is given by diagrams with equal colour structure
that have common factors, as it is illustrated in Fig.~\ref{fig:superamp}.
Accordingly, identical sub-amplitudes can be factored out and the 
cut graphs can be added, thereby reducing the number of complex 
multiplications to be carried out. The optimised helicity 
amplitudes for the particular process are stored in library files.

\medskip
The second main task, which needs to be accomplished by \Amegic, is
the generation of suitable phase-space integrators, which allow an
efficient cross-section evaluation and subsequent event generation.
The employed algorithm combines process-specific a priori knowledge 
about the integrand with self-adaptive Monte Carlo integration
techniques \cite{Kleiss:1994qy,*Berends:1994pv,Lepage:1980dq,Ohl:1998jn}.

The Feynman diagrams for a given process are analysed to create
specific phase-space mappings, which generate non-uniform momentum
distributions with weights that approximate the corresponding 
(squared) diagram. The phase space for a single diagram is
parametrised using invariant masses of the diagram's propagators and
variables related to the angles in particle splittings.
As exemplified in Fig.~\ref{fig:exchan}
the mapping (and the corresponding weight function) can be assembled
out of a few generic building blocks generating corresponding
weighted distributions of those parameters.
As the matrix elements themselves, all phase-space mappings are stored
in library files.
For the integration of a scattering process, all contributing channels are
combined in a self-adaptive multi-channel integrator, which automatically
adjusts to the relative importance of the single phase-space maps
to minimise the variance.

The efficiency of the integrator is further improved by applying
the self-adaptive \Vegas algorithm on single phase-space maps.
\Vegas \cite{Lepage:1980dq} is very efficient in the numerical adaptation to 
functions, which are factorisable into a product of one-dimensional
functions. Although this is clearly unlikely for a full matrix element,
the structure represented by a single phase-space mapping fulfills
this condition. \Vegas thus allows an adaptation to more detailed
structures generated e.g.\ by phase-space cuts or the spin of the
particles. These structures clearly are beyond the approximations
inherent in the mappings.

\medskip
Both the helicity amplitudes and the phase-space para\-metri\-sations, are
semi-automatically\footnote{In detail this means that after a first
  run generating all process libraries, these libraries have to be
  compiled and dynamically linked to the main code by executing the automatically
  generated ``makelibs'' script.}
compiled and linked to the code before the actual 
integration and event generation can start.

In recent versions of \Sherpa several improvements to the algorithm
outlined above have been implemented. They shall briefly be discussed
here.
\begin{itemize}
\item Besides calculating the full set of Feynman diagrams for a given
final state it is now possible to evaluate only certain resonant 
graphs from decays of unstable intermediate particles. 
Accordingly, the process is decomposed into the actual production 
and the decay(s) of the resonance(s). For a number of intermediate 
states, labelled $i$, the combined amplitude reads
\begin{equation}
  \mathcal{A}^{(n)}=\mathcal{A}^{(n_{\rm prod})}_{\rm prod}\otimes
  \prod\limits_{i\,\in\,\rm decays}\mc{P}_i\,\mathcal{A}_{{\rm dec}_i}^{(n_i)}\;.
\end{equation}
The $\otimes$ symbol represents the colour and spin correlations
between the production and decay amplitudes.
The factor $\mc{P}_i$ accounts for the intermediate propagator. 
Two options are realised: either the resonances are 
forced on their mass shell, or the full off-shell propagator is 
considered. Both approaches fully preserve colour and spin correlations, 
however, only the first method yields strictly gauge-invariant results, 
owing to the usage of on-shell matrix elements for production and decays. 
Although gauge invariance cannot be guaranteed when the full 
off-shell propagator is considered, this ansatz naturally incorporates 
finite-width effects.

This decay treatment can be applied iteratively, therefore, provides
a very elegant way to simulate entire decay chains as they appear for 
example in supersymmetric theories. Furthermore, the treatment allows 
for an easy incorporation of $n$-body decays as the decay amplitudes 
$\mathcal{A}_{{\rm dec}}$ are not restricted to $1\to2$ processes.
\item Despite the optimisation strategies employed in \\ \Amegic the
evaluation of Feynman diagrams for purely strong-interacting processes, 
in particular $n$-gluon scattering, is still quite cumbersome. For
this case, the matrix elements can be evaluated using Ca\-chazo--Svr\v{c}ek--Witten 
vertex rules \cite{Cachazo:2004kj}, which are based on maximal 
helicity violating (MHV) amplitudes as building blocks. For small and 
intermediate particle multiplicities, typically up to $2\to5$, these 
rules give the most compact expressions, cf. \cite{Gleisberg:2008ft}.
\end{itemize}

Algorithmically there is no upper limit on particle multiplicities
for which matrix elements can be generated. Practically, calculations
are however limited by the accessible computer power. Another
important criterion is given by comparing \Amegic's performance to
alternative approaches for matrix-element generation,
e.g. \cite{Mangano:2002ea,Alwall:2007st,Gleisberg:2008fv}.
Using the Feynman diagrammatic approach (including the phase-space
integration method described above), \Amegic performs reasonably and
can compete for scattering processes that typically involve up to a few
thousand diagrams. In the framework of the Standard Model this is
sufficient for processes with a total number of up to eight or
nine particles, or a maximum of six partons in the context of
evaluating hadronic cross sections. The latter limit is not only due
to the proliferation in the number of Feynman diagrams but also the number
of different parton-level processes contributing to a jet cross section,
which typically are all desired to be taken into account at once.
Accessible particle multiplicities can of course be increased by
employing the above mentioned decay treatment, since it restricts the
number of amplitudes by accounting for certain resonant graphs only.

\medskip
The built-in matrix-element generator \Amegic renders \Sherpa a very
powerful tool to study hard production processes, in particular
many-particle final states. The underlying tree-level calculations
thereby naturally account for aspects of multi-particle production
processes that often are only approximated in alternative approaches.
Thus, a clear strength of complete matrix-element calculations is the
proper incorporation of off-shell (finite-width) effects and
quantum interferences between different diagrams contributing to the
same final state. The correct treatment of angular correlations
between final-state particles can be achieved as well. All of these
aspects are important in approaching a realistic description of both
signal and background processes.

\subsection{Standard Model production processes}

All generic interactions stemming from the Standard Mo\-del Lagrangian
have been implemented, allowing the generation of matrix elements for
arbitrary processes involving all three generations of quarks and leptons,
all SM gauge bosons and the Higgs boson. 

In addition to the generic SM interactions effective interaction vertices
between a Higgs boson and massless gauge bosons are available in \Amegic. 
The coupling to gluons is mediated by a top-quark loop and modelled through 
the effective Lagrangian \cite{Dawson:1990zj}
\bea
  {\cal L}^{\rm eff}_{ggH}&=&g_{ggH}\frac{\alpha_S}{2\pi v} G^a_{\mu\nu}G^{\mu\nu}_a H\;,
\eea
where $G^a_{\mu\nu}=\partial_\mu A^a_\nu-\partial_\nu A^a_\mu-gf^{abc}A_\mu^bA_\nu^c$ 
is the gluon field-strength tensor. 
The effective coupling $g_{ggH}$ can be calculated either for finite 
top-quark mass or in the limit $m_t\to\infty$.

For the coupling of a Higgs boson to photons, mediated by top-quark and 
W-boson loops, the effective Lagrangian reads
\bea
\label{pphL}
{\cal L}^{\rm eff}_{\gamma\gamma H}&=&\frac{g_{\gamma\gamma H}}{v} 
F_{\mu\nu}F^{\mu\nu} H\;,
\eea
where $F_{\mu\nu}=\partial_\mu A^a_\nu-\partial_\nu A^a_\mu$ 
is the photon field-strength tensor.

The validation of \Sherpa's core -- providing the matrix-element
calculations -- is an important aspect in the development of the
\Sherpa Monte Carlo generator. \Amegic has successfully been tested
against various other programs for
a great variety of processes. This includes cross sections and
distributions for photon-photon collisions \cite{Bredenstein:2004ef}, 
$e^+e^-$ annihilations at different centre-of-mass energies 
\cite{Denner:1999gp,*Dittmaier:2002ap,Gleisberg:2003bi}, and 
particle production in hadron--hadron collisions \cite{MC4LHC:2003aa}. 
In Tab.~\ref{tab:mc4lhc_qcd} exemplary results
of a cross-section comparison for some LHC key processes are presented. 
For this comparison between \Amegic and the independent 
matrix-element generator \Comix, the calculational setup used
in \cite{MC4LHC:2003aa} has been employed. The results of both
codes agree very well within the statistical uncertainties, 
indicated by the numbers in parentheses, proving the correctness
of the respective calculations. 

\newcommand{\nt}{$^\dagger$}
\begin{table*}[!t]
\begin{center}
\begin{tabular}{|p{3cm}
  |a{0.9cm}{0.5cm}|a{0.9cm}{0.5cm}|a{0.7cm}{0.7cm}
  |a{0.7cm}{0.7cm}|a{0.5cm}{0.9cm}|}\hline
  $\sigma$ [pb]\vphantom{\Large P} & \multicolumn{10}{c|}{Number of jets}\\\hline
  $e^+\nu_e$ + QCD jets\vphantom{\Large P} & \multicolumn{2}{c|}{0} & \multicolumn{2}{c|}{1} 
  & \multicolumn{2}{c|}{2} & \multicolumn{2}{c|}{3} & \multicolumn{2}{c|}{4}\\\hline
  AMEGIC++ & 5432&(5) & 1279&(2)  & 466&(2) & 185&.2(5) & 77&.3(4) \\
  Comix    & 5434&(5) & 1274&(2)  & 465&(1) & 183&.0(6) & 77&.5(3) \\\hline
\end{tabular}\vspace*{3mm}
\begin{tabular}{|p{3cm}
  |a{0.7cm}{0.7cm}|a{0.7cm}{0.7cm}|a{0.5cm}{0.9cm}
  |a{0.5cm}{0.9cm}|a{0.5cm}{0.9cm}|}\hline
  $\sigma$ [pb]\vphantom{\Large P} & \multicolumn{10}{c|}{Number of jets}\\\hline
  $e^-e^+$ + QCD jets\vphantom{\Large P} & \multicolumn{2}{c|}{0} & \multicolumn{2}{c|}{1} 
  & \multicolumn{2}{c|}{2} & \multicolumn{2}{c|}{3} & \multicolumn{2}{c|}{4}\\\hline
  AMEGIC++ & 723&.0(8) & 188&.2(3) & 69&.6(2) & 27&.21(6) & 11&.1(1)  \\
  Comix    & 723&.5(4) & 187&.9(3) & 69&.7(2) & 27&.14(7) & 11&.09(4) \\\hline
\end{tabular}\vspace*{3mm}
\begin{tabular}{|p{3cm}|a{0.7cm}{0.7cm}|a{0.5cm}{0.9cm}|a{0.4cm}{1.0cm}|a{0.3cm}{1.1cm}|}\hline
  $\sigma$ [$\mu$b]\vphantom{\Large P} & \multicolumn{8}{c|}{Number of jets}\\\hline
  $jets$\vphantom{\Large P} & \multicolumn{2}{c|}{2} & \multicolumn{2}{c|}{3} 
  & \multicolumn{2}{c|}{4} & \multicolumn{2}{c|}{5}  \\\hline
  AMEGIC++ & 331&.0(4) & 22&.78(6) & 4&.98(1) & 1&.238(4) \\
  Comix    & 331&.0(4) & 22&.72(6) & 4&.95(2) & 1&.232(4) \\\hline
\end{tabular}\vspace*{3mm}
\begin{tabular}{|p{3cm}
  |a{0.7cm}{0.7cm}|a{0.7cm}{0.7cm}|a{0.7cm}{0.7cm}|}\hline
  $\sigma$ [pb]\vphantom{\Large P} & \multicolumn{6}{c|}{Number of jets}\\\hline
  $t\bar{t}$ + QCD jets\vphantom{\Large P} & \multicolumn{2}{c|}{0} & \multicolumn{2}{c|}{1} 
  & \multicolumn{2}{c|}{2} \\\hline
  AMEGIC++ & 754&.4(3) & 747&(1) & 520&(1) \\
  Comix    & 754&.8(8) & 745&(1) & 518&(1) \\\hline
\end{tabular}\vspace*{3mm}

\mytabcaption{0.99\textwidth}{
Comparison of Standard Model production cross sections at the LHC using the 
MC4LHC parameter setup \cite{MC4LHC:2003aa}. In parentheses the statistical 
error is stated in units of the last digit of the cross section.
\label{tab:mc4lhc_qcd}}
\end{center}
\end{table*}

\myfigure{t}{
  \includegraphics[width=0.48\textwidth]{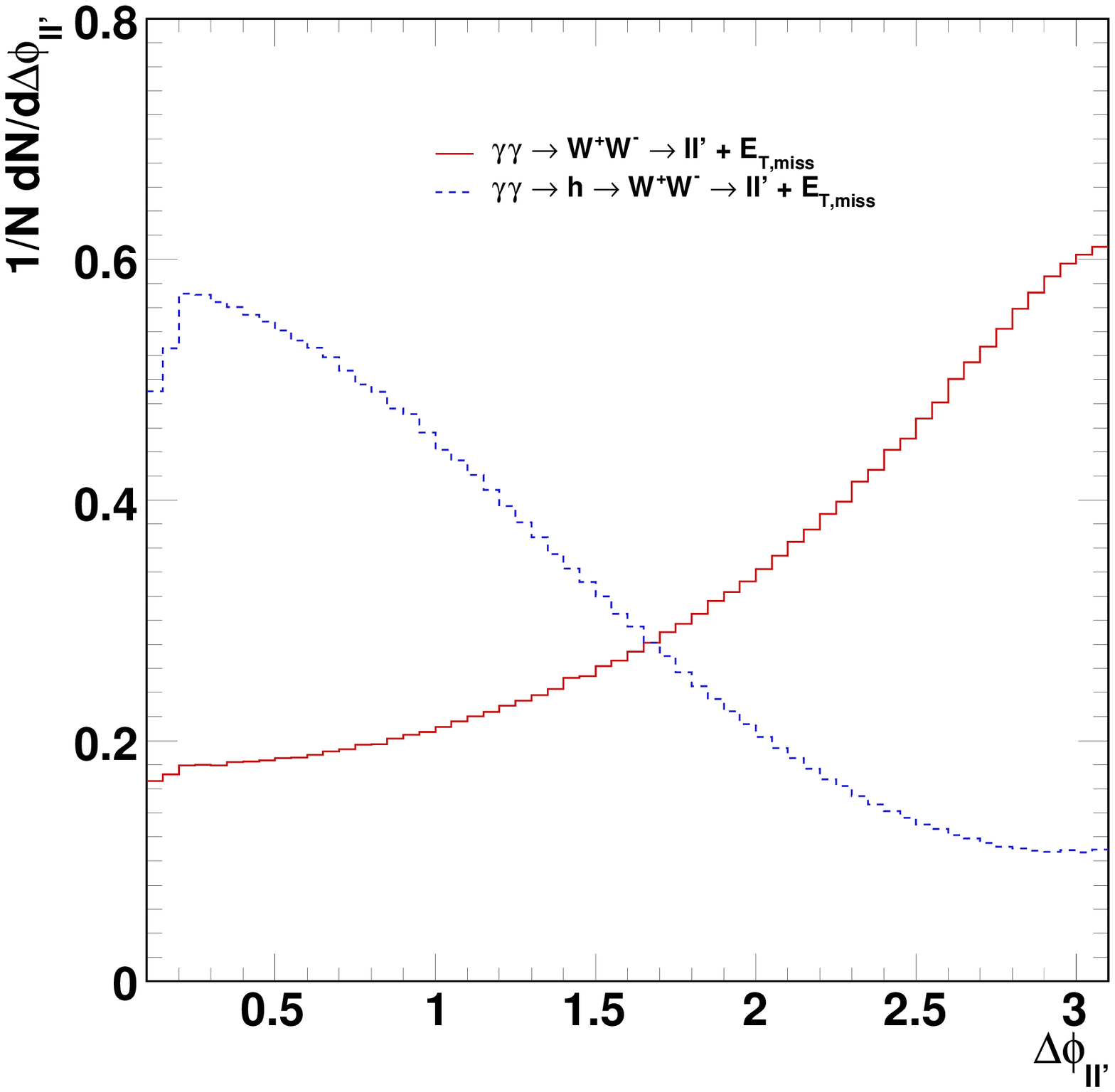}
  }{Lepton azimuthal separation in $\gamma\gamma\to W^+W^-\to ll'+\Miss{E}_T$
    with and without an intermediate Standard Model Higgs boson of
    $m_H=160$ GeV.\label{fig:dphi_ll}}

To illustrate the relevance of accounting for spin correlations we
consider the process 
$\gamma\gamma\to W^+W^-\to ll'+\Miss{E}_T$ at the CERN LHC, either 
in the continuum or mediated by a Standard Model Higgs 
boson of mass $m_H=160\,\rm GeV$. Figure~\ref{fig:dphi_ll} shows the 
azimuthal separation of all possible lepton combinations emerging 
from the $W$-decays. Apparently, this observable provides an excellent 
handle on the irreducible $W^+W^-$ background process. Note that both 
spectra have been normalised to unity in order to compare their 
respective shapes. To parametrise the initial-state photon energy, 
the spectra described in Sec.~\ref{sec:beampdf} have been employed.

\subsection[Matrix elements for BSM physics]{Matrix elements for physics beyond the Standard Model}

The implementation of new physics models into \Amegic proceeds in two 
steps. First, the new particles and the parameters of the model have to
be declared and corresponding setting routines need to be 
provided.\footnote{
  Essentially this is accomplished by reading numerical 
  values of masses and other parameters from input files.}
Second, the generic interaction vertices of the theory have to be
defined. For each new three- and four-point 
function, this includes the specification of the incoming and 
outgoing particles and the couplings, as well as the assignment of the 
$SU(3)$ colour structure of the interaction and its Lorentz structure.
Both steps are centralised in the module \Model such that the parametric 
quantities of the model can consistently be used in all 
\Sherpa modules.

The only limitation of the \Amegic approach, concerning its extension
to new physics scenarios, is the number of Lorentz structures or 
helicity-amplitude building blocks known to the program. Behind each 
Lorentz structure there resides a corresponding calculational 
method for the evaluation of a respective sub-amplitude. Although 
implementing a new building block is a well defined task, users that wish 
to do so should contact the authors for advice and help. However, there 
already exists a large variety of interaction operators for physics 
beyond the SM. In the following the new physics 
models available in \Amegic are briefly reviewed and some implementation 
details will be highlighted.

\subsubsection{The MSSM implementation}\label{sec:mssm}

For the Feynman rules of the $R$-parity conserving Minimal Supersymmetric 
Standard Model, the conventions of 
\cite{Rosiek:1989rs,*Rosiek:1995kg} are used. The general set of interaction 
vertices derived there, and as such implemented in \Amegic, include not 
only a full inter-generational mixing of the squark and slepton fields 
but also permit the inclusion of CP violating parameters. 
Furthermore, they include finite masses and Yukawa couplings for all 
three fermion generations. 

Necessary ingredients when dealing with the MSSM are specific Feynman
rules for Majorana fermions or fermion number violating interactions.  
To unambiguously fix the relative signs among Feynman diagrams involving 
Majorana spinors the algorithm described in~\cite{Denner:1992vza} is
used and the general set of fermion Feynman rules given therein has
been implemented.
Accordingly, the explicit occurrence of charge-conjugation matrices is 
avoided, instead a generalised fermion flow is employed that assigns
an orientation to complete fermion chains. This uniquely determines the 
external spinors, fermion propagators and interaction vertices involving
fermions. Furthermore, negative mass eigenvalues of the physical neutralino 
fields are taken into account at face value in the propagators and spinor 
products in the helicity-amplitude expressions. In this way a redefinition of 
the neutralino fields and couplings can be avoided. 

\begin{table*}[!t]
\begin{center}\hspace*{-7mm}
\begin{tabular}{|c||ll|ll|ll|}\hline
\multicolumn{7}{|c|}{$\bf{\sigma(W^+W^-\to X)}$ [fb]} \\\hline
 Final &  \multicolumn{2}{c|}{\Madgraph} & 
 \multicolumn{2}{c|}{\Whizard} & \multicolumn{2}{c|}{\Sherpa}\\
 state
  & \multicolumn{1}{c}{$\sqrt{s}=$0.5 TeV} 
  & \multicolumn{1}{c|}{$\sqrt{s}=$2 TeV} 
  & \multicolumn{1}{c}{$\sqrt{s}=$0.5 TeV} 
  & \multicolumn{1}{c|}{$\sqrt{s}=$2 TeV} 
  & \multicolumn{1}{c}{$\sqrt{s}=$0.5 TeV} 
  & \multicolumn{1}{c|}{$\sqrt{s}=$2 TeV} 
  \\\hline\hline 
  %%% madgraph .5+2 %%% Whizard .5+2 %%% sherpa .5+2
\tabc{\neua\neua}{\pth3.8822(2)}{\pten1.2741(4)}{\pth3.8824(1)}{\pten1.27423(8)}{\pth3.8821(2)}{\pten1.2741(1)}
\tabc{\neua\neub}{\pten121.29(1)}{\pone24.47(1)}{\pten121.2925(7)}{\pone24.472(3)}{\pten121.296(6)}{\pone24.477(1)}
\tabc{\neua\neuc}{\pth6.8936(7)}{\pone12.880(7)}{\pth6.8934(2)}{\pone12.8790(8)}{\pth6.8938(3)}{\pone12.8793(6)}
\tabc{\neua\neud}{\pth1.4974(1)}{\pten9.707(5)}{\pth1.4973(6)}{\pten9.7064(7)}{\pth1.49735(7)\hspace*{-12mm}}{\pten9.7078(4)}
\tabc{\neub\neub}{\pone5996.5(4)}{\pten1.0415(6)e3}{\pone5996.57(2)}{\pten1.04150(5)e3}{\pone5996.4(3)}{\pten1.04148(5)e3}
\tabc{\neub\neuc}{\mcent}{365.6(2)}{\mcent}{365.615(6)}{\mcent}{365.63(2)}
\tabc{\neub\neud}{\mcent}{467.8(2)}{\mcent}{467.775(8)}{\mcent}{467.77(2)}
\tabc{\neuc\neuc}{\mcent}{\pone82.35(3)}{\mcent}{\pone82.347(8)}{\mcent}{\pone82.352(4)}
\tabc{\neuc\neud}{\mcent}{138.20(5)}{\mcent}{138.18(1)}{\mcent}{138.205(7)}
\tabc{\neud\neud}{\mcent}{117.78(4)}{\mcent}{117.80(1)}{\mcent}{117.786(6)}
\hline
\tabc{\chap\cham}{\pone3772(1)}{944.3(8)}{\pone3771.6(4)}{944.2(1)}{\pone3771.8(2)}{944.32(5)}
\tabc{\chbp\chbm}{\mcent}{258.3(2)}{\mcent}{258.37(4)}{\mcent}{258.36(1)}
\tabc{\chap\chbm}{\mcent}{131.0(1)}{\mcent}{130.98(2)}{\mcent}{130.966(7)}
\hline
  \end{tabular}\vspace*{3mm}

\mytabcaption{0.99\textwidth}{
Sample cross sections (in fb) calculated using \Madgraph, \Whizard and \Sherpa 
for the production of neutral and charged gauginos in $WW$ scattering at fixed 
centre-of-mass energies of $0.5$ and $2.0$ TeV. The numbers in brackets reflect 
the absolute statistical error on the last digit. The considered supersymmetric 
spectrum corresponds to the SPS1a benchmark scenario \cite{Allanach:2002nj}, 
generated with \Softsusy \cite{Allanach:2001kg}. Additional details on 
the calculational setup can be found in \cite{Hagiwara:2005wg}.
\label{tab:catpiss}}
\end{center}
\end{table*}

Files that are conform with the SUSY Les Houches Accord (SLHA)
\cite{Skands:2003cj} are used to input the MSSM weak-scale parameters.
Such files are provided by external spectrum calculators, see for instance 
Ref.~\cite{Allanach:2001kg,*Djouadi:2002ze,*Porod:2003um}.
The parameters are translated to the conventions of 
\cite{Rosiek:1989rs} and accordingly fix the couplings of all 
interaction vertices. Note that so far \Sherpa only 
supports SLHA-v1 and the generalisations presented in 
\cite{Allanach:2008qq} have not been implemented yet. This
means that further assumptions on the physical model are implicit. 
The mixing of the squark and slepton fields is restricted to the 
third generation only and all SUSY breaking parameters and the 
Yukawa couplings are assumed to be real such that the model 
realises exact CP symmetry.

To verify the correctness of the 
implemented Feynman rules and the translation of the SLHA inputs 
to the conventions of \cite{Rosiek:1989rs}, detailed comparisons 
against the matrix-element generators \Madgraph 
\cite{Stelzer:1994ta,*Alwall:2007st} and \Whizard \cite{Kilian:2007gr}, 
have been carried out \cite{Hagiwara:2005wg}. Several hundred
cross sections for supersymmetric processes have been
compared and mutual agreement between the three independent 
codes has been found. As an example some of these results
are listed in Tab.~\ref{tab:catpiss}. Besides comparing explicit cross 
sections, thereby testing all phenomenologically relevant couplings, 
various unitarity and gauge-invariance checks were performed
and supersymmetric Ward- and Slavnov-Taylor identities were tested.

With the complete MSSM Lagrangian being available in a full-fledged 
matrix element generator like \Amegic detailed aspects of SUSY 
phenomenology can be studied. Examples include the impact of off-shell 
kinematics on MSSM signals \cite{Berdine:2007uv,Uhlemann:2008pm}, 
interference effects between various SUSY processes leading to the same 
final state (irreducible backgrounds) \cite{Hagiwara:2005wg}, or the 
impact of additional hard QCD radiation 
\cite{Plehn:2005cq,*Alwall:2008ve,Alwall:2008qv}. 
Furthermore, non-trivial production processes such as the weak-boson-fusion 
production of supersymmetric particles can easily be considered 
\cite{Cho:2006sx}.

\subsubsection{The ADD model of large extra dimensions}
The Arkani-Hamed--Dimopoulos--Dvali model \cite{ArkaniHamed:1998rs} 
extends the usual 3+1-dimensional SM by $\delta$ additional
compactified spatial dimensions where only gravity can propagate. 
In this model the usual 4-dimensional Planck scale
is related to a fundamental scale $M_S$ by
\bea
M_{\rm Pl}^2=8\pi R^\delta M_S^{\delta+2}
\eea
where $R^\delta$ is the volume of the compactified extra-di\-men\-sional space.
A sufficiently large compactification radius $R$ allows $M_S$ of the
order of $1$ TeV, providing an explanation of the mass hierarchy
between the electroweak and
the (4-dimensional) Planck scale. This has observable consequences for
TeV-scale colliders. An advantage of this model is that it introduces
only two new parameters, the scale $M_S$ and the number of extra dimensions $\delta$.

In a 4-dimensional picture the ADD model leads to the coupling of an
abundance of Kaluza-Klein graviton states to SM fields.
Feynman rules for those interactions can be derived from a linearised
gravity Lagragian \cite{Han:1998sg,Giudice:1998ck}.

\medskip
The \Amegic implementation includes all $3$- and $4$-point interactions
between gravitons and SM fields.
It features the generation of matrix elements, which include the exchange 
of virtual gravitons and the production of real gravitons.
To obtain physical cross sections this model additionally requires a 
summation (usually replaced by an integration) over the Kaluza-Klein states of 
the graviton. For the two cases this requires different approaches:
\begin{itemize}
\item Virtual graviton production:\\
The integration is performed analytically at the level of the graviton propagator.
A number of parametisations appeared in the literature,
which include a necessary cut-off for masses above some scale of the order of $M_S$.
Three widely used conventions have been implemented 
\cite{Han:1998sg,Giudice:1998ck,Hewett:1998sn}.
\item Real graviton emission:\\
The sum over accessible Kaluza-Klein states
is evaluated together with the phase-space integral using Monte Carlo techniques.
\end{itemize}
For details of the implementation, cf.~\cite{Gleisberg:2003ue}.

\myfigure{t}{
  \includegraphics[width=0.48\textwidth]{figures/add.eps}
  }{Dependence of the total cross section on the ADD scale $M_S$ for
  jet-graviton production. The SM background at the LHC is also shown. 
  A phase-space cut of $p_T^{\rm miss}> 1$ TeV is imposed. The dashed lines
  are for $M_{G_n} < M_S$, while the solid lines are for $\sqrt{\hat s}<M_S$.
  \label{fig:addexpl}}

\medskip
A simple example for the calculation of an ADD cross section 
with \Amegic is given in Fig.~\ref{fig:addexpl}, which presents
the LHC monojet cross section from ADD graviton production 
for various numbers of extra dimensions as a function
of the scale $M_S$ in the notation of \cite{Giudice:1998ck}.
Figure~\ref{fig:addexpl} also shows the SM background from $Z$ bosons
decaying into a neutrino pair.\footnote{Note that the results
  presented in Fig.~\ref{fig:addexpl} can be compared to those of
  Fig.~5(a) in \cite{Hagiwara:2008iv}.}

All new Lorentz structures implemented for the ADD model
are completely generic. This allows to use them in
other models that feature couplings to gravitons. 

\subsubsection{Anomalous interactions of the SM gauge bosons}

Besides the complete set of SM weak-gauge boson interactions \Amegic includes 
a number of effective Lagrangians describing anomalous triple and
quartic gauge 
interactions. To account for the missing UV completion of the effective theory
approach, a simple unitarisation method can be applied.

\begin{itemize}

\item\underline{$WWV$\/ interactions:}\\[1mm]
  The general set of operators describing the interaction of two charged vector 
  bosons and a neutral one,  
  \bea
      {\cal L}_{WWV}/g_{WWV}&=& 
      i g_1^V( W^\dagger_{\mu\nu}W^\mu V^\nu-W^\dagger_{\mu}V_\nu W^{\mu\nu})\nnb\\
      &&+\,i\kappa_V W^\dagger_{\mu}V_\nu W^{\mu\nu}
      +\frac{i \lambda_V}{m_W^2} W^\dagger_{\lambda\mu}W^\mu_\nu V^{\nu\lambda}\nnb\\
      &&-g_4^V W^\dagger_{\mu}W_\nu(\partial^\mu V^\nu+\partial^\nu V^\mu)
      + g_5^V \epsilon^{\mu\nu\rho\sigma}(W^\dagger_{\mu}
      \overleftrightarrow{\vphantom !}\!\!\!\!\!\partial_\rho W_\nu)V_\sigma\nnb\\
      &&+\frac{i \tilde\kappa_V}{2}\epsilon^{\mu\nu\rho\sigma}W^\dagger_{\mu} W_\nu V_{\rho\sigma}
      +\frac{i\tilde\lambda_V}{2m_W^2}\epsilon^{\mu\nu\rho\sigma}W^\dagger_{\mu\lambda} W^\lambda_\nu V_{\rho\sigma}\;,
      \label{anomalousL}
  \eea
  cf.\ \cite{Hagiwara:1986vm}, is implemented. Here $V^\mu$ denotes either the photon or the $Z$ field, 
  $W^\mu$ is the $W^-$ field, $W_{\mu\nu}=\partial_\mu W_\nu-\partial_\nu W_\mu$, 
  $V_{\mu\nu}=\partial_\mu V_\nu-\partial_\nu V_\mu$,
  $\tilde{V}_{\mu\nu}=\frac12 \epsilon_{\mu\nu\rho\sigma}V^{\rho\sigma}$ and
  $(A\overleftrightarrow{\partial_\mu} B)=A(\partial_\mu B)-(\partial_\mu A)B$.
  The overall coupling constants are given by 
  \bea
  g_{WW\gamma}=-e\quad {\rm and}\quad g_{WWZ}=-e\cot\theta_W\;.
  \eea
  The SM values of the individual couplings are $g^{Z/\gamma}_1=\kappa_{Z/\gamma}=1$
  while all others vanish.\\

\item\underline{Quadruple interactions:}\\[1mm]
  The following $SU(2)$ custodial symmetry conserving quartic interactions 
  are available:
  \bea
      {\cal L}_4&=&\alpha_4e^4\left(\frac12 W^\dagger_{\mu}W^{\dagger\mu}W_{\nu}W^{\nu}
      +\frac12(W^\dagger_{\mu}W^\mu)^2
      +\frac1{c_W^2}W^\dagger_{\mu}Z^{\mu}W_{\nu}Z^{\nu}
      +\frac1{4c_W^4}(Z^{\mu}Z^\mu)^2\right),
  \eea
  \bea
      {\cal L}_5&=&\alpha_5\left((W^\dagger_{\mu}W^\mu)^2
      +\frac1{c_W^2}W^\dagger_{\mu}Z^{\mu}W_{\nu}Z^{\nu}
      +\frac1{4c_W^4}(Z^{\mu}Z^\mu)^2\right)\,,
  \eea
  cf.\ \cite{Gangemi:1999gt}. In the SM limit the parameters $\alpha_4$ and $\alpha_5$
  are identical zero.\\

\item\underline{$\gamma$-$Z$\/ interactions:}\\[1mm]
  The general anomalous coupling of two on-shell neutral gauge bosons ($V_1$ and $V_2$) to 
  an off-shell boson ($V_3$) has been given in \cite{Hagiwara:1986vm}. The corresponding 
  Feynman rule is displayed in Fig.~\ref{nagcv}. The vertex functions $\Gamma_{V_1V_2V^*_3}$ 
  for $V_2=Z$ and $V_2=\gamma$ are given in Eqs.~(\ref{eq:GZZV}) and (\ref{eq:GZPV}), 
  respectively. In the SM all coupling parameters, i.e.\ $f_4^V,\,f_5^V,\,h_i^V$, equal zero.
  \begin{equation}
    \begin{split}
      &\Gamma_{ZZV^*}^{\alpha\beta\mu}(q_1,q_2,P)=\frac{i\left(P^2-m_V^2\right)}{m_Z^2}
      \left[f_4^V(P^\alpha g^{\mu\beta}+P^\beta g^{\mu\alpha})
      +f_5^V\epsilon^{\mu\alpha\beta\rho}(q_1-q_2)_\rho\right]\label{eq:GZZV}\\
    \end{split}
  \end{equation}
  \begin{equation}
    \begin{split}
      &\Gamma_{Z\gamma V^*}^{\alpha\beta\mu}(q_1,q_2,P)=\frac{i\left(P^2-m_V^2\right)}{m_Z^2}
      \left\{\vphantom{\frac{h_4^V}{m_Z^2}}\;h_1^V(q_2^\mu g^{\alpha\beta}-q_2^\alpha g^{\mu\beta})\right.\\
      &\qquad\quad+\left.\frac{h_2^V}{m_Z^2}P^\alpha\left[(Pq_2)g^{\mu\beta}-q_2^\mu P^\beta\right]
      +h_3^V\epsilon^{\mu\alpha\beta\rho}q_{2\rho}
      +\frac{h_4^V}{m_Z^2}P^\alpha\epsilon^{\mu\beta\rho\sigma}P_\rho q_{2\sigma}
      \right\}\label{eq:GZPV}
    \end{split}
  \end{equation}
  \begin{figure}[!t]
    \begin{center}
      {
        \SetScale{1.}
        \begin{picture}(300,100)
          \Vertex(100,50){3}
          \Photon(60,50)(100,50){2}{5}
          \Photon(100,50)(130,20){2}{5}
          \Photon(100,50)(130,80){2}{5}
          \ArrowLine(85,55)(86,55)\Line(75,55)(85,55)
          \ArrowLine(114,71)(115,72)\Line(107,64)(114,71)
          \ArrowLine(114,29)(115,28)\Line(107,36)(114,29)
          \Text(20,50)[l]{$V_{3\mu}(P)$}
          \Text(120,90)[l]{$V_{1\alpha}(q_1)$}
          \Text(120,10)[l]{$V_{2\beta}(q_2)$}
          \Text(150,50)[l]{$=ie\Gamma_{V_1V_2V^*_3}^{\alpha\beta\mu}(q_1,q_2,P)$}
        \end{picture}
      }\\
      \myfigcaption{0.75\textwidth}{Feynman rule for the anomalous triple neutral gauge boson vertex.\label{nagcv}}
    \end{center}
  \end{figure}
  \\
  It should be noted that the most general anomalous coupling between three off-shell
  neutral gauge bosons allows more coupling terms \cite{Gounaris:2000dn}, which 
  are, however, not implemented in the current version. To account for this a symmetrised 
  version of the above vertex is used outside the on-shell limit of two of the vector bosons. 
  For example, the interaction of three off-shell $Z$ bosons is modelled through 
  $\Gamma_{Z^*Z^*Z^*}=\Gamma_{Z_1Z_2Z^*_3}+\Gamma_{Z_2Z_3Z^*_1}+\Gamma_{Z_3Z_1Z^*_2}$.\\

\item\underline{Unitarisation:}\\[1mm]
  Owing to the effective nature of the anomalous couplings unitarity 
  might be violated for coupling parameters other than the SM values.
  For very large momentum transfers, such as those probed at the LHC, this
  will lead to unphysical results. As discussed in \cite{Baur:1988qt}
  this can be avoided by introducing energy dependent form factors for the 
  deviation of coupling parameters from their Standard Model values:
  \bea
  a(\hat s)=\frac{a_o}{(1+\hat s/\Lambda^2)^n}
  \eea
  where $\hat s$ is the partonic centre-of-mass energy, $\Lambda$ represents 
  the new physics scale, and the exponent $n$ is typically chosen to be $n=2$.

\end{itemize}

\subsubsection{Further BSM models}

Besides the major model implementations described above \Amegic 
offers various other physics scenarios that shall briefly be
listed here:

\begin{itemize}
\item \underline{Two-Higgs-Doublet model:}\\[1mm]
	As byproduct of the MSSM implementation the THDM of type II can be studied. 
	In contrast to the MSSM, the relevant input parameters of the extended Higgs 
	sector (masses of the Higgs bosons $h^0$, $H^0$, $A^0$, and $H^\pm$, 
	the mixing angle between the two doublets, $\alpha$, and the ratio of their
	vevs, $\tan\beta$) have to be specified explicitly.\\
\item \underline{Phantom-Higgs model:}\\[1mm]
	This model has been discussed in \cite{Dedes:2008bf}.  It emerges by adding a
	complex scalar gauge singlet to the SM, which interacts with the SM particle
	through a mixed quartic coupling with the original Higgs doublet of the SM.
	After symmetry breaking, a massless pseudoscalar and a massive scalar augment
	the original SM particle spectrum.  The latter will undergo a mixing with the
	original Higgs boson, therefore dilute its couplings to the other SM 
	particles through a mixing parameter.  On the other hand, the massless pseudoscalar
	does not interact with the SM world, apart from the Higgs boson, leading to
	a sizable invisible partial width of both scalars.  The relevant parameters in
	this model are the masses of the two scalars, their mixing angle, $\tan\theta$,
	and the ratio of vevs, $\tan\beta$.\\
\item \underline{4th generation:}\\[1mm]
	In this model a fourth generation family has been included, i.e.\ a lepton $\ell_4$,
	a neutrino $\nu_4$, an up-type and a down-type quark $u_4$ and $d_4$, respectively.
	They are parametrised by their masses, which also serve as input parameters.
	In addition, both may decay to the Standard Model particles through the charged
	current weak interaction where the mixing to the other generations is parametrised
	by one mixing angle each.  In the lepton case, this implies that all $\ell_4$ and
	$\nu_4$ must decay into a $\tau$-neutrino or a $\tau$-lepton, respectively, while
	the pattern in the quark case is a bit more involved.  There the mixing matrix
	emerges by multiplying the original $3\times 3$ CKM matrix with a 
	rotation matrix in the $34$-sector, cf.\ \cite{Botella:1985gb,*Hou:1987ya,*Hou:1987vd}.\\
\item \underline{Axigluons:}\\[1mm]
	Here, an axigluon, i.e.\ a massive axial vector with the colour quantum numbers
	and interactions of the gluon is added to the Standard Model, cf.\
	\cite{Pati:1975ze,*Hall:1985wz,*Frampton:1987dn,*Frampton:1987ut,*Bagger:1987fz}.
	The only input parameter is the mass of this particle.  The implications of such 
	a particle have been discussed, e.g., in \cite{Antunano:2007da}.
\end{itemize}

There exist further (though private) implementations of alternative
physics scenarios that are not publicly available yet but can be
obtained from the respective authors on request, see e.g.\ Refs.~\cite{Agashe:2006hk,Kilic:2008ub}. Note that to further alleviate the 
implementation of new physics ideas and to allow for an easy distribution an interface 
to the \FeynRules package \cite{Christensen:2008py} is currently being developed.
\section{Initial- and final-state radiation}
\label{sec:shower}
QCD parton evolution and the occurrence of jets can be understood theoretically when the 
structure of perturbative amplitudes is examined in the kinematical
regime of intrajet evolution, i.e.\ where two or more partons get
close to each other in phase space. Whenever this happens,
any QCD matrix element squared factorises into a matrix element squared containing the combined ``mother'' 
parton and a universal function describing the splitting into the ``daughters''. 
In this limit, the theory becomes semi-classical and can be understood in a Markovian approach, 
where a single initiating parton develops a cascade of independent branchings.
This is the basic concept of any shower Monte Carlo. 
Potential differences then arise in the factorisation scheme only.
The default shower generator within \Sherpa is \Apacic~\cite{Kuhn:2000dk,Krauss:2005re}.
It is essentially based on virtuality ordered DGLAP parton evolution with superimposed angluar
ordering constraints.

\subsection{The parton cascade \tops{\Apacic}{APACIC}}
\label{sec:apacic}
Final-state showering in \Apacic proceeds along the lines of 
\cite{Bengtsson:1986et}. $Q^2$ evolution of QCD partons is simulated by
$1\to2$ splittings, which occur with differential probability 
\begin{equation}\label{eq:diff_prob_apacic_fs}
  \begin{split}
  \frac{\done\mc{P}^{(F)}_{{\rm branch},a}}{\done t}=\frac{\done}{\done t}\,
    \exp\left\{-\int_t^{t'}\frac{\done \bar t}{\bar t}
      \int_{\tilde{z}_{\rm min}(\bar t\,)}^{\tilde{z}_{\rm max}(\bar t\,)}
      \done\tilde{z}\;
    \frac{\alpha_s({\rm k}_\perp(\tilde{z},\bar t\,))}{2\pi}\,
      \sum_{b=q,g}P_{ab}(\tilde{z})\right\}\;.
  \end{split}
\end{equation}
Here $a$ denotes the flavour of the splitting parton and $P_{ab}(z)$ 
are the unregularised Altarelli--Parisi kernels in four dimensions for 
the splitting $a\to bc$. The transverse momentum ${\rm k}_\perp$ is given
with respect to the axis defined by the direction of the decayer.
The condition ${\rm k}_\perp>{\rm k}_{\perp,\rm min}$, with ${\rm k}_{\perp,\rm min}$
some cutoff value for ${\rm k}_\perp$, yields the boundaries of the 
$\tilde{z}$-integral, $\tilde{z}_{\rm min}$ and $\tilde{z}_{\rm max}$, cf.~\cite{Krauss:2005re}.
Evolution and splitting variable are defined by%~\cite{Bengtsson:1986et}
\begin{align}\label{eq:definition_variables_apacic_fs}
  t\,=&\;p_a^2-m_a^2\;,
  &&{\rm and}
  &\tilde{z}\,=&\;\frac{E_b}{E_a}\;,
\end{align}
respectively, $m_a$ being the on-shell mass of parton $a$. The splitting variable $\tilde{z}$ 
is related to the light-cone momentum fraction $z=p_b^+/p_a^+$ (with the 
``+'' direction defined by $p_a$) through
\begin{align}
  2\,\tilde{z}\kappa E_a^2\,+z\,(t_a-\kappa^2 E_a^2)\,=&\;t_a+t_b-t_c
  &&{\rm where}
  &\kappa\,=&\;1+\sqrt{1-\frac{t_a}{E_a^2}} \;\;.
\end{align}
Although defined in an apparently non-covariant way, the splitting variable 
actually is Lorentz invariant~\cite{Bengtsson:1986et}. 
Colour coherence during evolution is taken into account by an 
explicit angular veto, which means that a branching is rejected if 
the opening angle of the emission is larger than the one of the 
previous branching.

Initial-state showering proceeds in the backward-evolution picture
along the lines of \cite{Sjostrand:1985xi}. The differential branching
probability reads
\begin{equation}\label{eq:diff_prob_apacic_is}
  \begin{split}
  \frac{\done\mc{P}^{(I)}_{{\rm branch},a}}{\done t}=\frac{\done}{\done t}\, 
    \exp\left\{-\int_t^{t'}\frac{\done \bar t}{\bar t} 
      \int_{x}^{\bar{z}_{\rm max}(\bar t\,)}
      \frac{\done\bar{z}}{\bar{z}}\;
      \frac{\alpha_s({\rm k}_\perp(\bar{z},\bar t\,))}{2\pi}\, 
      \sum_{b=q,g}P_{ba}(\bar{z})\,\frac{f_b(x/\bar{z},\bar t)}{f_a(x,\bar t)}
      \right\}
  \end{split}
\end{equation} 
where the ratio $f_b(x/\bar{z},\bar t)/f_a(x,\bar t)$ accounts for the change 
of parton distributions in each shower step. The splitting variable $\bar{z}$ 
can be reinterpreted as 
\begin{equation}\label{eq:definition_z_apacic_is}
  \bar{z}\,=\;\frac{\hat{s}'}{\hat{s}}\;,
\end{equation}
$\hat{s}$ and $\hat{s}'$ being the partonic center of mass energies before
and after the branching, respectively. This immediately yields the relation
$x'=x/\bar{z}$, thus is partially defining the kinematics after the
branching.

An important issue for DGLAP shower algorithms is the convention to implement
kinematic constraints once a splitting generates recoil owing to the branching 
parton going off mass-shell. The recoil strategy seems ambiguous 
because the branching equations are independent of it. In fact, for the 
derivation of the DGLAP equation, it may be assumed that there is a spectator 
parton aligned along the same axis as the splitter, but with opposite
direction. This leads to the following approach for \Apacic:
\begin{itemize}
\item In final-state branchings the parton, which originates from the
  same splitting as the branching parton, takes the recoil. If this
  parton has been already decayed, the decay products are boosted
  accordingly. This amounts to redefining the splitting variable of
  the respective branching by
  \begin{equation}
    \begin{split}
    &\tilde{z}\to\tilde{z}^{\,\prime}\,=\;
      \rbr{\tilde{z}-\frac{t_a+t_b-t_c}{2\,t_a}}
      \sqrt{\frac{(t_a-t_b'-t_c')^2-4\,t_b'\,t_c'}{(t_a-t_b-t_c)^2-4\,t_b\,t_c}}
      +\frac{t_a+t_b'-t_c'}{2\,t_a}
    \end{split}
  \end{equation}
  where $t$ and $t'$ denote original and reassigned virtualities, 
  respectively.
\item In initial-state branchings all remaining partons take the recoil.
  For any splitting $b\to a$, the process is redefined with parton $b$ 
  rather than $a$ aligned along the beam axis and $\hat s\to\hat
  s/\bar z$.
\end{itemize}
The Altarelli--Parisi splitting functions, $P_{ab}(z)$, are taken in
the quasi-collinear limit and include mass effects of the emitting
parton, cf.\ \cite{Catani:2000ef},
\begin{equation}\label{eq:dglap_kernels_apacic}
  \begin{split}
  P_{qq}(z)\,=&\;C_F\,\sbr{\,\frac{1+z^2}{1-z}-2\mu_{qg}^2\,}\,,\\
  P_{gq}(z)\,=&\;T_R\,\sbr{\,z^2+(1-z)^2-\mu_{q\bar{q}}^2\,}\,,\\
  P_{gg}(z)\,=&\;C_A\,\sbr{\,\frac{z}{1-z}+\frac{1-z}{z}+z\,(1-z)\,}\;.
  \end{split}
\end{equation}
Quark-mass dependencies are given in terms of the dimensionless variable
\begin{equation}\label{eq:def_mass_terms_apacic}
  \mu_{ij}^2=\frac{m_i^2+m_j^2}{(p_i+p_j)^2-m_{ij}^2}\;.
\end{equation}
\Apacic has been thoroughly tested and validated~\cite{Krauss:2005re}.
Modifications to the shower algorithm, which were necessary for a
proper implementation of the CKKW merging procedure, have been
implemented in full generality, see Sec.~\ref{sec:ckkw}.

\subsection{Showering off heavy resonances}
\Apacic is also equipped with the possibility to generate radiation 
off intermediate heavy resonances, once these are described by
separable production and decay processes, cf. Sec.~\ref{sec:mes}. An
example is top-quark pair production, which plays a significant role,
both as a signal for a better measurement of Standard Model parameters
and as a background to new physics searches.

The QCD radiation pattern in heavy-flavour decays has been thoroughly 
investigated in Refs.~\cite{Corcella:1998rs,*Hamilton:2006ms}.
Within \Apacic a rather simple strategy is employed. QCD radiation 
off the decaying heavy particle is described by the standard parton shower
with massive splitting functions, except for two modifications:
\begin{enumerate}
\item In ordinary final-state parton showering, the mother particle
  goes off-shell, while the daughters retain their respective on-shell 
  masses. In showering off decaying heavy particles, on the contrary, the
  mother particle retains its on-shell mass, while the daughter of the
  same flavour goes off-shell with decreased virtuality.
\item The maximally allowed phase-space volume in a branching process
  of a decaying particle is reduced by the factor 
  \begin{equation} 
    w_{\rm PS}=\frac{\done\Phi^{(2)}(t',t_b,t_c)}{\done\Phi^{(2)}(t,t_b,t_c)}
      =\sqrt{\frac{t^3}{t^{\prime 3}}\,\frac{\lambda(t',t_b,t_c)}{\lambda(t,t_b,t_c)}}
  \end{equation} 
  where $\lambda(a,b,c)=(a-b-c)^2-4bc$, $t$ and $t'$ are the virtualities of the decaying particle 
  before and after the emission, respectively, and $t_b$ and $t_c$ are
  the virtualities of the decay products. This correction weight
  corresponds to a decrease in phase-space volume $\done\Phi^{(2)}$ owing to a decrease
  in three-momentum $|{\rm p}_{a/b}^{\rm cm}|$ in the centre-of-mass
  frame of the decayed parton.
\end{enumerate}
To correctly describe the decay process of the heavy flavour, it is vital
to respect spin correlations between production and decay amplitudes.
Within \Sherpa, this is done by firstly computing the full matrix element 
for production and decay of the heavy flavour and subsequently adding
in the parton evolution of the intermediate state.
In this respect, it has to be defined, how the kinematics of the decay
products is to be reconstructed, once a parton emission has occurred in
the shower. The following strategy is used:
\begin{enumerate}
\item If the decaying heavy particle keeps its mass, i.e.\ if the 
  radiation occurs in the production part of the process, the decay
  products are simply boosted into the new centre-of-mass frame of the
  decayer.
\item If the decaying heavy particle does not keep its mass, 
  i.e.\ if the radiation occurs in the decay part of the process, the
  decay products are reconstructed such that in the centre-of-mass
  frame of the daughter the momenta point into the same direction as
  they did in the centre-of-mass frame of the mother before. 
\end{enumerate}
This procedure largely retains the correlations between
the final-state particles initially described by the matrix element.

\section{Combining LO matrix elements and parton showers}
\label{sec:ckkw}
One of the most important challenges for simulating events at 
modern high-energy collider experiments is the accurate 
theoretical description of multijet final states. They constitute
the testbed for many new-physics searches, whose success will
largely depend on our understanding of the Standard Model multijet 
production mechanisms. The best theoretical tools available are
therefore needed to approach this problem as accurately as possible
and obtain reliable estimates of the Standard Model backgrounds to
experimental analyses. Underlying hard processes should be accounted
for by full matrix-element calculations and the subsequent evolution
and conversion of hard partons into hadronic jets should be modelled
by QCD parton cascades and phenomenological hadronisation models.
However, there are several scales involved that determine the thorough
development of an event which makes it difficult to unambiguously
disentangle the components belonging to the hard process and the
hard-parton evolution. Given an $n$\/ jet event of well separated
partons, its jet structure is retained when emitting a further
collinear or soft parton only. An additional hard, large-angle
emission, however, gives rise to an extra jet changing the $n$\/ to an
$n+1$ jet final state. Accordingly a merging scheme for matrix element
calculations and parton showers has to be defined, that determines on
an event-by-event basis which possibility has to be followed. The
primary goal is to avoid double counting by preventing events to
appear twice, i.e.\ once for each possibility, as well as dead regions
by generating each configuration only once and using the appropriate
path.

Basically two different solutions for this problem have been proposed and 
implemented in the past few years, the MLM scheme 
\cite{Mangano:2001xp,*Mangano:2006rw}, relying on a geometric analysis 
of the unconstrained radiation pattern in terms of cone jets, and the 
CKKW scheme or CKKW merging algorithm \cite{Catani:2001cc,Krauss:2002up}, 
employing an analytical reweighting of the matrix elements supplemented 
by a constrained parton-shower evolution. A reformulation of CKKW to a 
merging procedure in conjunction with a dipole shower (CKKW-L) has been 
presented in \cite{Lonnblad:2001iq,*Lavesson:2005xu}. Common to all
schemes is that sequences of tree-level multileg matrix elements with
increasing final-state multiplicity are merged with parton showers to 
yield a fully inclusive sample with no double counting and correct at 
leading-logarithmic accuracy. On the theoretical side there have been 
various studies to compare the different approaches, see e.g. 
Refs.~\cite{Mrenna:2003if,*Lavesson:2007uu,Hoche:2006ph,Alwall:2007fs}, 
and the first dedicated experimental analyses on comparing these improved 
Monte Carlo predictions with actual data have been presented 
\cite{Abazov:2006gs,*Chung:2006dv,*Voutilainen:2006qq,*Hesketh:2006qu,%
  *Hegeman:2007zz,*Abazov:2008ez,*Neu:2008cr,%
  *Harel:2008px,*Abazov:2008yg,DOnote5066,Aaltonen:2007ip}.
However, there is more work needed, on the one hand to understand the
systematic uncertainties of the various schemes, and, on the other
hand further experimental inputs are needed to validate the available
tools. This is crucial in prospect of the LHC providing first
measurements soon.

The \Sherpa generator provides a general and largely process-independent
implementation of the CKKW algorithm \cite{Schalicke:2005nv}, which has
extensively been validated for $e^+e^-$ collisions into hadrons 
\cite{Schalicke:2005nv,Krauss:2005re}, the production of single vector 
bosons at the Fermilab Tevatron \cite{Krauss:2004bs} and the CERN LHC 
\cite{Krauss:2005nu} and for $W^+W^-$ production at hadron colliders, 
see \cite{Gleisberg:2005qq}. The combined treatment of matrix elements
and parton showers is supplemented by a multiple-interactions 
description, which respects the jet-production scales of the primary 
process. In the following, these key features of \Sherpa shall be
discussed in more detail. The actual merging algorithm employed will
be presented in Sec.~\ref{merge_algo}. Aspects of its application will
be discussed in Sec.~\ref{merge_appl}. 

\subsection{The algorithm implemented in \protect\Sherpa}
\label{merge_algo}

The idea underlying CKKW is to divide the phase space of partonic 
emissions according to a $k_T$ measure 
\cite{Catani:1991hj,*Catani:1993hr,*Catani:1992zp} into a regime of 
jet production, described by appropriate matrix elements,
and a regime of jet evolution, described by parton showering. The
separation is defined by the merging scale, denoted by $Q_{\rm cut}$.
The matrix elements are reweighted by $\alpha_s$ coupling factors and
terms that arise from analytic Sudakov form factors. The
acceptance or rejection of jet configurations is realised according
to this reweighting. Then, each hard parton of the reweighted
matrix-element final state undergoes vetoed parton showering, i.e.\
starting from the scale where this parton appeared first, any new
emission that would give rise to an extra jet is vetoed. In this way,
the dependence on the, in principle, arbitrary separation scale 
$Q_{\rm cut}$ regularising the matrix elements is mostly eliminated 
and the accuracy of the parton shower is preserved. However, the
cancellations are exact only up to at most next-to-leading logarithmic order.
This leaves some unavoidable residual dependence, which can be used 
to tune the procedure and obtain optimal agreement with data.

In \Sherpa the CKKW merging of matrix elements and parton showers
is accomplished as follows:
\begin{enumerate}
\item\label{ckkw_step_1} All cross sections $\sigma_k$ for processes with
  $k=0,1,\ldots,N$\/ extra partons are calculated with the constraint
  that the matrix-element final states pass the jet criteria. They
  are determined by a $k_T$ measure and the minimal distance is set by
  the actual merging scale $Q_{\rm cut}$. Beyond that, $Q_{\rm cut}$
  also acts as a regulator setting the factorisation (PDF) as well as
  the renormalisation ($\alpha_s$) scales of the matrix-element
  calculations. The $k_T$ measure used for jet identification in
  electron-positron collisions can be written as~\cite{Catani:1991hj}
  \begin{equation}
    Q_{ij}^2\;=\;2\min\{E^2_i,E^2_j\}\,\rbr{\vphantom{\sum}1-\cos\theta_{ij}}\,,
  \end{equation}
  and quantifies the $k_T$ distance between the final-state particles
  $i$\/ and $j$. The jet cuts are satisfied if $Q_{ij}>Q_{\rm cut}$. 
  For hadron--hadron collisions, a $k_T$
  scheme is employed, which defines two final-state particles to
  belong to two different jets, if their relative transverse momentum
  squared, defined as
  \begin{equation}
    \begin{split}
    Q^2_{ij}\;=&\;2\min\left\{m_T^{(i)},\,m_T^{(j)}\right\}^2
    \;\frac{\cosh(y^{(i)}-y^{(j)})-\cos(\phi^{(i)}-\phi^{(j)})}
	 {D^2}\label{eq:sherpakt}
    \end{split}
  \end{equation}
  is larger than the critical value $Q^2_{\rm cut}$. In addition, the
  transverse momentum of each jet has to be larger than the merging
  scale $Q_{\rm cut}$. The magnitude $D$, which is of order $1$, is a
  parameter of the jet algorithm, see \cite{Blazey:2000qt}.
\item Processes of fixed parton multiplicity are chosen with
  probability $\sigma_k/\sum^N_{l=0}\sigma_l$. The event's hard
  process is picked from the list of partonic processes having the
  desired multiplicity and according to their particular cross-section
  contributions. All particle momenta are distributed respecting the
  correlations encoded in the matrix elements. Merged samples
  therefore fully include lepton-jet and jet--jet correlations up to
  $N$\/ extra jets.
\item The parton configuration of the matrix element has to be
  analysed to eventually accomplish the reweighting. The partons are
  clustered backwards according to the same $k_T$ jet clustering
  algorithm used for the regularisation of the final-state phase space
  of the matrix elements (cf.\ step~\ref{ckkw_step_1}). The clustering is 
  guided by the physically allowed parton combinations and automatically
  yields the nodal $k_T$ values $Q_{ij}$ of each parton emission. It
  is stopped after a $2\to2$ configuration (a core process) has been
  identified. In fact, this backward clustering constructs
  a limit of leading logarithmic accuracy of the full radiation
  pattern, i.e.\ determines a possible parton-shower history. The
  sequence of clusterings can thus be taken as a pseudo shower
  configuration, off which the event's evolution will be properly
  continued by parton showering.
\item The reweighting proceeds according to the reconstructed shower
  history. For the strong-coupling weight, the identified nodal $k_T$
  values are taken as scales in the strong-coupling constants and
  replace the predefined choice of the initial generation. The Sudakov
  weight attached to the matrix elements accounts for having no
  further radiation resolvable at $Q_{\rm cut}$. The NLL Sudakov form
  factors employed, cf.\ \cite{Catani:1991hj}, are defined by
  \begin{eqnarray}
    \Delta_q(Q,Q_0) &=&
    \exp\left\{-\int\limits_{Q_0}^{Q} dq \,\Gamma_q(Q,q)\right\} \;,\\
    \Delta_g(Q,Q_0) &=&
    \exp\left\{-\int\limits_{Q_0}^{Q} dq
    \left[ \Gamma_g(Q,q) + \Gamma_f(q) \right]\right\} \;,\label{eq:sud}
  \end{eqnarray}
  where $\Gamma_{q,g,f}$ are the integrated splitting functions for
  $q\to qg$, $g\to gg$\/ and $g\to q\bar q$ ($f\bar f$), given by
  \begin{eqnarray}
    \Gamma_q(Q,q) &=& \displaystyle\frac{2 C_F}{\pi}\frac{\alpha_s(q)}{q}
    \left( \ln \frac Q q - \frac 3 4 \right) \;,\\
    \Gamma_g(Q,q) &=& \displaystyle\frac{2 C_A}{\pi}\frac{\alpha_s(q)}{q}
    \left( \ln \frac Q q - \frac{11}{12} \right) \;,\\
    \Gamma_f(q)   &=& \displaystyle\frac{N_f}{3\pi}\frac{\alpha_s(q)}{q}\;,
  \end{eqnarray}
  respectively. $\Gamma(Q,q)$ is cut off at zero, such that
  $\Delta_{q,g}(Q,Q_0)$ retains its probability interpretation for
  having no emission resolvable at scale $Q_0$ during the evolution
  from $Q$ to $Q_0$. Hence, $\Delta$-factors are used to reweight
  in accordance to the appearance of external parton lines. The ratio
  of two Sudakov form factors $\Delta(Q,Q_0)/\Delta(q,Q_0)$ accounts
  for the probability of having no emission resolvable at $Q_0$ during
  the evolution from $Q$ to $q$. Ratio factors thus are employed for
  the reweighting of internal parton lines. In both cases the lower limit
  is taken to be $Q_0=Q_{\rm cut}$ or $Q_0=D\,Q_{\rm cut}$ for partons
  that are clustered to a beam or to another final-state parton,
  respectively.
  \mywidefigure{t!}{
    \includegraphics[width=0.45\textwidth]{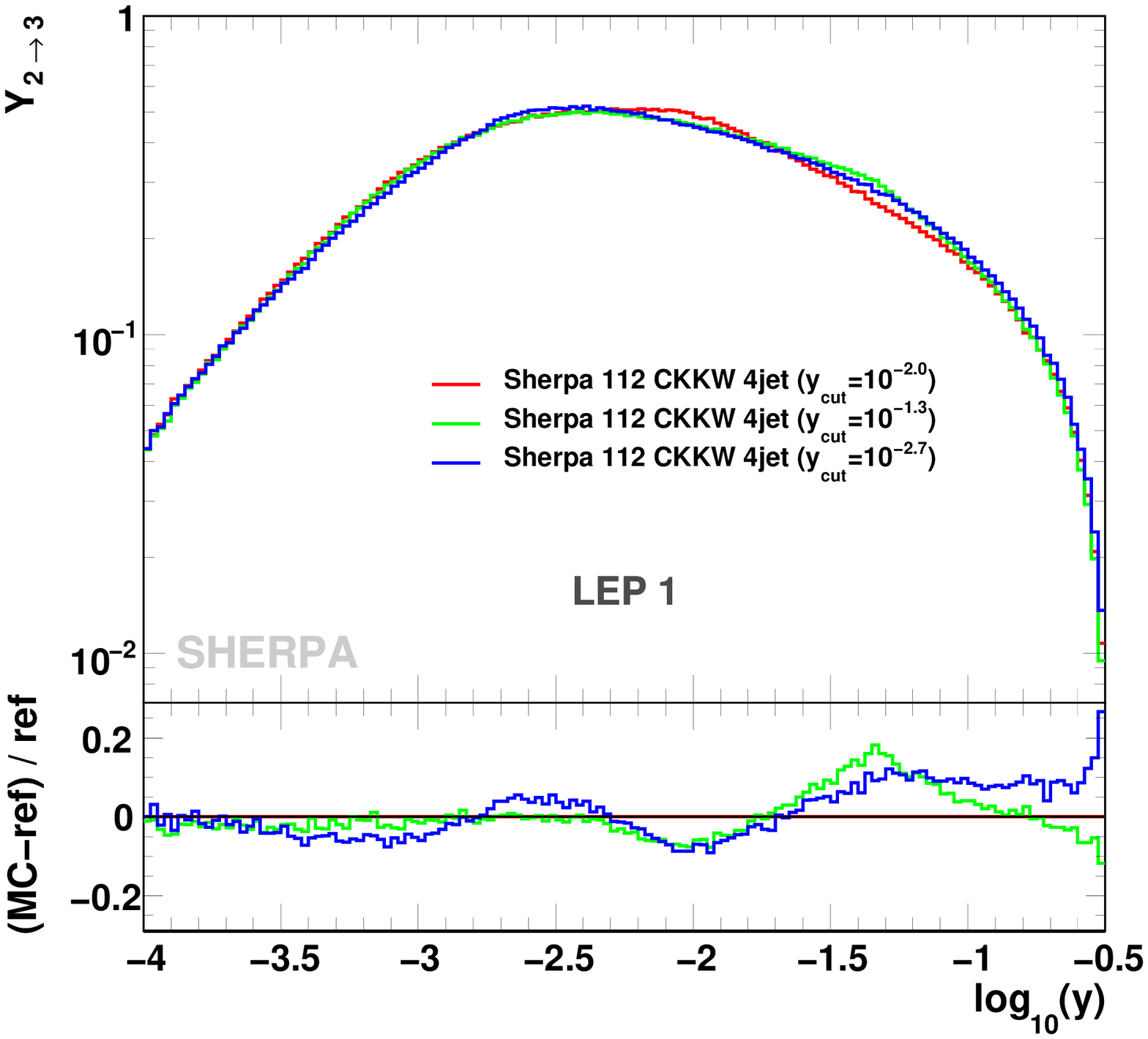}\hspace*{5mm}
    \includegraphics[width=0.45\textwidth]{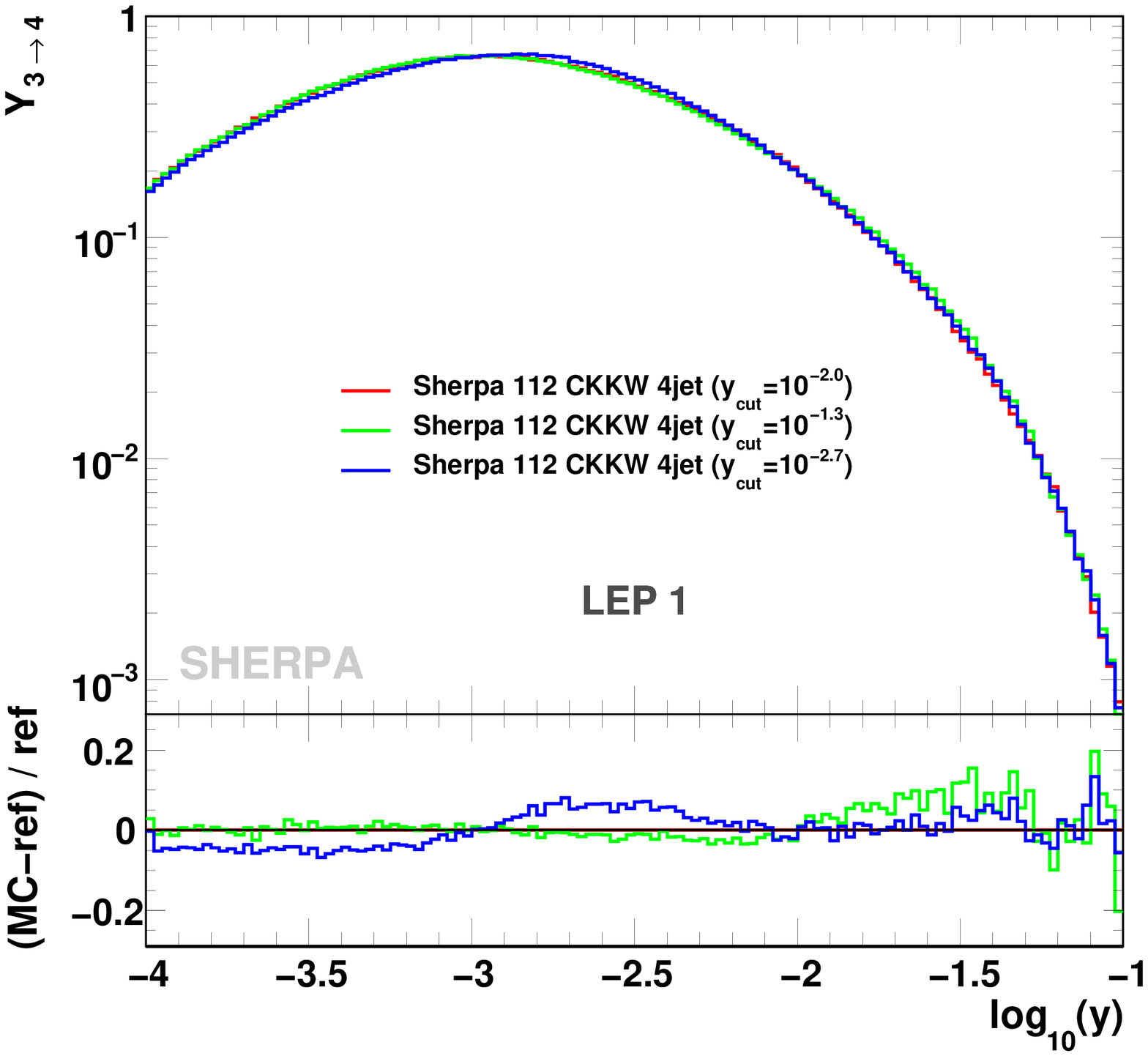}
  }{Durham differential $2\to3$ (left) and $3\to4$ (right) jet rates as
    a function of the jet-resolution parameter $y$ for different
    merging-scale choices $y_{\rm cut}$.\label{fig:y234}}
\item After the reweighting of the matrix-element configuration, the
  parton shower is invoked starting from the pseudo history
  constructed before. Each parton continues its evolution at the scale
  where it was produced. For the virtuality-ordered parton shower of
  \Apacic, this scale is given by the invariant mass of the mother
  parton belonging to the QCD splitting, through which the considered
  parton has been initially formed. Recall that this QCD splitting is
  identified by the ``cluster-backwards'' procedure. For the cases
  where the considered leg originates from the core process, the (hard)
  scale still needs to be determined. For example, all four partons
  resulting from a clustering that terminated in a pure QCD $2\to2$
  process will commence their evolution at the corresponding hard QCD
  scale.
\item In all circumstances parton-shower radiation is subject to the
  condition that no extra jet is produced. Any emission that turns out
  to be harder than the separation cut $Q_{\rm cut}$ is vetoed. The
  exception to this veto -- called highest-multiplicity treatment --
  is for matrix-element configurations with the maximal number $N$\/
  of extra partons. These cases require the parton shower to cover the
  phase space for more jets than those produced by the matrix
  elements. To obtain an inclusive $N$-jet prediction, the veto
  therefore is on parton emissions at scales harder than the softest
  clustering scale, $Q_{\rm softest}\ge Q_{\rm cut}$. Of course,
  correlations including the $N+1$st jet are such only approximately 
  taken into account.
\end{enumerate}
As a final remark it is worthwhile to note that the procedure described
here has been fully automated, i.e.\ in \Sherpa the merging of the
different jet multiplicities is handled on the fly.

\subsection{Applications}
\label{merge_appl}
\subsubsection{\tops{$e^+e^-\to$}{e+ e- ->} hadrons}
To exemplify the working of the algorithm described above one may
consider the simple example of jet production in electron--positron
annihilation.

The lowest-order matrix elements account for the $k=2$ processes
$e^+e^-\to q\bar q$. They all have the same topology and the
corresponding Sudakov weights read
\begin{equation}
  W_2\;=\;\left[\Delta_q(E_{\rm cm},Q_{\rm cut})\right]^2\,.
\end{equation}
The vetoed parton showers for both the quark and the antiquark
start at the scale equal to the centre-of-mass energy $E_{\rm cm}$ of
the $e^+e^-$ annihilation. A veto is applied for emissions above 
$Q_{\rm cut}$. There are two topologies for the $k=3$ subprocesses, 
i.e.\ for $e^+e^-\to q\bar qg$, corresponding to the emission of a
gluon off the quark or antiquark, respectively. The related Sudakov 
weights read
\begin{eqnarray}
  W_3 &=& \Delta_q(E_{\rm cm},Q_{\rm cut})\;
          \frac{\Delta_q(E_{\rm cm},Q_{\rm cut})}{\Delta_q(q,Q_{\rm cut})}\;
      \Delta_q(q,Q_{\rm cut})\;\Delta_g(q,Q_{\rm cut})\nonumber\\
      &=& \left[\Delta_q(E_{\rm cm},Q_{\rm cut})\right]^2
          \Delta_g(q,Q_{\rm cut})\,,
\end{eqnarray}
where $q$\/ denotes the scale at which the gluon has been resolved.
Again the parton showers for the quark and antiquark start at
$E_{\rm cm}$. The shower for the gluon is initiated at scale $q$. All
showers are subject to a veto on $Q_{\rm cut}$. Note that in the case
$N=k=3$ the weight reduces to
\begin{equation}
  W_3\;=\;\left[\Delta_q(E_{\rm cm},q)\right]^2
\end{equation}
and shower vetoes are performed w.r.t.\ $q$.
\mywidefigure{t}{
  \includegraphics[width=0.32\textwidth]{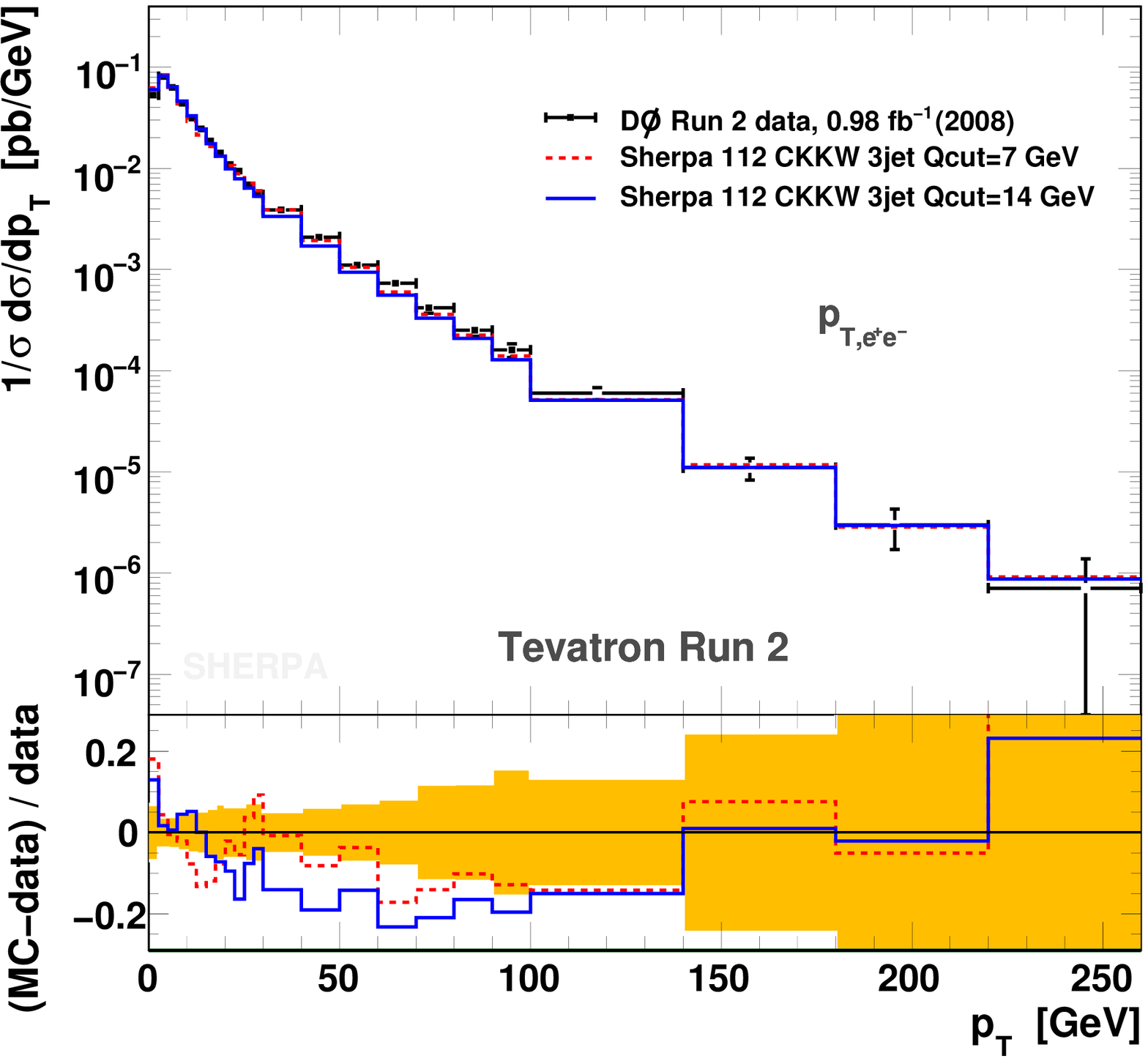}
  \includegraphics[width=0.32\textwidth]{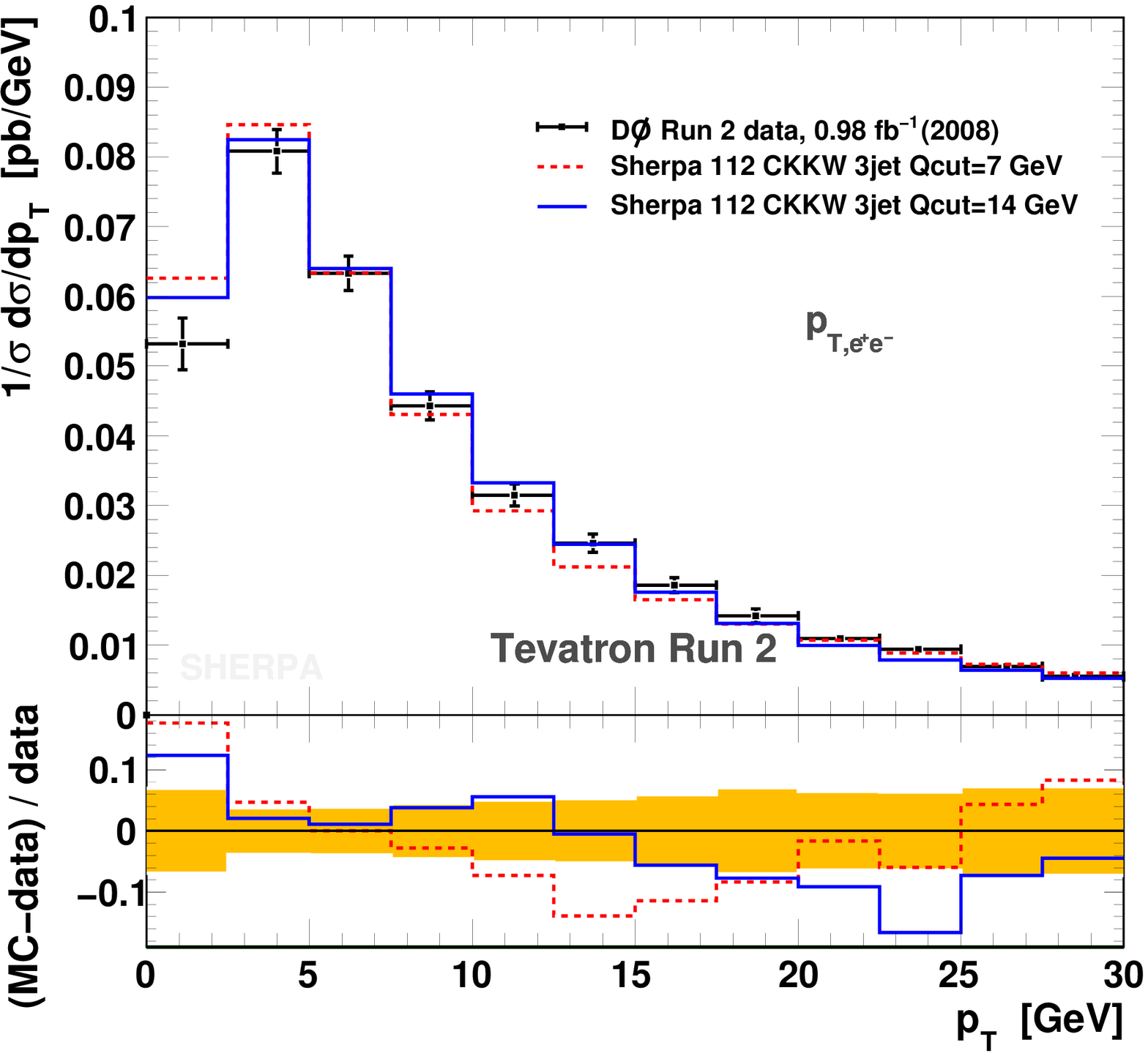}
  \includegraphics[width=0.32\textwidth]{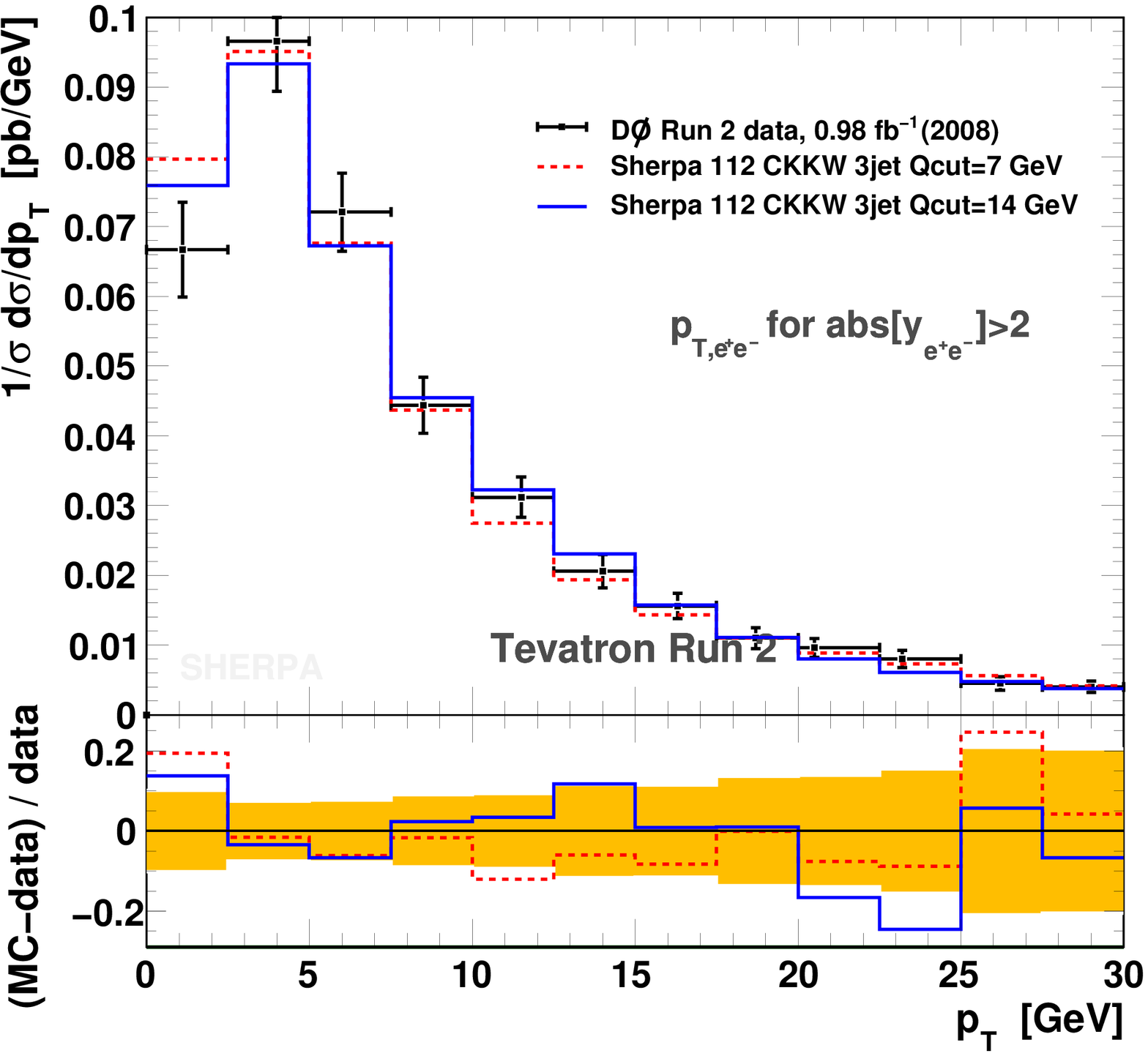}
}{\Sherpa CKKW shape predictions for two different merging-scale
  choices and various $p_T$ spectra of the vector boson in $Z$+jets
  production at Tevatron Run II. The left part shows
  the distribution for the full $p_T$ range as measured by D\O,
  whereas the other parts focus on the peak region. The rightmost plot
  has been obtained by requiring forward-rapidity lepton pairs. Data
  are taken from \cite{:2007nt} and the shaded bands visualize the sum
  of statistical and systematic uncertainties on the data.\label{fig:ptz}}
\mywidefigure{t}{
  \includegraphics[width=0.45\textwidth]{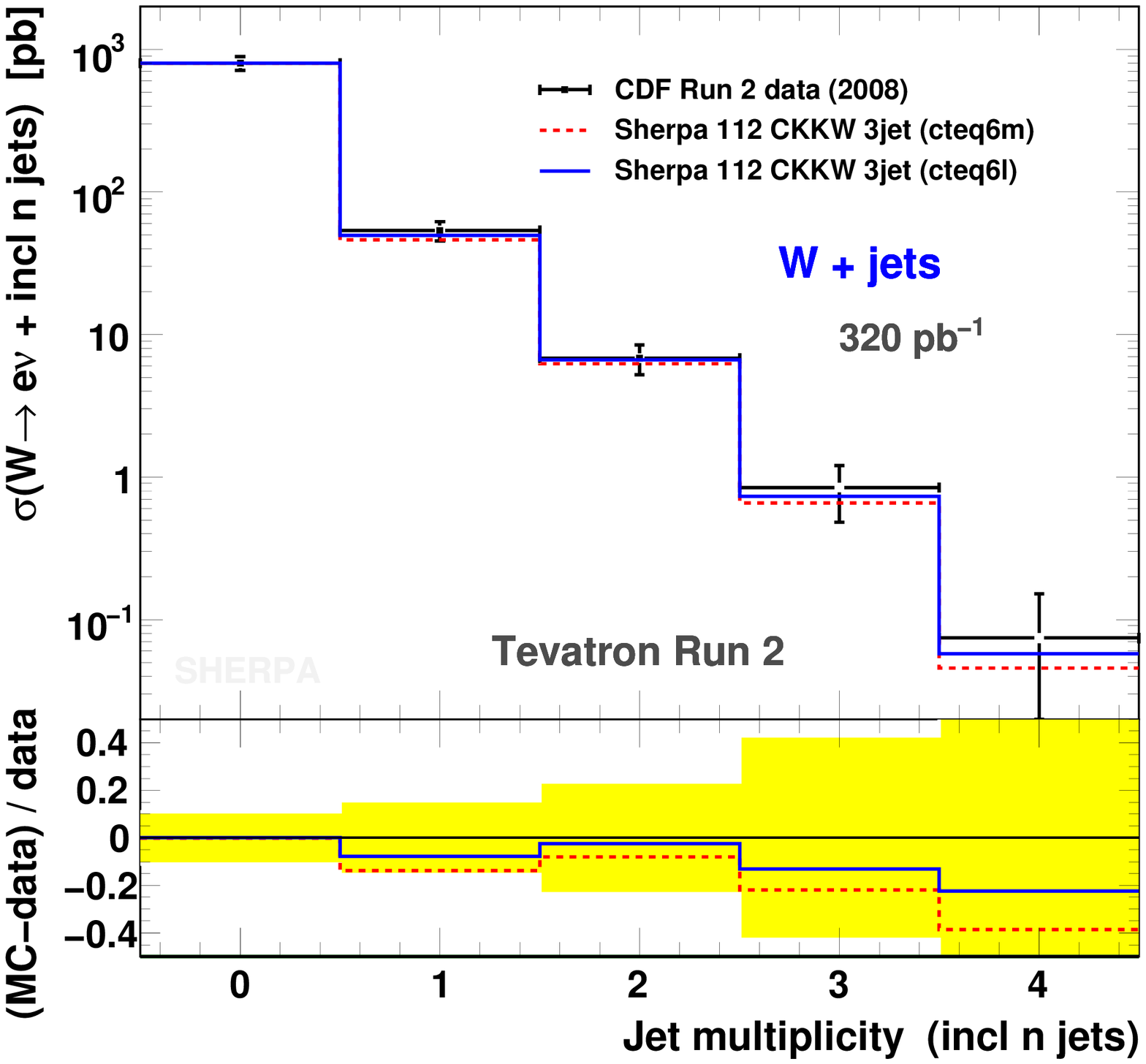}\hspace*{5mm}
  \includegraphics[width=0.45\textwidth]{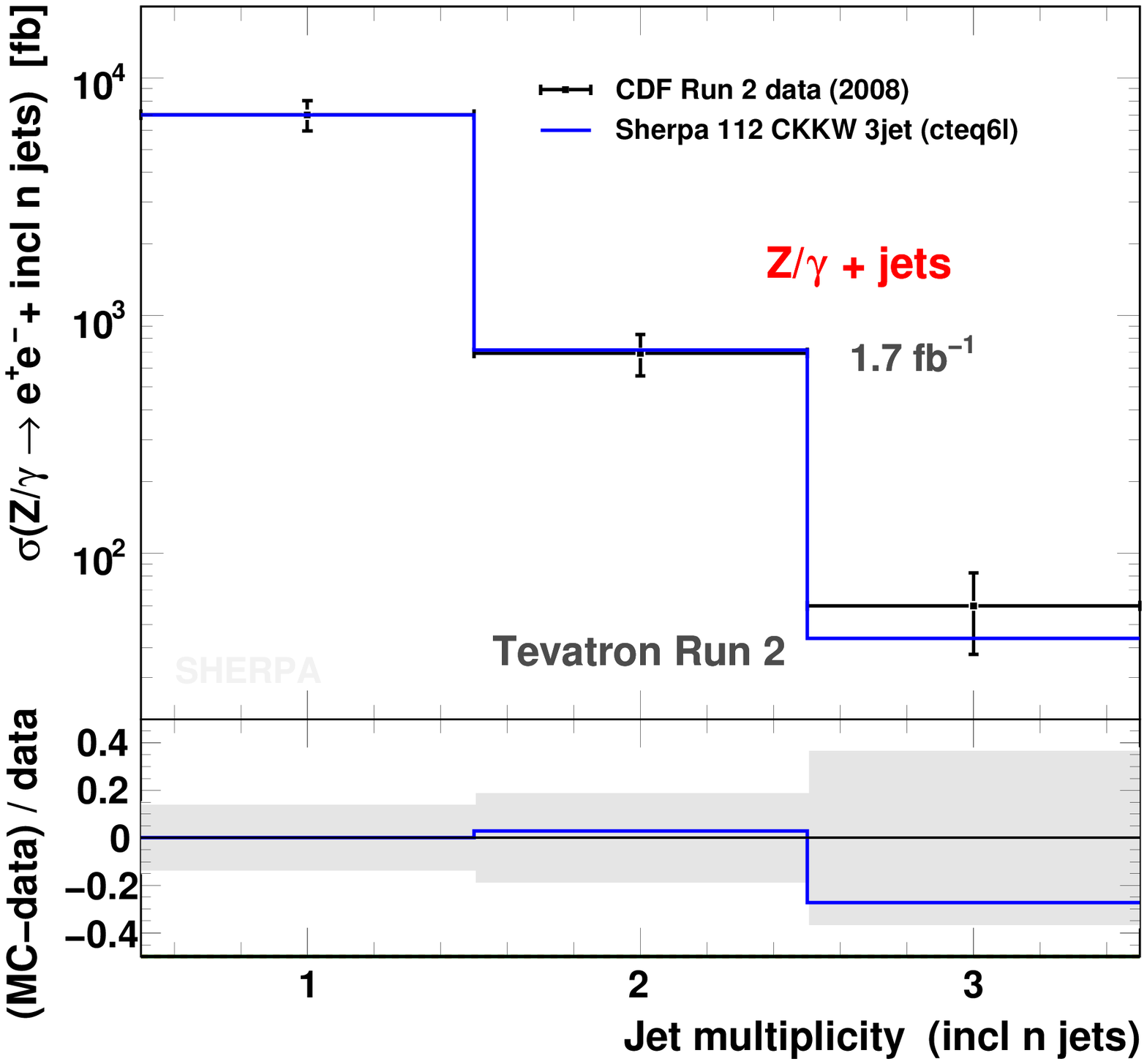}
}{Inclusive $n$\/ jet cross sections compared to CDF Tevatron Run II
  data for $W$+jets production \cite{Aaltonen:2007ip} (left panel) and $Z$+jets
  production \cite{:2007cp} (right panel). The shaded bands represent the
  full error on the data, except for the luminosity error being not
  included for $W$+jets. In both cases the blue solid line labels
  \Sherpa CKKW predictions obtained with the CTEQ6L PDF set; for the
  red dashed curve, the CTEQ6M set was used. The predictions
  for $W$\/ and $Z$+jets are normalized to the inclusive $n=0$ and
  $n=1$ cross sections, resulting in constant $K$-factors of $1.33$
  ($1.14$ for the dashed curve) and $1.71$, respectively. For cuts and
  details of both analyses, the reader is referred to the respective
  publication.\label{fig:jetmulti}}

The quality of the procedure can properly be studied by means of the
differential jet rates, which show the distributions of $y=Q^2/S$
values at which an $n+1$ jet event is merged into an $n$\/ jet event
according to the jet clustering scheme employed by the algorithm. In
particular they allow a thorough inspection of the transition region
$y\approx y_{\rm cut}$. The merging procedure provides a good
description of both the hard and the soft part of these
distributions. Large values of $y$\/ indicate the region of hard
emissions, where the matrix elements improve over any given
parton-shower estimate. For decreasing values of $y$, the
real-emission matrix elements diverge and descriptions based on them
become unreliable. Instead kinematically enhanced logarithms need to be
resummed to all orders as accomplished by parton showers.

Shower-level results as given by the procedure for the Durham
differential $n\to n+1$ jet rates are depicted in Fig.~\ref{fig:y234}
for $n=2,3$. They clearly show the features described above: predictions
for different values of $y_{\rm cut}$ vary at a level of 15\% only. The
transitions between shower and matrix-element domains for the various
$y_{\rm cut}$ choices are smooth. The plots emphasise that the merging
procedure works well and at the level of accuracy it claims to provide
for the parton-shower tree-level matrix-element matching.
\begin{figure*}[t]
  \begin{center}
    \includegraphics[width=0.32\textwidth]{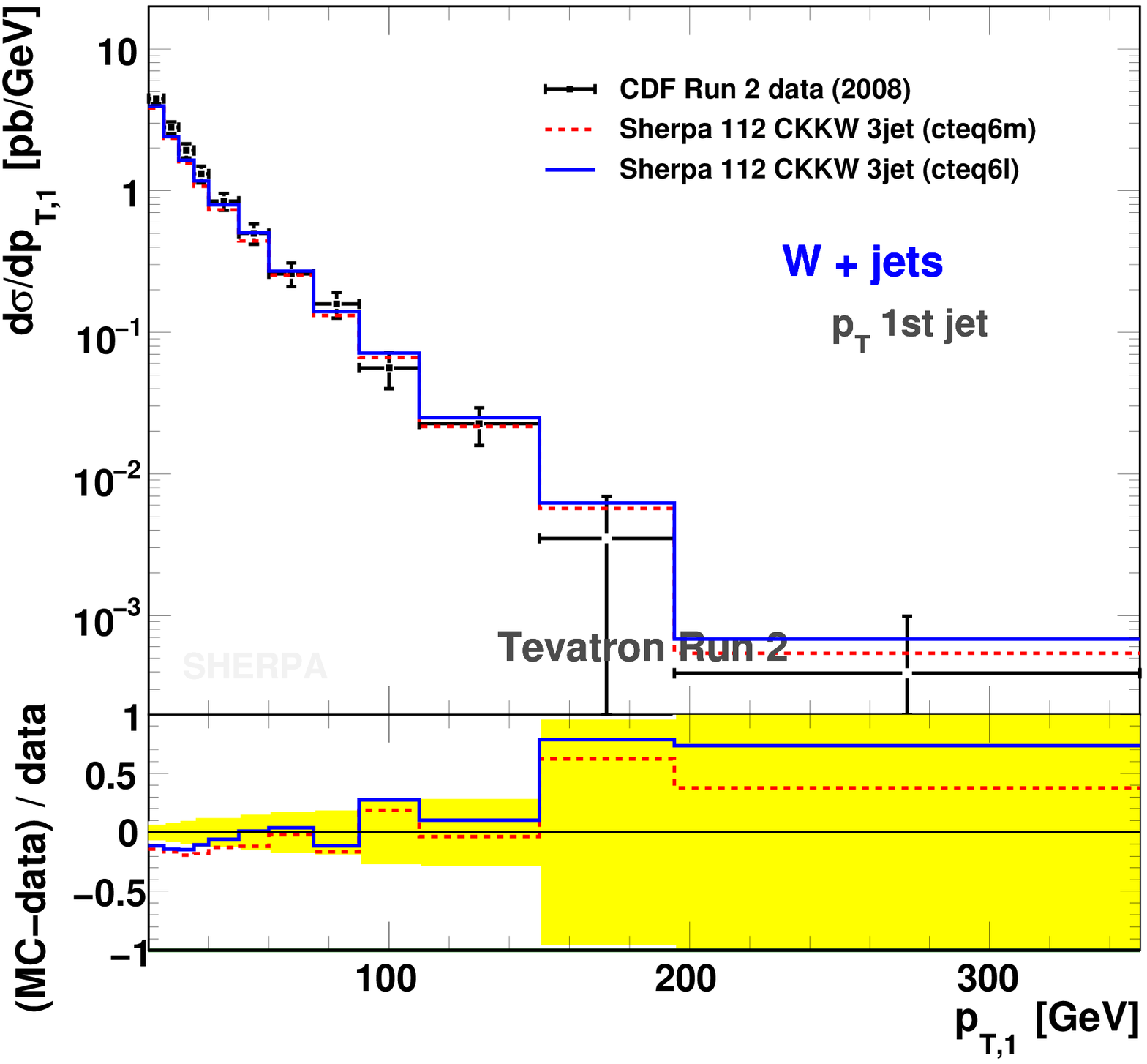}
    \includegraphics[width=0.32\textwidth]{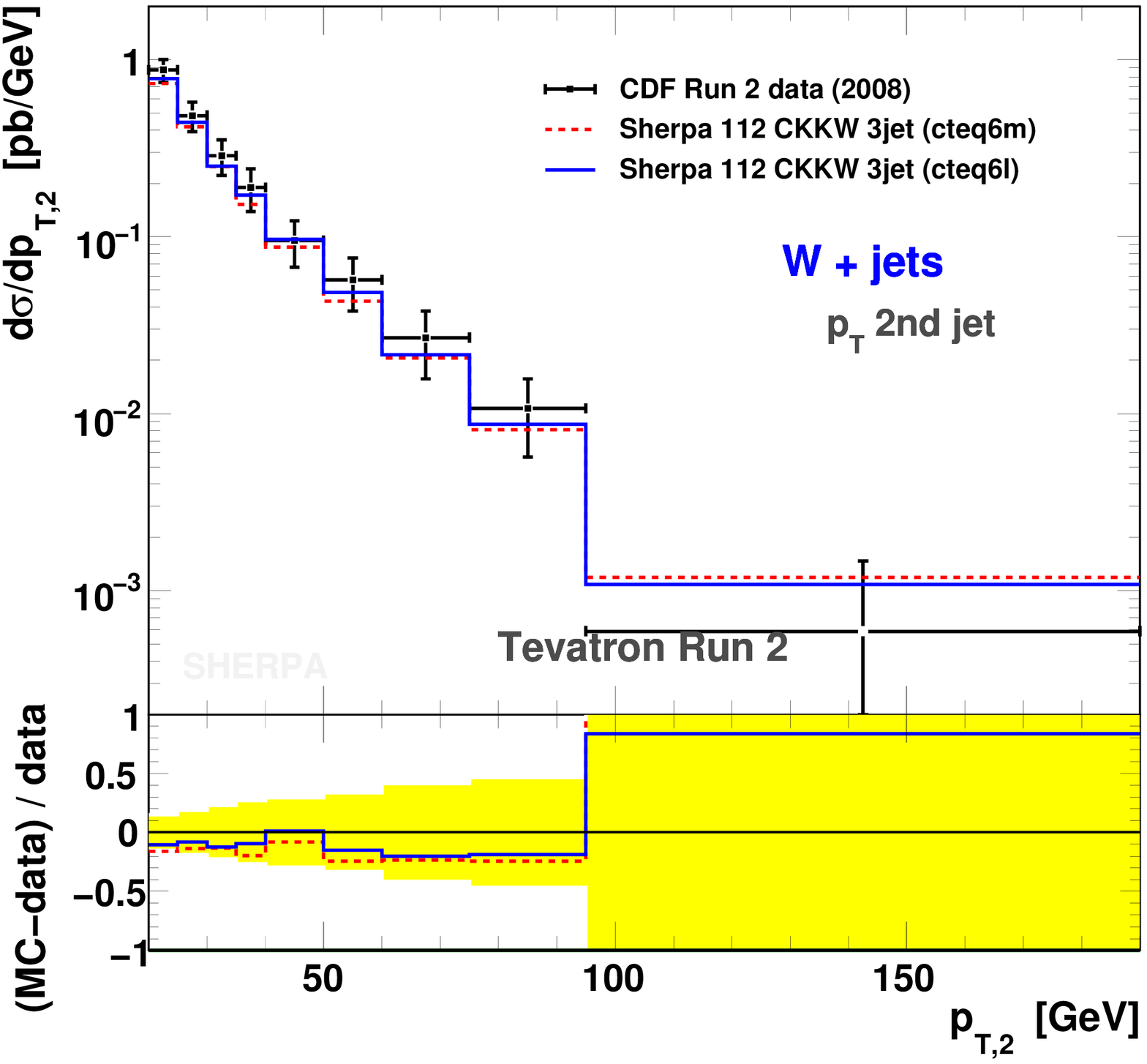}
    \includegraphics[width=0.32\textwidth]{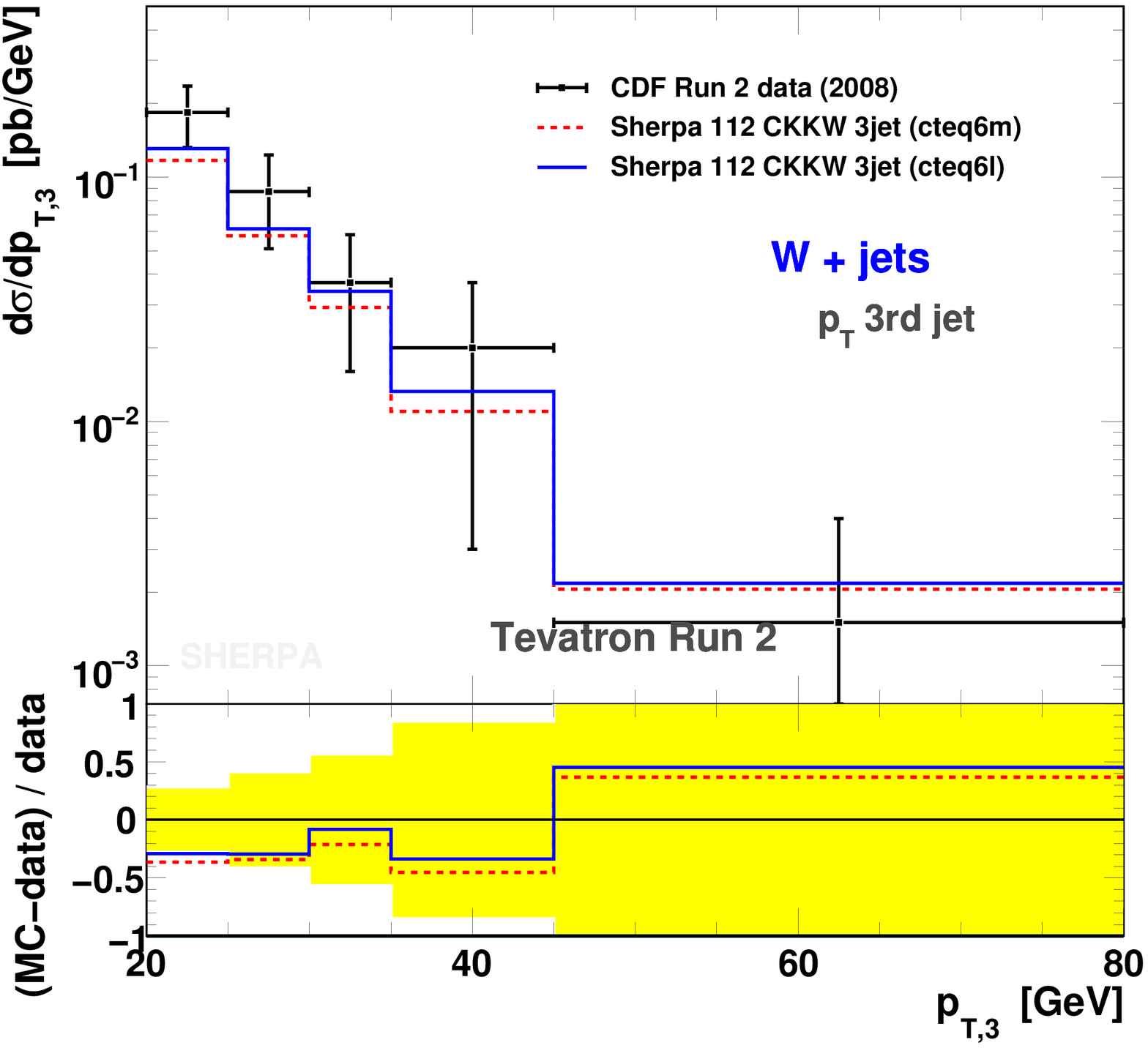}
    \\
    \includegraphics[width=0.32\textwidth]{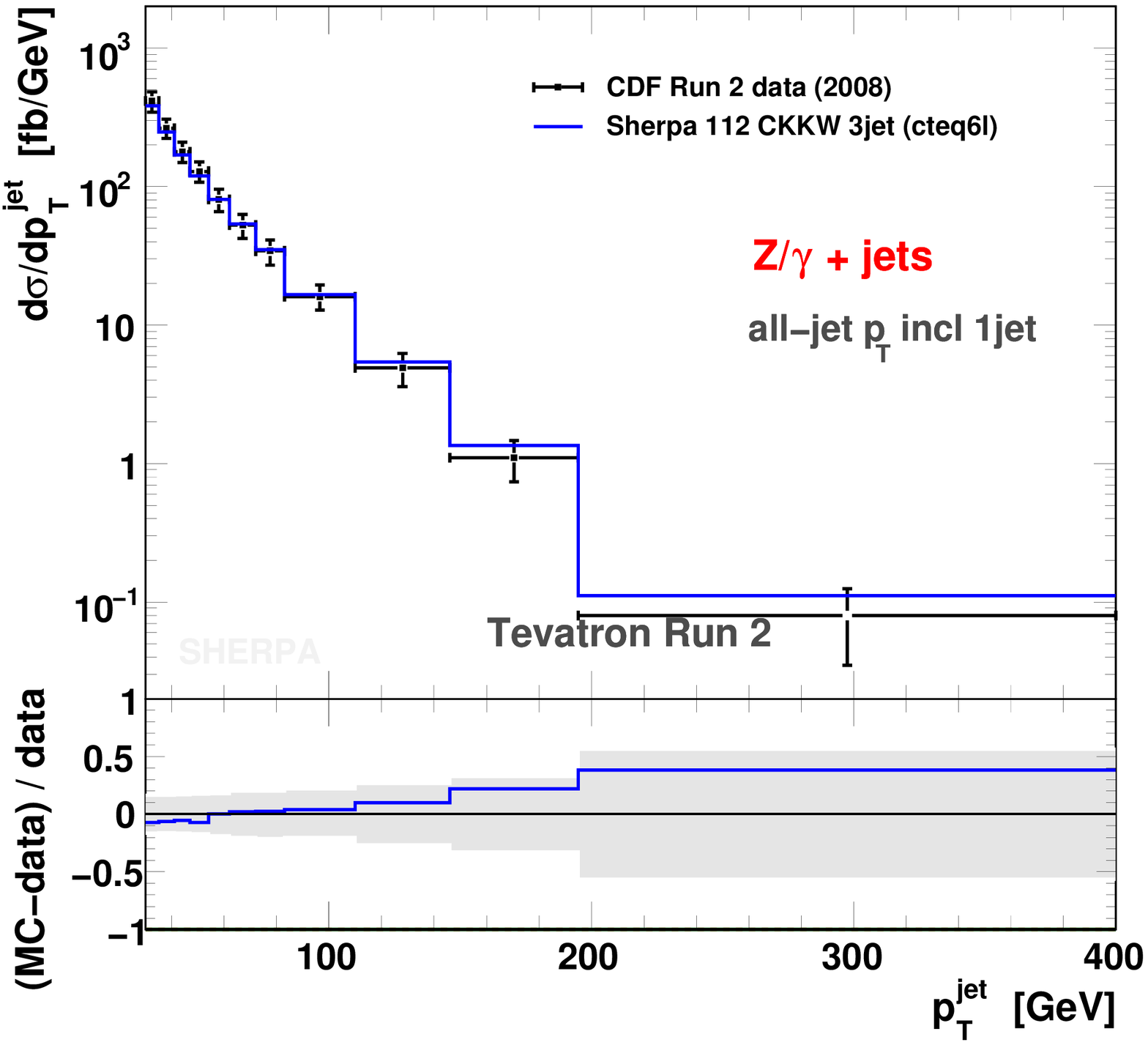}\hspace*{10mm}
    \includegraphics[width=0.32\textwidth]{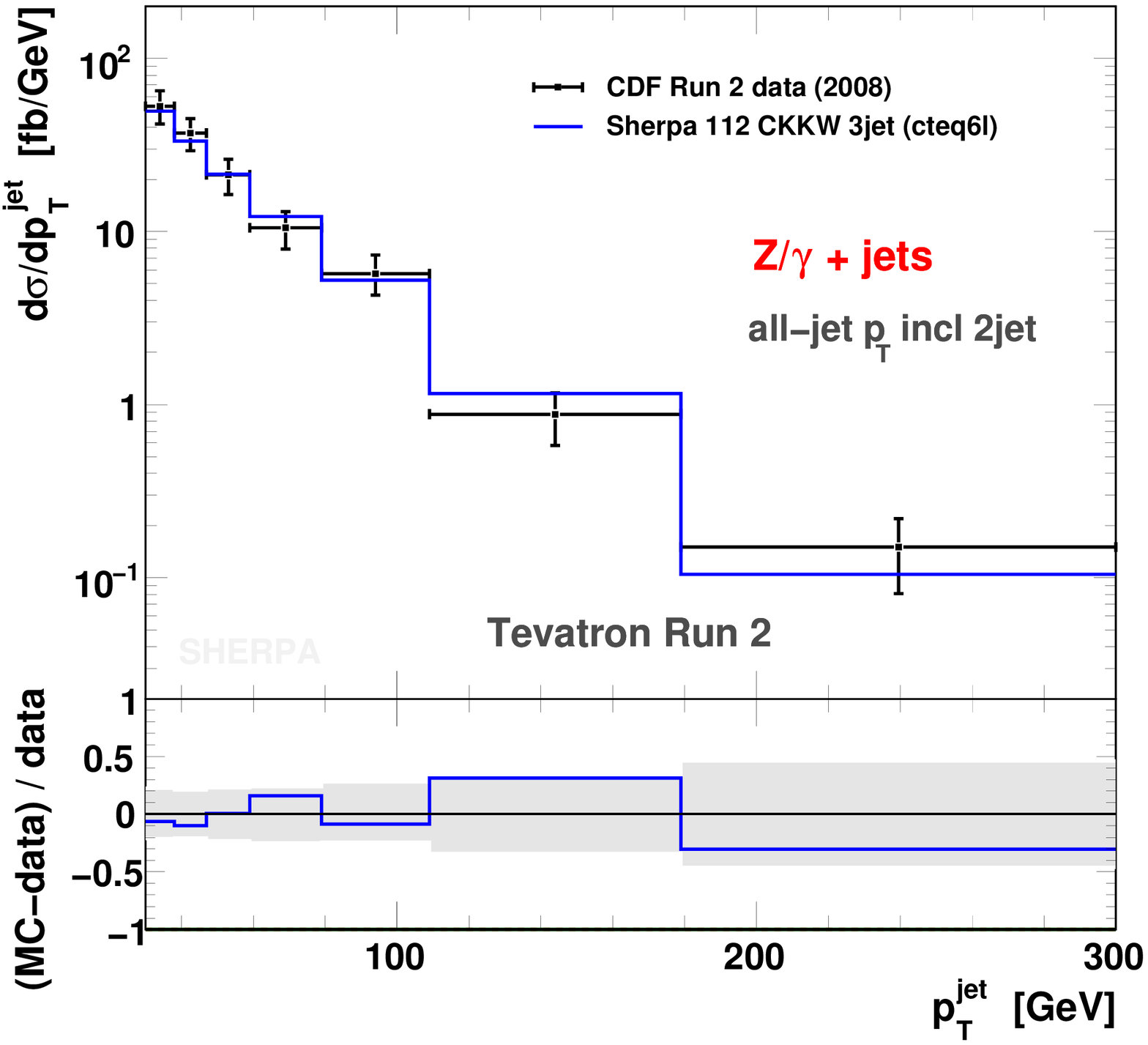}
    \\\myfigcaption{0.96\textwidth}{Jet $p_T$ spectra compared to CDF
      Tevatron Run II data for $W$+jets production
      \cite{Aaltonen:2007ip} (top row)
      and $Z$+jets production \cite{:2007cp} (bottom row). Labelling,
      $K$-factors and cuts are the same as given in the caption of
      Fig.~\ref{fig:jetmulti}. Note that the top row gives inclusive
      jet $p_T$ spectra of the $n$th jet; the bottom row depicts
      all-jet $p_T$ spectra for events with at least $n$\/ jets, i.e.\
      a $3$ jet event gives $3$ entries to all distributions up to
      $n=3$. \label{fig:jetpt}}
  \end{center}
\end{figure*}
\mywidefigure{t}{
  \includegraphics[width=0.45\textwidth]{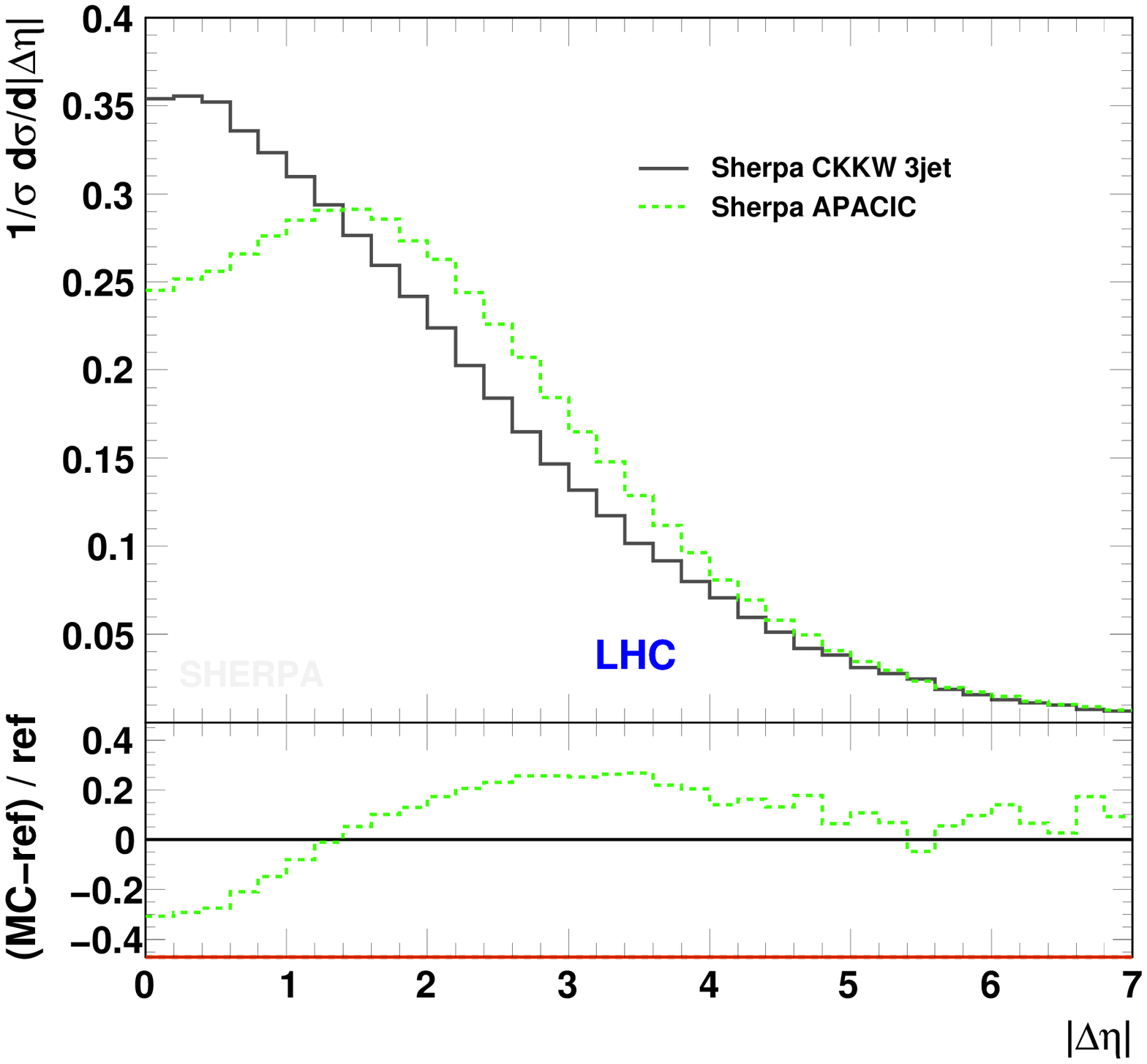}
  \hspace*{5mm}
  \includegraphics[width=0.45\textwidth]{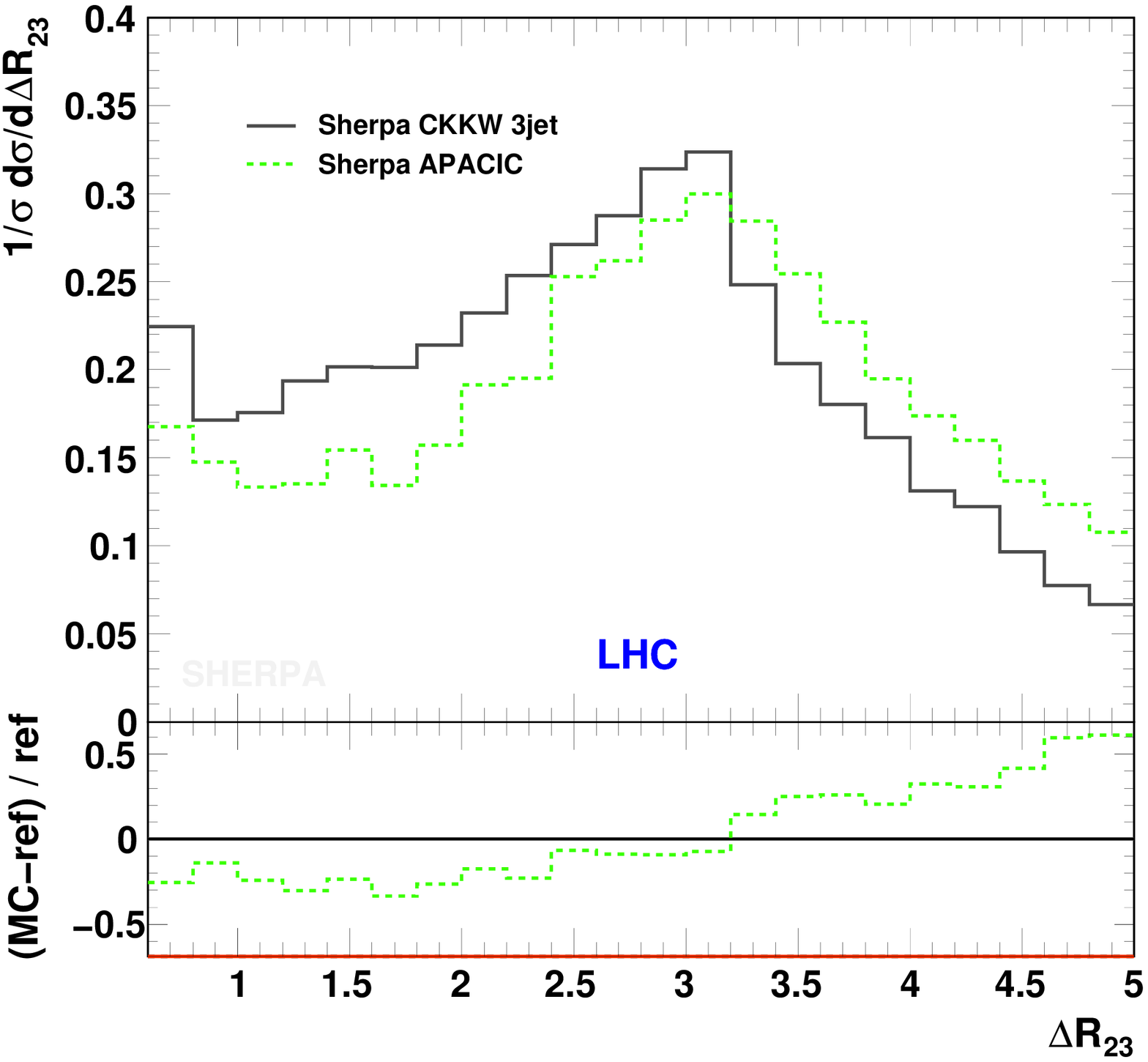}
}{Left panel: $|\Delta\eta|$ distributions between the vector boson
  and the first jet in $Z/\gamma^\ast$+jets production at the LHC.
  Right panel: Spatial separation between the second and third hardest
  jet, $\Delta R_{23}$, in $Z/\gamma^\ast$+jets production at the LHC.
  Dark solid curves represent CKKW predictions including matrix
  elements up to three extra partons. The bright dashed curves are
  pure shower predictions generated by \Apacic. Jets are defined by
  the Run II $k_\perp$-algorithm with parameter $D=0.4$
  \cite{Blazey:2000qt}, and the cuts are $p^{\rm jet}_T>20$ GeV and
  $|\eta^{\rm jet}|<4.5$.
  \label{fig:correl}}

\subsubsection{\tops{$W/Z$}{W/Z}+jets}
Weak boson production in association with jets is one of the most
prominent backgrounds in many new-physics searches with signals of the
type multijets plus missing transverse energy plus leptons. The
validation and tuning of Monte Carlo tools against newest Tevatron
data on weak boson plus jets production therefore has to be considered
a crucial prerequisite for successfully applying the available tools
in LHC analyses. The extrapolation to higher energies can only be more
meaningful once it starts off a solid and well understood test
scenario. With more and more Tevatron data coming in, its
discriminating power gradually increases and data can thus be used to
refine the multijet-production algorithms, even rule out choices or
falsify assumptions that have been made in constructing them.

Here, new comparisons are presented using very recent experimental
$W/Z$+jets data taken during Run II at the Fermilab Tevatron and
published by the D\O\ and CDF collaborations in \cite{:2007nt} and
\cite{Aaltonen:2007ip,:2007cp}, respectively. All CKKW predictions in
this subsection have been obtained from inclusive samples where matrix
elements including up to three extra (light-flavour) partons were merged with
the parton showers. The PDF set used was CTEQ6L. Note that effects of
the underlying event and hadronisation have not been taken into
account for this study.

Figure~\ref{fig:ptz} shows transverse momentum distributions of the
Drell--Yan lepton pair. On a rather inclusive level these spectra
allow for insight to the nature of the occuring QCD radiation, which
the boson is finally recoiling against. The data are taken from a D\O\
shape measurement of $Z$\/ boson events with mass $40<M<200$ GeV and
integrated luminosity of almost $1\ fb^{-1}$ \cite{:2007nt}. The overall
agreement is satisfactory. In the left panel of Fig.~\ref{fig:ptz} the
full $p_T$ range measured up to $260$ GeV is depicted. The two CKKW
predictions differing in their choice of the merging scale $Q_{\rm cut}$
tend to undershoot the data between $30<p_T<80$ GeV. This may be
improved further by tuning the merging scale and fully including
underlying-event plus hadronisation effects. The center and right
panels respectively show the region of soft $p_T$ without and with the
additional constraint of reconstructing the rapidity of the Drell--Yan
pair to be larger than $2$. For both plots, the results in the very
first bin may be corrected by tuning the intrinsic $k_T$ smearing
parameters. The good agreement, in particular for forward lepton-pairs
events, confirms that resummation effects are reasonably accounted
for.

The TeV centre-of-mass energies of the Tevatron/LHC provide
sufficiently/enormously large phase space to be filled by relatively
hard QCD radiation giving rise to (hard) jets accompanying the vector
boson. Such topologies cannot be generated by parton showers.
Figure~\ref{fig:jetmulti} displays inclusive cross sections for
various jet multiplicities. These cross sections are very sensitive to
the correct description of QCD multijet final states and hence provide
excellent probes for \Sherpa's merging approach.
Figure~\ref{fig:jetpt} shows the transverse momentum spectra of jets
produced in association with the Drell--Yan lepton pairs. These
observables allow a more detailed study of the QCD structure of the
events. In both figures CKKW predictions have been compared to CDF
data published in \cite{Aaltonen:2007ip} for $W$+jets and \cite{:2007cp} for
$Z/\gamma^\ast$+jets production. \footnote{The new data has been used
  to further validate and tune \Sherpa by revising and finally
  changing some of the default choices of the implemented merging
  procedure; the new settings will be made available to the public.}
The predictions are normalized by a constant $K$-factor to the lowest
available $n$\/ jet cross sections, $n=n_0$. The description of the
$n>n_0$ cross sections then is very satisfactory and the predictions
are low for the highest jet only. Merging in matrix elements with even
more extra partons will certainly improve on this situation. The same
constant $K$-factors were employed to generate the jet $p_T$ spectra.
As a result rate and shape of all $p_T$ distributions is in quite
remarkable agreement with the data. There is a tendency of
overestimating the hardness of the leading jet, yet the predictions
are consistent with data. For the case of $W$+jets production, a
second CKKW curve was added obtained by employing the CTEQ6M PDF set.
This led to a smaller $K$-factor and softer spectra in general. It
also provides an estimate of the uncertainty that comes along with the
use of different PDF sets.

A clear deficiency of the parton shower is the limitation in correctly
accounting for boson--jet (lepton--jet) and jet--jet correlations.
This drawback can be overcome by using the CKKW merging approach.
Examples for $Z/\gamma^\ast$+$X$\/ production at the LHC are given in
Fig.~\ref{fig:correl}. The left panel depicts the pseudo-rapidity
difference between the lepton pair arising from the decay of the
vector boson and the hardest jet accompanying the boson. The
prediction obtained by solely adding parton showering to the
Drell--Yan lepton-pair production (here given by the \Apacic shower)
shows a suppression at low $|\Delta\eta|$. The correct correlation is
encoded in the QCD real-emission matrix elements for $Z/\gamma^\ast$
production and can hence be accounted for by the CKKW approach. As a
consequence the suppression has disappeared in the CKKW merging
prediction. The distributions of the spatial separation between the
second and third hardest jet are given in the right panel of
Fig.~\ref{fig:correl}. The pure shower approach predicts separations
that are too wide. This again is corrected for in the CKKW procedure
by including matrix elements here up to three extra parton emissions.

As demonstrated by these examples CKKW predictions lead to
considerable improvements where the application of parton showers
merely gives approximations or is even misleading. An improved ISR
description as well as a correct treatment of correlations involving
jets can only be achieved by including the corresponding matrix
elements for extra QCD radiation. Merging algorithms are therefore an
ideal approach to provide inclusive $W/Z$+jets samples up to a certain
number of jets including effects of hard ISR and jet correlations. The
comparison to recent Tevatron data proves the relevance of \Sherpa's
merging approach in describing multijet events realistically.

\subsubsection{\tops{$t\bar t$}{t tbar}+jets}
It is worthwhile to study the implications of the CKKW formalism 
for relevant processes at future colliders. An example for such a process,
which plays a significant role, both as a signal for a better measurement
of Standard Model parameters and as a background to new physics searches,
is top quark pair production. In this publication it serves as an example
for the merging of matrix elements and parton showers in processes with
strongly interacting heavy resonances. Properly implemented real 
next-to-leading order corrections can be crucial for this kind
of interaction, as pointed out e.g.\ in~\cite{Alwall:2008qv}.
The combination of matrix elements and parton showers in this respect
is based on using a (potentially Breit-Wigner improved) narrow width 
approximation to compute the corresponding hard matrix elements, 
cf.\ Sec.~\ref{sec:mes}.
Through the identification of the decaying intermediate particle, a 
parton shower can independently be assigned to its production and decay.  
Upon application of the CKKW merging algorithm and when integrating over 
the phase space of the outgoing particles, jet measures $Q$ between strongly 
interacting particles must be larger than a critical value $Q_{\rm cut}$.  
Due to the factorised structure of the process, this critical value might
be chosen separately per subprocess, i.e.\ there may be different values 
$Q_{\rm prod} = Q_{\rm cut}({\rm prod})$ and 
$Q_{{\rm dec},i} = Q_{\rm cut}({\rm dec}_i)$.

Technically the merging proceeds as follows:
Once a particular momentum configuration is chosen for the hard matrix element, 
the parton-shower history is identified as in the standard merging prescription.  
Since the full amplitude factorises over time-like propagators, only particles 
belonging to the same subamplitudes $\mathcal{A}_{\rm prod}$ or 
$\mathcal{A}_{{\rm dec},i}$ can be combined. When reweighting the hard
matrix element with Sudakov form factors, the potentially different cutoff 
values $Q_{\rm cut}$ must be employed. Sudakov reweighting for the resonant
intermediate particles takes places according to the original prescription.
The jet veto is applied separately within each part of the parton 
showers related to a different subamplitude, employing the corresponding 
veto scale $Q_{\rm cut}$. The highest multiplicity treatment is applied 
separately in each subamplitude. 
To summarise, the above prescription translates into the CKKW method
being applied separately and completely independent for each part of the
process under consideration.
\myfigure{t}{
  \includegraphics[width=0.48\textwidth]{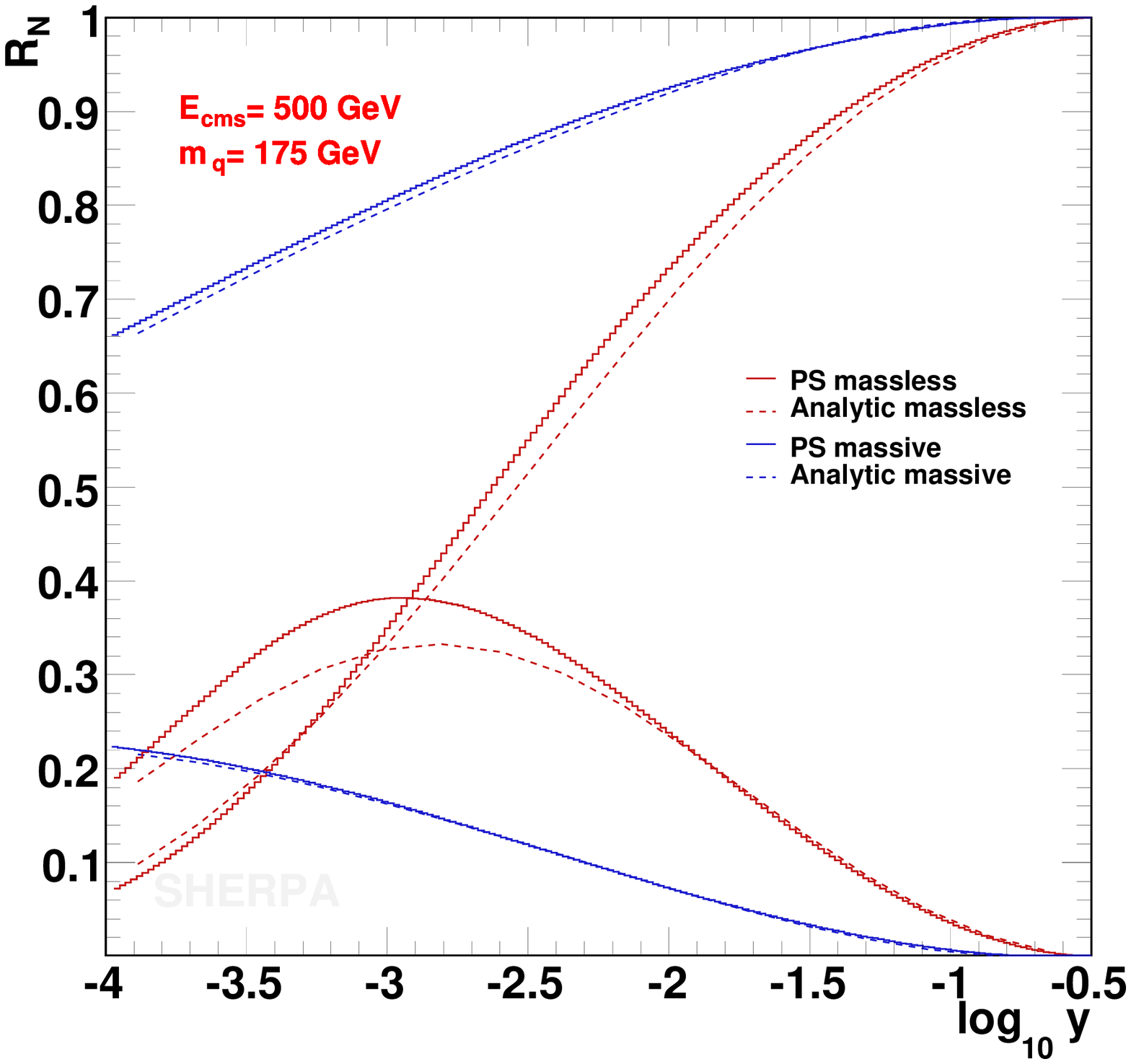}
  }{Comparison of the Durham 2- and 3-jet rates for massive quarks, 
  calculated according to~\cite{Rodrigo:2003ws}, with results 
  from \Apacic. Good agreement is found between the parton-shower result 
  (solid histogram) and the analytical calculation (dashed line). 
  For reference, results with massless quarks are also shown.
  \label{fig:massive_jetrates_ee}}

For this procedure, reweighting with analytic Sudakov form factors 
for heavy quarks is necessary.
It must therefore be assured that those match the respective distributions 
generated by the parton shower. A corresponding comparison is shown in 
Fig.~\ref{fig:massive_jetrates_ee}. 
Predictions from \Apacic and an analytic calculation performed 
in~\cite{Rodrigo:2003ws} are displayed for the Durham $k_T$-jet rates at a 
fictious electron positron collider operating at 500 GeV.

Figure~\ref{fig:lhc_tt_pt_psme_comparison} 
shows the impact of the
description of additional hard radiation through appropriate matrix
elements in the case of $t\bar t$-production at the LHC. Subsamples
from a given matrix element configuration are displayed in colour with
the colour code explained above the figure. The following notation is
employed
\begin{itemize}
\item The first number denotes the additional jet multiplicity in the $t\bar t$
  production process.
\item The second number denotes the additional jet multiplicity
  in the $t$ and $\bar t$ decay process.
\end{itemize}
It can be observed that the 
transverse momentum spectrum of the produced heavy quark pair is significantly 
enhanced at high values when employing the CKKW prescription. This is easily 
explained by the failure of the parton-shower approach to correctly describe 
hard initial-state radiation.
Also,
it is clearly seen that subsamples including an additional hard jet in the production 
process (1-0 and 1-1) deform
the pseudorapidity spectrum of the first additional hard jet (identified through 
$k_T$-clustering and Monte Carlo truth based $b$-tagging). The pure
parton-shower approach predicts a dip at central rapidity, while the merged sample 
does not show this feature.
\myfigure{t}{
  \includegraphics[width=0.48\textwidth]{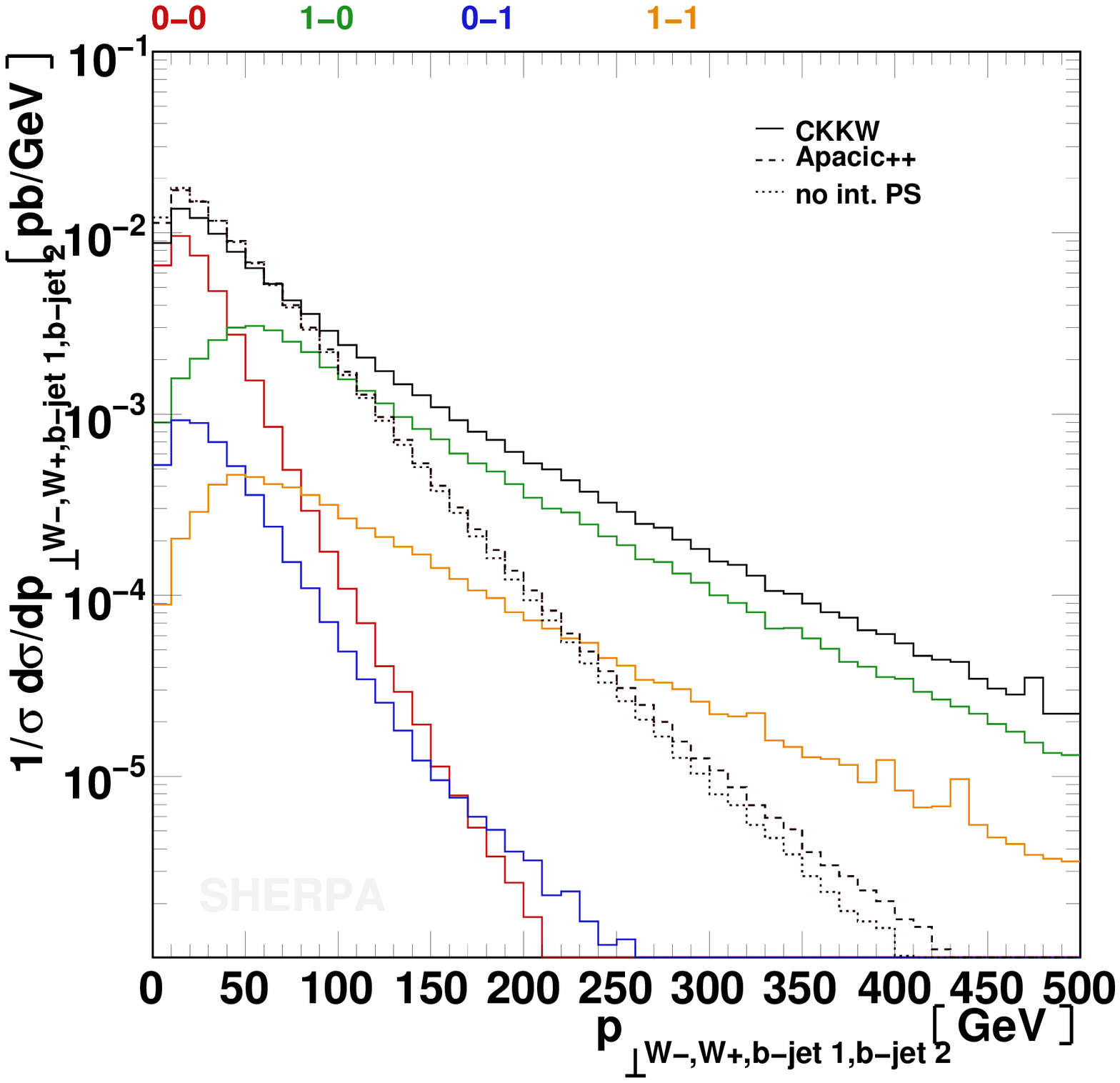}\hspace*{3mm}
  \includegraphics[width=0.48\textwidth]{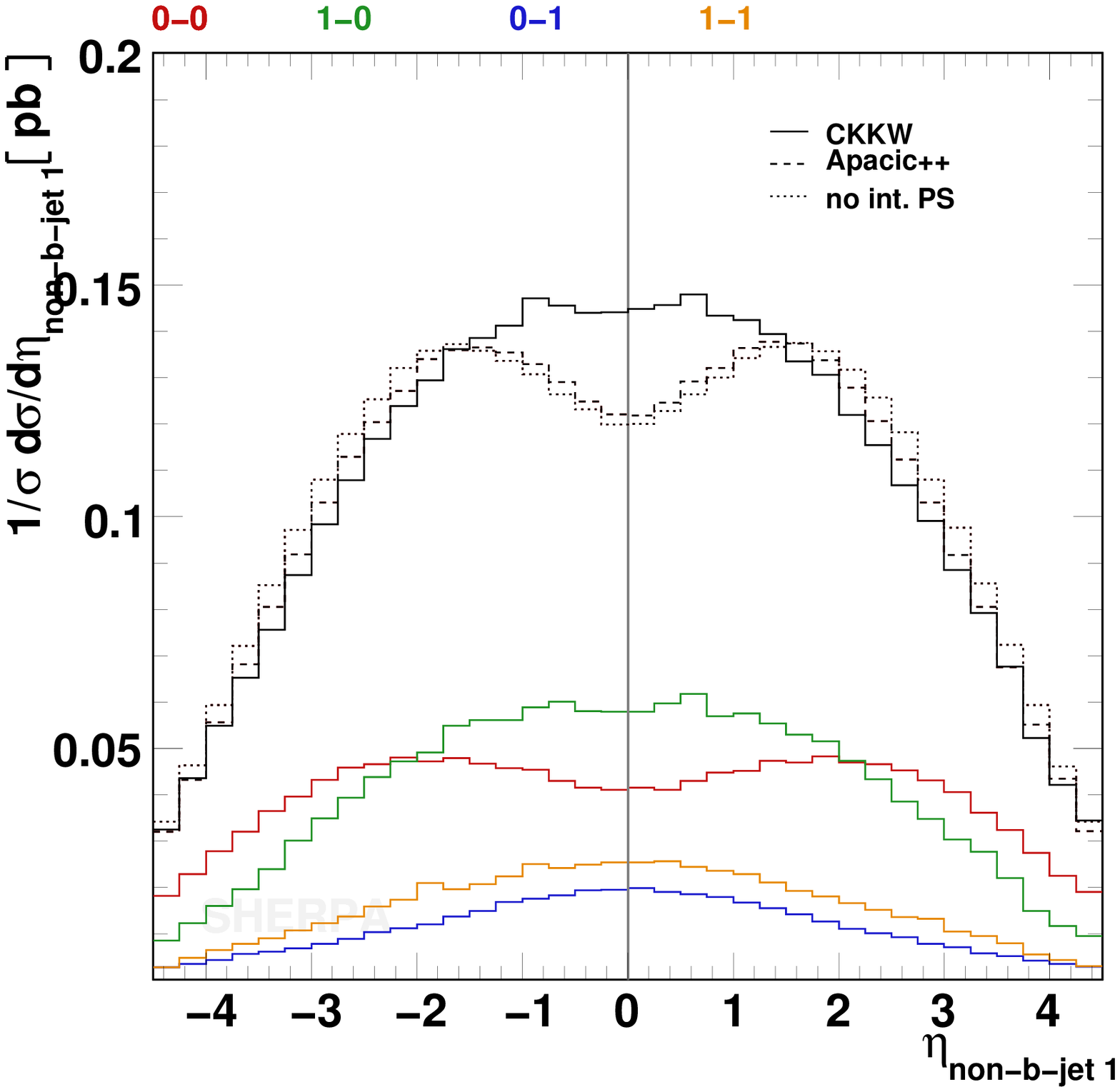}
  }{Transverse momentum spectrum of the $t\bar{t}$-pair (left panel) and 
  rapidity spectrum of the first extra jet (right panel) in 
  $pp\to t\bar{t}\to W^+W^-b\bar{b}+jets$ events at $\sqrt{s}=14\;{\rm TeV}$. 
  Shown is a comparison between the pure parton-shower result (no int. PS), 
  parton shower with radiation off intermediate top quarks (Apacic++) 
  and a CKKW-merged sample. Contributions from various matrix element 
  configurations are highlighted in colour for the merged sample, 
  with the colour code indicating the additional jet multiplicity in 
  the production $\sst\otimes$ decay process. Up to one extra jet 
  has been simulated through matrix elements in the production 
  and each decay.\label{fig:lhc_tt_pt_psme_comparison}}

\section{Multiple parton interactions}
\label{sec:mpi}
The hardest (primary) partonic interaction in hadronic collisions
may be accompanied by softer (secondary) ones, which may, however,
still fall in the realm of perturbative QCD. This is a valid
assumption and clear evidence has come from experimental studies
carried out by the CDF collaboration at the Fermilab Tevatron during
Run~I and
Run~II~\cite{Abe:1997bp,*Abe:1997xk,*Acosta:2004wqa,Affolder:2001xt}.
It was shown that a correct description at the hadron level of particle
multiplicities and jet activities can only be achieved by Monte Carlo
event generators that incorporate a model for multiple parton
scatterings.

The basic idea of such a model is to postulate the probability 
distribution for the occurrence of multiple scatterings. According to
the model of \cite{Sjostrand:1987su} this is given as a function of the
non-diffractive cross section $\sigma_{\rm ND}$, the hard perturbative
cross section $\sigma_{\rm hard}$ and the outgoing transverse momentum
$p_{\perp}$ in the scattering by
\begin{equation}\label{mi_probability}
  p(p_\perp,b)\,=\;f_cf(b)\,\frac{1}{\sigma_{\rm ND}}
  \frac{{\rm d}\sigma_{\rm hard}}{{\rm d}p_{\perp}}\;,
\end{equation}
where the prefactors $f_c$ (normalisation) and $f(b)$ (proton shape
function) incorporate an additional impact-parameter dependence of the
distribution on an event-by-event basis. In the collinear
factorisation approach a phase-space cut on the hard matrix elements,
e.g.\ a minimum jet resolution cut $Q_{\rm cut}$, has to be introduced
to obtain a well-defined differential cross section in perturbation
theory. Then ${\rm d}\sigma_{\rm hard}/{\rm d}p_\perp$
receives its dominant contributions from $2\to 2$ processes and,
hence, the definition of $p_{\perp}$ in
Eq.~\eqref{mi_probability} is unambiguous. Moreover it suffices to
consider hard $2\to 2$ QCD processes only.

The multiple-interactions model of \Sherpa is implemented in the
module \Amisic, which generates multiple scatterings according to
the basic ideas listed above. There are however important details where
the \Sherpa approach deviates from the original formalism presented in
\cite{Sjostrand:1987su}. Firstly, all secondary interactions are
undergoing parton-shower corrections and, therefore, \Amisic has been
interfaced to the parton-shower module \Apacic. A second class of
extensions concerns the combination of the CKKW merging approach with
the modelling of the multiple interactions, which essentially are
independent of the treatment of the hardest scattering in the
collision. It is vital that the parton showers related to secondary
interactions respect the initial $p_\perp$ distribution of
the hard scatterings. In particular, partons with a $k_T$-separation
from other partons larger than $p_\perp$ of the respective (secondary)
interaction must not be radiated. The appropriate way
to incorporate this constraint into the parton-shower formalism is in
fact identical to the realisation of the highest-multiplicity
treatment in the CKKW approach. The corresponding algorithm then works
as follows:
\begin{enumerate}
\item\label{step1}
  Create a kinematic configuration of the final-state particles
  according to the hard matrix element of the primary interaction and
  run its parton-shower evolution according to the CKKW formalism
  (cf.\ Sec.~\ref{sec:ckkw}).
\item\label{step2} 
  Employ a $k_T$-type algorithm in the $E$-scheme to cluster the complete 
  final state of the previous scattering into a $2\to 2$ process, similar 
  to how this is done to define starting conditions for showers in the
  CKKW approach. Set the starting scale for multiple interaction evolution 
  to the maximum of
  \begin{enumerate}
  \item\label{step21}
    the largest relative transverse momentum between QCD partons 
    that have been combined, or
  \item\label{step22}
    the transverse momentum in the $2\to 2$ process, if this process 
    contains a QCD parton in the final state.
  \end{enumerate}
\item\label{step3}
  Select the $p_\perp$ of the next secondary interaction according to
  Eq.~\eqref{mi_probability}. If this happens for the first time in the
  event, select the impact parameter $b$ of the collision as described
  in \cite{Sjostrand:1987su}. Create the kinematic configuration of
  the secondary interaction.
\item\label{step4} 
  Set the jet veto scale of the parton shower to the transverse momentum 
  selected in step \ref{step3}. Start the parton shower of the secondary 
  interaction at the QCD scale
  \begin{equation}
    \mu^2_{\rm QCD}\,=\;\frac{2\, stu}{s^2+t^2+u^2}\;.
  \end{equation}
\item\label{step5} 
  Return to step \ref{step2}.
\end{enumerate}

\mywidefigure{t}{
  \includegraphics[width=0.48\textwidth]{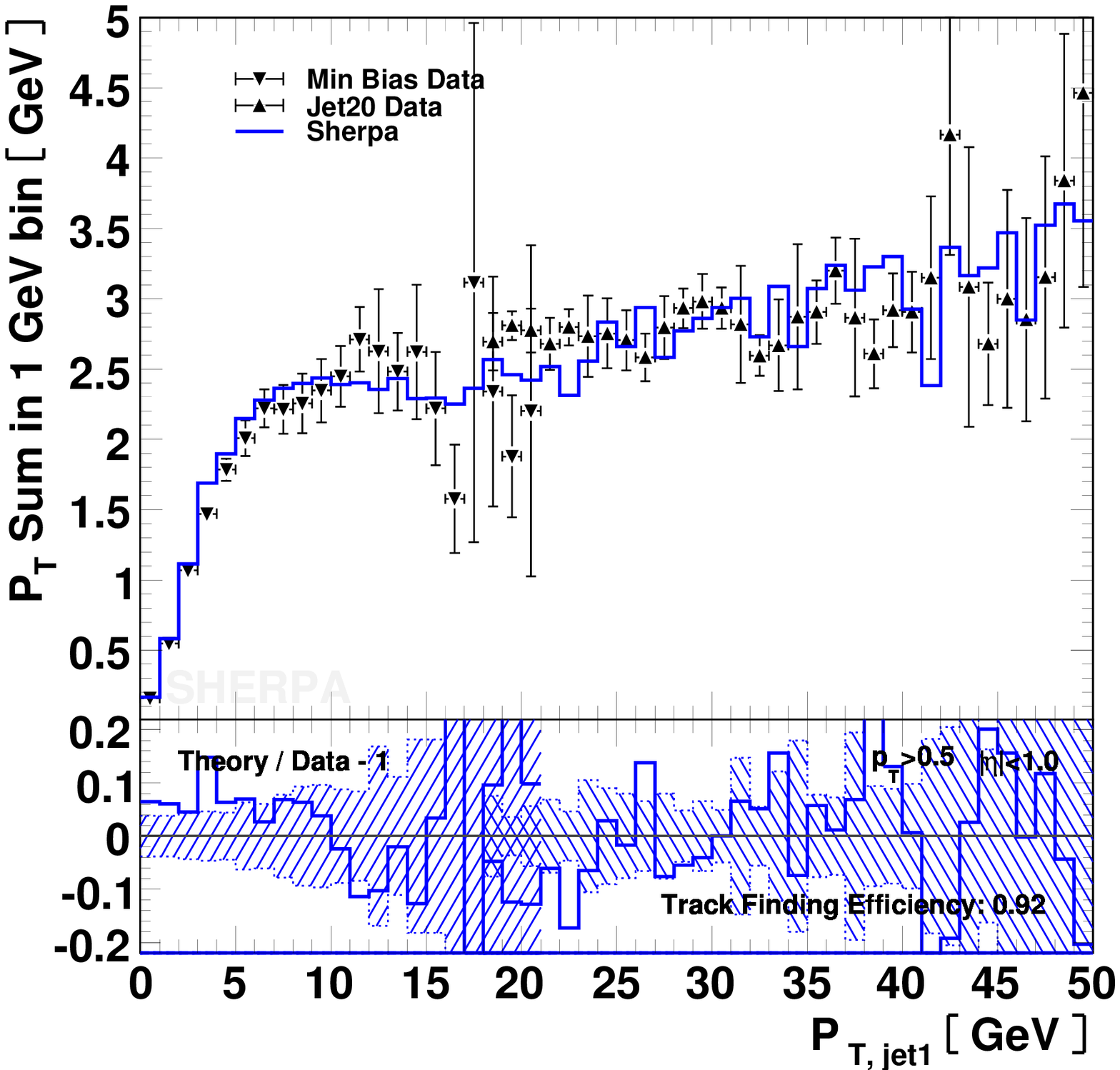}
  \includegraphics[width=0.48\textwidth]{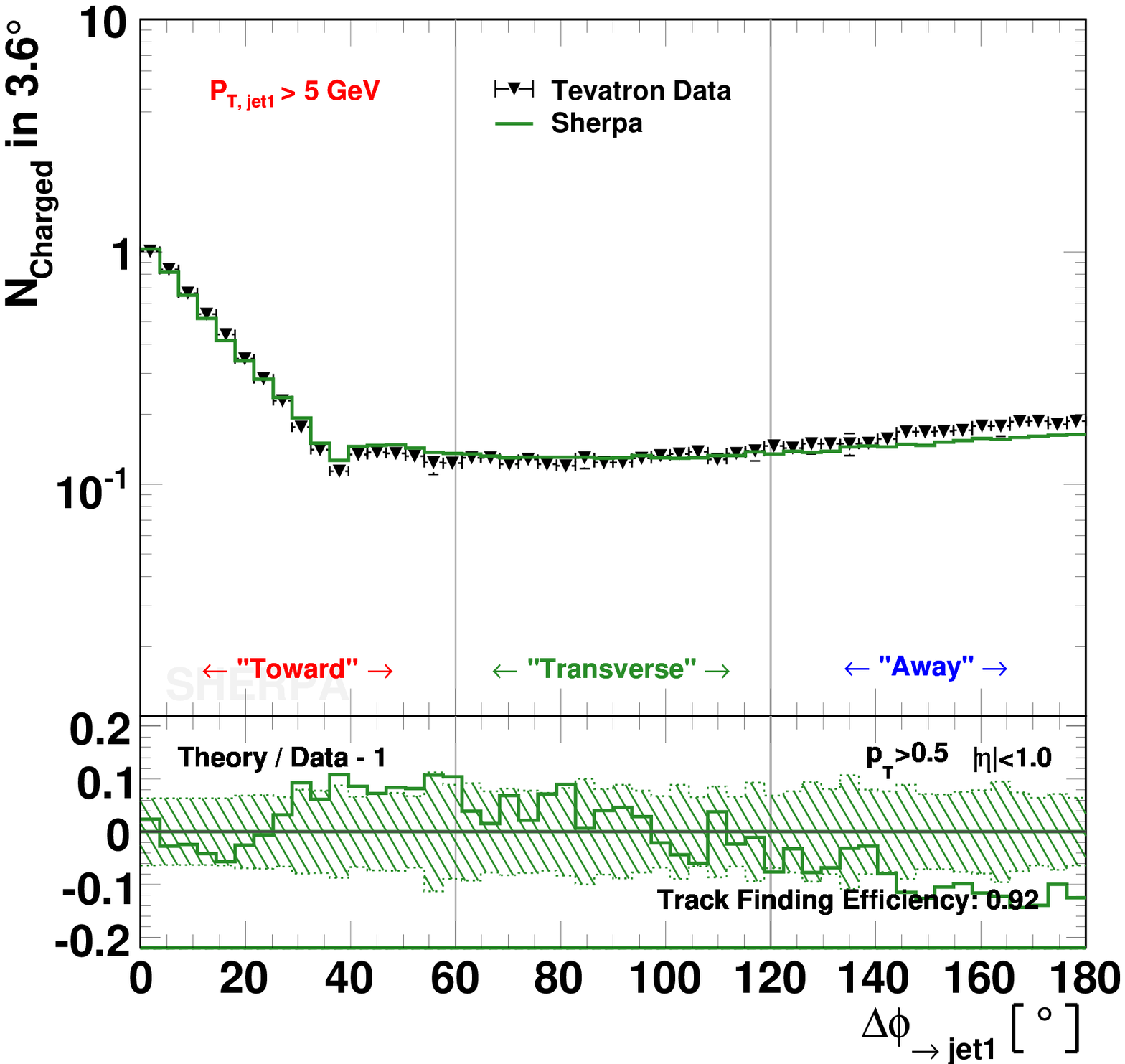}
  }
  {The left panel displays the average scalar $p_T$ sum versus $p_T$ 
   of the leading charged particle jet while the right panel displays 
   the average charged particle multiplicity 
   versus $\Delta\phi$ with respect to the leading charged particle 
   jet for $p_T>5$ GeV of this jet. In both cases Sherpa results are 
   contrasted with data from~\cite{Affolder:2001xt}.
   \label{fig:rick_tevi_spt_vs_jetpt_and_rick_tevi_nch_vs_dphi}
  }

The above algorithm works for pure QCD hard matrix elements as well as 
for electroweak processes being the primary hard scattering in the event. 
In the QCD case the selected starting scale for the determination of the
first additional interaction reduces to the transverse momentum in 
Eq.~\eqref{mi_probability} and is thus equal to the original 
ordering parameter. For electroweak core processes, e.g.\ $W$ or $Z$
boson production, there is no such unique identification. However,
it is required that the multiple scatterings in the underlying
event do not spoil the jet topologies generated by the primary hard 
scattering. Regarding the
electroweak bosons as being radiated off QCD partons during the parton
shower evolution of a hard QCD event, it is appropriate to
reinterpret this hard matrix element as a QCD+EW process, whereof the
simplest is an EW+1-jet process. In this way all primary jet topologies
can be preserved.

An appropriate method to include an impact-parameter dependence in the 
multiple-interactions model is described in \cite{Sjostrand:1987su}.
In \Amisic the default (anti-)proton shape in position space is chosen
to follow a Gaussian matter distribution. \Amisic has primarily been 
validated using the data presented in~\cite{Affolder:2001xt}.
These tests revealed that effects owing to the variation of the
particular shape function are of minor importance for the description 
of the underlying event. It might however turn out that the shape function 
becomes more important when facing more precise data.

Finally, it has to be stressed that \Amisic relies on complete
factorisation of the hard (primary) process and the additional
scatterings in the underlying event. Corrections, which arise from
multi-parton PDFs as discussed in \cite{Kirschner:1979im,*Snigirev:2003cq},
are not taken into account.

In Figure~\ref{fig:rick_tevi_spt_vs_jetpt_and_rick_tevi_nch_vs_dphi} 
predictions from \Sherpa are compared with experimental data 
from~\cite{Affolder:2001xt}. Both observables are very sensitive to the 
correct description of the underlying event.

\section{Parton-to-hadron fragmentation}
\label{sec:fragmentation}

After the parton shower has terminated, a configuration 
of coloured partons at some low scale of the order of a few GeV in transverse momentum emerges.  These partons, in order to match experiments, have to be translated into hadrons. 
Since there is no first-principles approach yielding quantitative results,
hadronisation is achieved by phenomenological models only.  Usually, they are based on
some qualitative ideas on how the parton-to-hadron transition proceeds, like, e.g., 
local parton-hadron duality or infrared safety, defining the model's coarse 
properties.  However, many of the important finer details, often related to how flavour
is created and distributed in the procedure, are entirely undefined and subject to
phenomenological parameters only.  These are essentially free and must
be fixed by extensive comparisons with data, as done for instance in \cite{Hamacher:1995df,*Abreu:1996na}.  

\subsection{Beam-remnant handling}
In collisions involving hadronic initial states, there is the additional problem that
one or more final-state particles will be colour-\-connected to the hadron remnants.  
The breakup of the initial hadron into partons then needs to be modelled suitably. 
This is achieved in such a way that only a minimal set of particles (quarks and diquarks, 
the latter as carriers of baryon number) is produced in order to reconstruct the constituent 
flavour configuration of the corresponding hadron.  Once their colour
connections are established, the complete set of final-state partons can be
hadronised affecting energy flows and other, similar properties of the
event at hadron level.

The distribution of colour in the hadron remnants cannot be inferred from first principles.  
Hence phenomenological models have to be employed. In \Sherpa the default distribution of 
colour is guided by the idea of minimising the relative transverse momentum of colour dipoles 
spanned by outgoing partons.  When including multiple parton interactions in the simulation, 
it is not always possible to accomplish free colour selection in the hard process and 
minimisation of relative transverse momenta simultaneously.  In such cases the colour configurations 
of the matrix elements are kept but the configuration of the beam remnants is shuffled at 
random until a suitable solution is found.  The shuffling is implemented also as an 
alternative choice if no multiple interactions are present.  So far, no significant 
differences have been found in the predictions from the two models on
the level of physical observables.

In addition to the issues related to colour neutralisation with the
beam remnants, all shower initiators and beam partons obtain a mild
$k_\perp$-kick according to a Gaussian distribution parametrising the
Fermi motion of these particles inside the incoming hadron.  It should be stressed that within \Sherpa mean and width of this 
distribution are rather small (about $0.2-0.8$ GeV), such that the
physical interpretation indeed holds.

\subsection{Cluster-hadronisation model}
\label{sec:ahadic}
For a long time, for the hadronisation of partons, \Sherpa has relied
on the implementation of
the Lund string model~\cite{Andersson:1983ia,*Andersson:1981ce,*Andersson:1998tv}
in \Pythia, accessible through a corresponding interface.  Since version 1.1, \Sherpa 
employs its own module \Ahadic~\cite{ahadicprep}, which implements a cluster-hadronisation 
model as described in~\cite{Gottschalk:1982yt,*Gottschalk:1983fm,*Webber:1983if,*Gottschalk:1986bv}
and extended by ideas presented in~\cite{Winter:2003tt}.  The basic assumption underlying
this class of models is local parton-hadron duality, i.e.\ the idea that quantum numbers
on the hadron level follow very closely the flow of quantum numbers on the parton level.
In this framework, the mass spectrum of the emerging colour-neutral
clusters is dominated by typical hadron masses or masses slightly above.  It is therefore natural to think of 
them as some kind of ``hadron matter'', carrying the flavour and momentum quantum numbers 
of hadrons.  This motivates to translate the light clusters directly
into hadrons or, if they are too heavy, to treat them like hitherto
unknown heavy hadron resonances, which decay further into lighter ones.  The idea underlying \Ahadic is to take this interpretation
very literal, to compose clusters out of all possible flavours, including 
diquarks, and to have a flavour-dependent transition scale between
clusters and hadrons.  This results in solely translating the very
light clusters directly into hadrons, whereas slightly heavier clusters experience a competition between being either translated 
into heavy hadrons or being decayed into lighter clusters, see below.  In addition, for all 
decays, a QCD-inspired, dipole-like kinematics is chosen.

In more detail, in \Ahadic, the hadronisation of quarks and gluons proceeds as follows:

First of all, the four-momenta of all partons are boosted and scaled such that all of them 
reside on constituent-mass shells ($m_u=m_d=0.3$ GeV, $m_s=0.5$ GeV, $m_c=1.8$ GeV, 
$m_b=5.2$ GeV), whereas the gluon remains massless in contrast to the cluster model 
implemented in \Herwig and \Herwigpp.  In addition to the quarks, further states will emerge, 
diquarks, which effectively carry baryonic quantum numbers.  Their
masses are typically given by the sum of the constituent masses of the
two respective quarks making them up, i.e.\ $m_{qq'} = m_q+m_{q'}$.
In the following, quarks and diquarks, will commonly be called ``flavours''.

At the onset of hadronisation, in cluster models the gluons must decay.  This is because in 
all hadronisation models, all hadrons are made of flavour constituents only.  This forced 
decay is the reason, why an effective gluon mass enters as parameter in \Herwig and 
\Herwigpp, where the gluon decays with no spectator involved.  In \Ahadic, the gluon's decay 
is facilitated in a dipole framework, and there will always be a colour-connected spectator 
to compensate the recoil when the massless gluon decays into massive flavours.  In these 
non-perturbative, enforced decays, the transverse momentum measure $p_\perp$ is bounded from 
above by a parameter $p_\perp^{\rm max}$, typically of the order of the parton-shower 
cut-off scale.

\mywidefigure{t}{
  \includegraphics[width=0.4\textwidth]{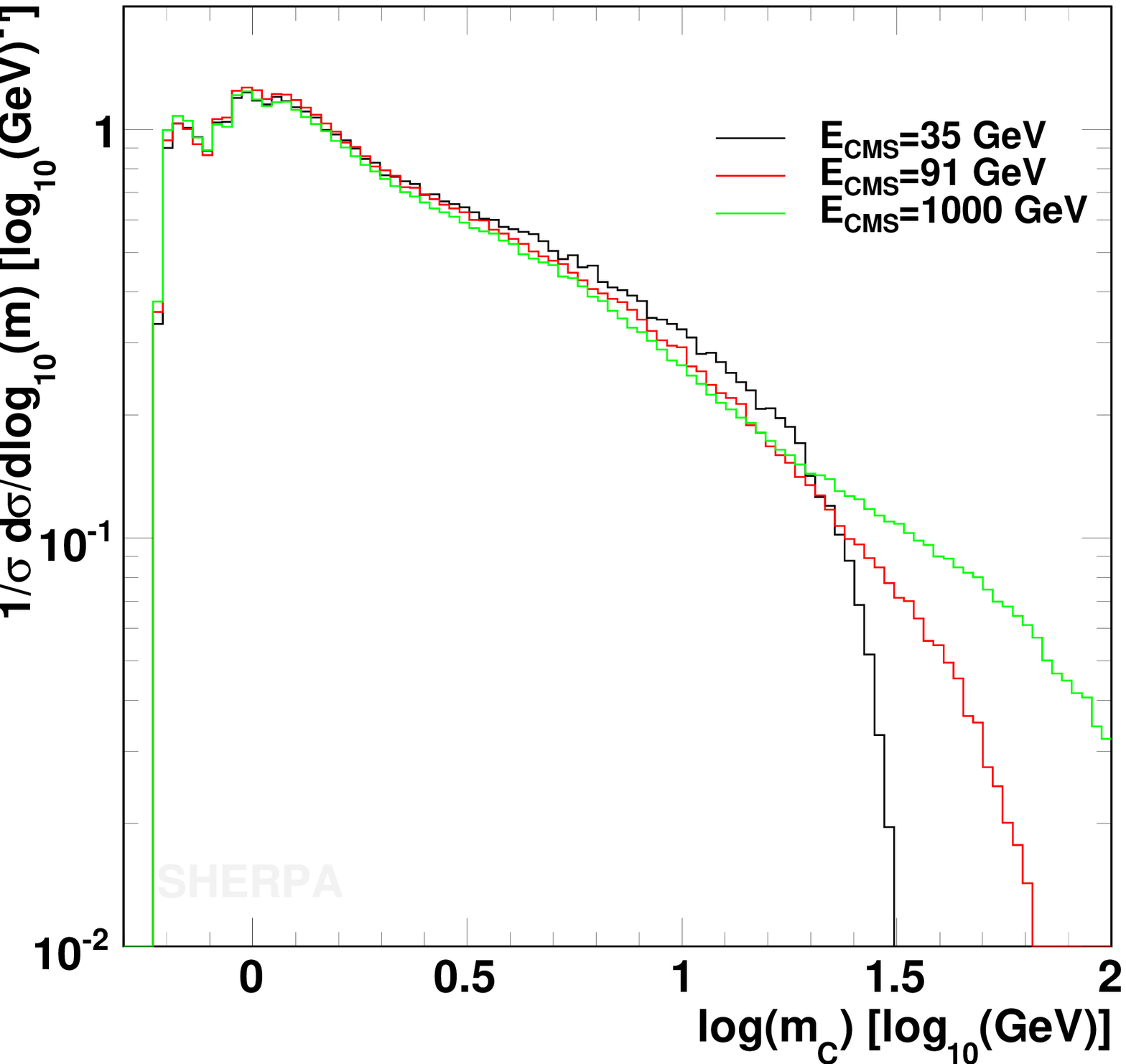}
  \hspace*{5mm}
  \includegraphics[width=0.4\textwidth]{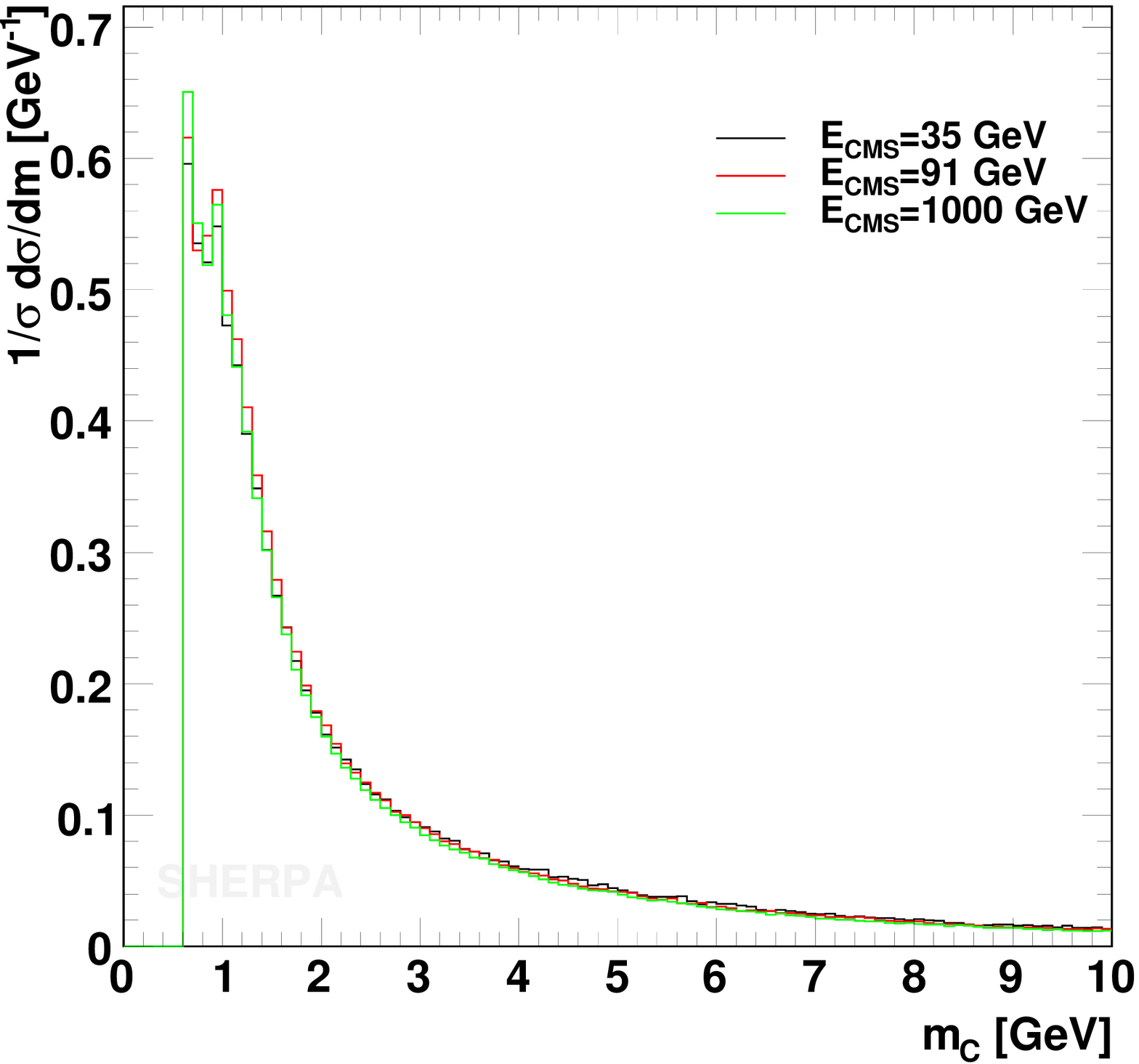}
}{Cluster-mass distribution in $e^+ e^-$ collisions at various center-of-mass energies.
  \label{fig:clustermass}}

In addition, in \Ahadic a non-perturbative $\alpha_s$ coupling has
been encoded for all decays (gluons and clusters).  It is chosen such that it agrees with a 
measurement from the GDH sum rule. The corresponding form given in
\cite{Deur:2008rf} has been implemented with all parameters.  In
addition to this default setting, constant-value choices are also
available.

For the production of the flavours, the available phase space leads to kinematic enhancements 
of lighter-flavour pairs in dependence on the mass of the dipole (i.e.\ pair of gluon and 
spectator).  It is supplemented by constant weight factors, ${\cal P}_q$, also used in cluster 
decays to hadrons.  When all gluons have been decayed, colour-connected flavours are merged
into colour-neutral clusters.  In contrast to the original version of the cluster model, 
also diquarks are allowed as cluster constituents from the beginning such that in every stage
of the hadronisation baryonic quantum numbers can be produced.

In order to guarantee the universality of this hadronisation procedure, it is important
that the cluster mass distribution is almost independent of the centre-\-of-\-mass-\-energy in otherwise 
identical processes.  This has been checked, e.g.\ for the case of $e^+e^-$ annihilations
into light quarks, cf.\ Fig.\ \ref{fig:clustermass}.
The structures in the peak region appear owing to masses of the
constituent quarks.

At every step, all clusters are checked whether they fall into one or both of two regimes, 
namely 
\begin{itemize}
\item either the regime where the mass of the cluster is of the order of the mass of the heaviest 
        hadron with a flavour wave component identical to the flavour content of the 
        cluster, i.e.\ where $M_c < m_h+\delta_h$,
\item or the regime where the mass of the cluster is lighter than the summed mass of the two 
	heaviest hadrons it can decay into by adding a flavour--antiflavour pair to its 
	flavour content, i.e. where $M_c < m_h+m_{h'}+\delta_{hh'}$.  
\end{itemize}
In order to steer this classification further, two offsets $\delta_h$ and $\delta_{hh'}$ have been introduced.

If a cluster is outside both regimes, it will decay further into two clusters.  This proceeds 
by the flavour--antiflavour pair firstly emitting a gluon with transverse-momentum choices 
as in the case of gluon decays discussed above, before then the heavier of the two colour
dipoles, made of one of the two flavours and the gluon, decays further, exactly as above.

If in contrast a cluster falls into one or both transition regimes, probabilities for the 
correspondingly allowed transitions into one or two clusters are calculated.  For the 
direct $C\to H$ transitions, this probability is given by
\bea
	{\cal P}_{M_c\to m_h} & = & \frac{\kappa}{2M_c}
	\frac{(M_c\Gamma_h)^2}{(M_c^2-m_h^2)^2+(m_h\Gamma_h)^2}
	\;|\Psi_{q\bar q}(h)|^2\,,
\eea
where $m_h$ and $\Gamma_h$ respectively are the mass and width of the hadron in question, and where 
$\Psi_{q\bar q}(h)$ denotes the corresponding flavour wavefunction component of the hadron 
(e.g.\ $\Psi_{u\bar u}(\pi^0) = 1/\sqrt{2}$).  For decays $C\to H_1H_2$, the probablity reads
\bea
	{\cal P}_{M_c\to m_1m_2} \, = \,
	\frac{\sqrt{[M_c^2-(m_1+m_2)^2][M_c^2-(m_1-m_2)^2]}}{16\pi M_c^3}
	\left[\frac{4m_1m_2}{M_c^2}\right]^\alpha 
	|\Psi_{q\bar q'}(h_1)|^2|\Psi_{q'\bar q}(h_2)|^2\cdot {\cal P}_{q'}\,.\,
\eea
The parameter $\kappa$ can be used to change the relative rate of
transitions and decays.  In the decay weights, the exponent $\alpha$
steers the preference of the decays for heavier hadrons.  ${\cal P}_q$
is the probability to create an additional $q$-flavour pair in the
field of the decaying cluster.  In each channel, the weights are further modified by including weights for 
the various hadron multiplets.  In addition, there is a singlet suppression factor to reduce 
transition probabilities to singlet-octet mixed states in the meson sector.

Once a decay into two hadrons is chosen, the decay kinematics has to be fixed.  This is 
achieved by giving the hadrons a transverse momentum w.r.t.\ the original flavour axis in
the cluster rest frame.  It is distributed according to 
\begin{equation}
p_\perp\;=\;|\vec p|\sin\theta\qquad\mbox{\rm where}\qquad
\sin\theta\;=\;\#^\eta\,.
\end{equation}
The parameter $\eta$ is adjustable and $\#$ is a uniformly chosen random number 
between 0 and 1.

\section{Decays of unstable particles}
\label{sec:decays}

Unstable particles can be produced in various phases of the event generation,
for example in the signal process, or after fragmentation, or even from decays
themselves.

For unstable particles generated by the hard process, \Sherpa provides the
possibility to specify the decay chain directly 
(cf. Sec.~\ref{sec:mes}). Since the complete process is translated into Feynman diagrams,
full correlations for all propagators are respected. This approach however is
not ideal for particles with many decay channels as appearing for example in BSM models.
In these cases one often wants to decay a particle from the signal process
inclusively. Currently this is only possible for the $\tau$-leptons,
which owing to their hadronic decay channels are treated by the hadron-decay module
(cf. Sec.~\ref{sec:decays:taus}). A more general solution for other fundamental
particles is being worked on for a future version of \Sherpa.

On the other hand, unstable hadrons, generated e.g. during fragmentation,
and the $\tau$-lepton can be inclusively decayed in cascade-like processes. Features
of this implementation shall be described in the following.

A natural starting point for the description of such cascades is the 
branching ratios collected in the PDG tables \cite{Amsler:2008zz}, and
to merely choose according to them the decay channel for individual
hadrons.  This is not always possible, since in many cases the
respective branching ratios do not add up to one.  There are several
options to restore a complete decay table.
For many of the light hadrons, especially in the cases of the
pseudoscalar and vector-meson multiplets and the octet and decuplet
baryons, it is viable to slightly rearrange the branching ratios, such
that they add up to 100\%.
Introducing new decay channels, ``guesstimated'' from flavour symmetries and
phase-space arguments is an option for some of the higher, not well-measured
resonances made of light quarks (like, e.g., the tensor mesons).
But especially for hadrons involving heavy quarks, the sheer number of
potentially open decay channels forbids such a treatment. Thus, a
mixture of explicit hadron decays
and transitions of the heavy-hadron state to quarks and gluons, 
invoking again the hadronisation model, turns out to be a much more 
powerful strategy.  However, care has to be taken that 
the hadronic final state emerging after such a partonic decay has not 
already been covered by an explicit hadron-decay channel coexisting
with the partonic mode.  In the hadron-decay framework
of \Sherpa this is realised by explicitly vetoing all unwanted hadronic 
final states emerging after the hadronisation of a partonic decay.  

Having selected a specific decay mode, its decay kinematics needs to be
modelled.  In order to go beyond an isotropic distribution of decay
products in the phase space, explicit matrix elements are
employed.  They are usually inferred from the spin structure and similar
symmetries of the initial and final state, giving rise to only 
a few amplitudes.  This simple picture can be further refined by invoking 
form factors (FFs) for certain transitions -- notable examples are the 
weak transitions of heavy quarks of the type $b\to c$ or $b\to u$, 
typically used in semileptonic decays of heavy hadrons
(cf. Sec.~\ref{sec:decays:heavymesons}), or FFs emerging 
for hadronic currents in $\tau$-decays (cf. Sec.~\ref{sec:decays:taus}).

The complexity of the emerging hadronic final-states and the multitude
of hadron decays make it impossible to implement and calculate matrix
elements and phase space for a full final state consisting of stable
particles only.  It is necessary to resort to the construction of
``chains'' of subsequent decays.  Naively, it looks like a reasonable
as well as feasible approximation to deal with each decay of such a
chain completely independently.  Nevertheless, looking a bit more
carefully reveals that complete independence is not always a valid
assumption.  Clearly, the spin structure of decaying and intermediate
particles potentially leads to non-trivial correlations among them and
other intermediate particles possibly emerging in the primary decay.
The algorithm presented in \cite{Richardson:2001df} has been
implemented to take such non-trivial correlations into account.  To
illustrate it, consider the decay $B\to D\rho$ with a subsequent
$\rho\to\pi\pi$ decay.  Naively, neglecting form factors, the full
matrix element for $B\to D\pi\pi$ can be described by
\bea \label{eq:cuttingpropagators}
\mathcal{M}
& \sim & i  (p^\mu_B + p^\mu_D) \,\cdot\,
  \sbr{\,g_{\mu\nu}-\frac{p_\mu p_\nu}{p^2}\,}\,\cdot\, 
  (p^\nu_{\pi_1} - p^\nu_{\pi_2}) \nnb\\
& \to & i
    \sum\limits_{\lambda_\rho}  
    \left(p^\mu_B + p^\mu_D\right)\,\cdot\, \varepsilon_\mu^*(\lambda_\rho)\,
    \varepsilon_\nu(\lambda_\rho)\,\cdot\, \left(p^\nu_{\pi_1} - p^\nu_{\pi_2}\right)\,.
\eea
This then allows to write the matrix element for $B\to D\pi\pi$ in
factorised form as
\bea
\mathcal{M} & \sim &
\sum\limits_{\lambda_\rho}\mathcal{M}\left(B\to D\rho\,;\;\lambda_\rho\right)
                       \mathcal{M}\left(\rho\to\pi\pi\,;\;\lambda_\rho\right).
\eea
Here the only difference to the decay cascade is that the sum over
helicities $\lambda_\rho$ is not computed independently for the two
subprocesses.  The spin-correlation algorithm takes this into account.
\myfigure{htbp}
{\scalebox{0.7}{% GNUPLOT: LaTeX picture with Postscript
\begingroup
  \makeatletter
  \providecommand\color[2][]{%
    \GenericError{(gnuplot) \space\space\space\@spaces}{%
      Package color not loaded in conjunction with
      terminal option `colourtext'%
    }{See the gnuplot documentation for explanation.%
    }{Either use 'blacktext' in gnuplot or load the package
      color.sty in LaTeX.}%
    \renewcommand\color[2][]{}%
  }%
  \providecommand\includegraphics[2][]{%
    \GenericError{(gnuplot) \space\space\space\@spaces}{%
      Package graphicx or graphics not loaded%
    }{See the gnuplot documentation for explanation.%
    }{The gnuplot epslatex terminal needs graphicx.sty or graphics.sty.}%
    \renewcommand\includegraphics[2][]{}%
  }%
  \providecommand\rotatebox[2]{#2}%
  \@ifundefined{ifGPcolor}{%
    \newif\ifGPcolor
    \GPcolortrue
  }{}%
  \@ifundefined{ifGPblacktext}{%
    \newif\ifGPblacktext
    \GPblacktextfalse
  }{}%
  % define a \g@addto@macro without @ in the name:
  \let\gplgaddtomacro\g@addto@macro
  % define empty templates for all commands taking text:
  \gdef\gplbacktext{}%
  \gdef\gplfronttext{}%
  \makeatother
  \ifGPblacktext
    % no textcolor at all
    \def\colorrgb#1{}%
    \def\colorgray#1{}%
  \else
    % gray or color?
    \ifGPcolor
      \def\colorrgb#1{\color[rgb]{#1}}%
      \def\colorgray#1{\color[gray]{#1}}%
      \expandafter\def\csname LTw\endcsname{\color{white}}%
      \expandafter\def\csname LTb\endcsname{\color{black}}%
      \expandafter\def\csname LTa\endcsname{\color{black}}%
      \expandafter\def\csname LT0\endcsname{\color[rgb]{1,0,0}}%
      \expandafter\def\csname LT1\endcsname{\color[rgb]{0,1,0}}%
      \expandafter\def\csname LT2\endcsname{\color[rgb]{0,0,1}}%
      \expandafter\def\csname LT3\endcsname{\color[rgb]{1,0,1}}%
      \expandafter\def\csname LT4\endcsname{\color[rgb]{0,1,1}}%
      \expandafter\def\csname LT5\endcsname{\color[rgb]{1,1,0}}%
      \expandafter\def\csname LT6\endcsname{\color[rgb]{0,0,0}}%
      \expandafter\def\csname LT7\endcsname{\color[rgb]{1,0.3,0}}%
      \expandafter\def\csname LT8\endcsname{\color[rgb]{0.5,0.5,0.5}}%
    \else
      % gray
      \def\colorrgb#1{\color{black}}%
      \def\colorgray#1{\color[gray]{#1}}%
      \expandafter\def\csname LTw\endcsname{\color{white}}%
      \expandafter\def\csname LTb\endcsname{\color{black}}%
      \expandafter\def\csname LTa\endcsname{\color{black}}%
      \expandafter\def\csname LT0\endcsname{\color{black}}%
      \expandafter\def\csname LT1\endcsname{\color{black}}%
      \expandafter\def\csname LT2\endcsname{\color{black}}%
      \expandafter\def\csname LT3\endcsname{\color{black}}%
      \expandafter\def\csname LT4\endcsname{\color{black}}%
      \expandafter\def\csname LT5\endcsname{\color{black}}%
      \expandafter\def\csname LT6\endcsname{\color{black}}%
      \expandafter\def\csname LT7\endcsname{\color{black}}%
      \expandafter\def\csname LT8\endcsname{\color{black}}%
    \fi
  \fi
  \setlength{\unitlength}{0.0500bp}%
  \begin{picture}(7200.00,5040.00)%
    \gplgaddtomacro\gplbacktext{%
      \csname LTb\endcsname%
      \put(990,660){\makebox(0,0)[r]{\strut{} 0}}%
      \put(990,1248){\makebox(0,0)[r]{\strut{} 0.1}}%
      \put(990,1836){\makebox(0,0)[r]{\strut{} 0.2}}%
      \put(990,2424){\makebox(0,0)[r]{\strut{} 0.3}}%
      \put(990,3012){\makebox(0,0)[r]{\strut{} 0.4}}%
      \put(990,3600){\makebox(0,0)[r]{\strut{} 0.5}}%
      \put(990,4188){\makebox(0,0)[r]{\strut{} 0.6}}%
      \put(990,4776){\makebox(0,0)[r]{\strut{} 0.7}}%
      \put(1122,440){\makebox(0,0){\strut{} 0}}%
      \put(2030,440){\makebox(0,0){\strut{} 0.5}}%
      \put(2938,440){\makebox(0,0){\strut{} 1}}%
      \put(3845,440){\makebox(0,0){\strut{} 1.5}}%
      \put(4753,440){\makebox(0,0){\strut{} 2}}%
      \put(5661,440){\makebox(0,0){\strut{} 2.5}}%
      \put(6569,440){\makebox(0,0){\strut{} 3}}%
      \put(220,2718){\rotatebox{90}{\makebox(0,0){\strut{}$\frac{1}{\Gamma}\frac{d\Gamma}{d\phi}$}}}%
      \put(3974,110){\makebox(0,0){\strut{}$\phi$}}%
    }%
    \gplgaddtomacro\gplfronttext{%
      \csname LTb\endcsname%
      \put(5839,4603){\makebox(0,0)[r]{\strut{}No spin correlations}}%
      \csname LTb\endcsname%
      \put(5839,4383){\makebox(0,0)[r]{\strut{}Spin correlations with scalar Higgs}}%
      \csname LTb\endcsname%
      \put(5839,4163){\makebox(0,0)[r]{\strut{}Spin correlations with pseudoscalar Higgs}}%
    }%
    \gplbacktext
    \put(0,0){\includegraphics{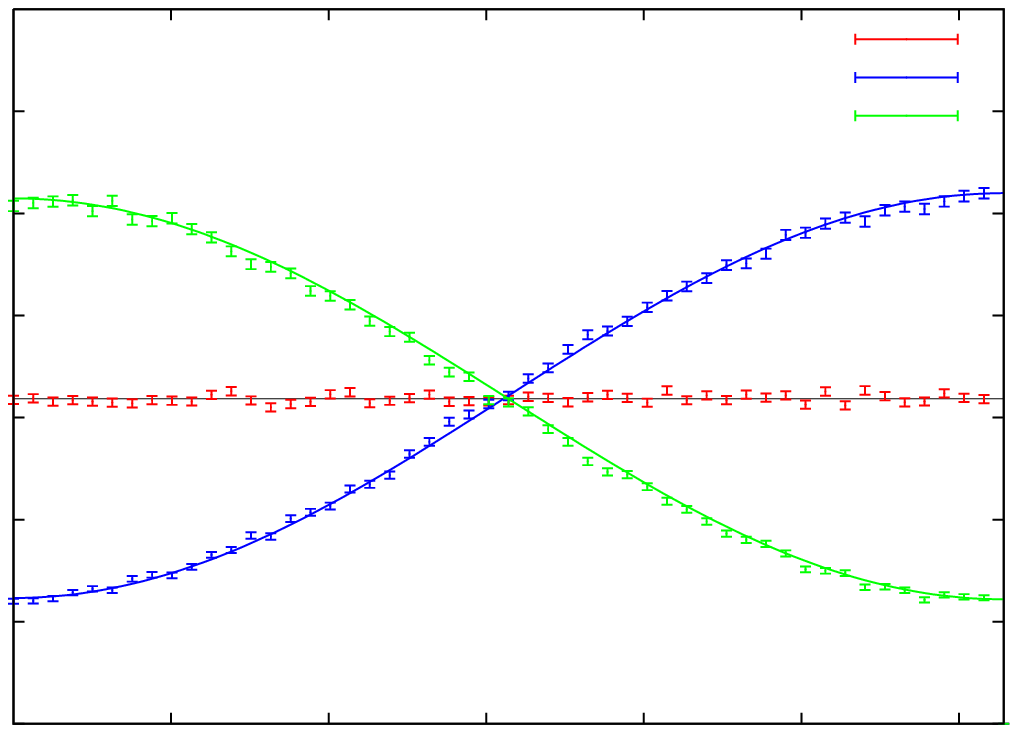}}%
    \gplfronttext
  \end{picture}%
\endgroup
}}
{
  Effect of spin correlations for the angle between the decay planes in
  $h \to \tau^- \tau^+ \to \pi^- \nu_\tau \pi^+ \bar{\nu}_\tau$
  \label{fig:decays:sc_higgstautau}
}

Spin correlations for $\tau$-leptons produced by the hard matrix element are
also taken into account when decaying the $\tau$. An example for this can be
seen in Fig.~\ref{fig:decays:sc_higgstautau} displaying the angle
between the decay planes of $\tau^-$ and $\tau^+$ produced from a
Higgs boson decay.  The solid lines represent the analytical
predictions taken from \cite{Kramer:1993jn}.

Unstable particles such as the $\rho$ meson have a finite width, which
may be quite large.  In order to accomodate 
for this effect, the invariant mass of the decaying particle can a posteriori, 
i.e.\ after the particle has been produced on its mass shell, be smeared
according to a relativistic Breit--Wigner distribution.  When doing so, 
however, kinematic bounds must be respected.  It is quite obvious that there 
is an upper bound given by the mass of the decaying particle
subtracted by the masses of the other decay products, which themselves
may have to be smeared.  Hence, a strategy is needed concerning the sequence of
the mass modifications.  Apart from the upper limit, there is also a lower
limit for the invariant mass of the decaying particle, given by the minimal
mass of its decay products.  This inevitably will vary with the decay
channel -- fixing the mass before fixing the decay channel could therefore
lead to unwanted biases, modifying the branching ratios in an unacceptable 
way.  In \Sherpa, these two problems are solved by fixing the decay channels
of the secondaries before applying the Breit--Wigner smearing on their
masses, the latter with rising width of the particles.

This still leaves some room for improvement.  Currently, the mass of any
unstable particle is distributed according to a superposition of 
Breit--Wigner distributions with different weights given by the respective
branching ratios and different lower cut-offs given by the minimal masses
of the respective final state.  This simple picture may be refined by
adding in threshold effects emerging when new decay channels are becoming 
kinematically accessible.  This can be done by reweigthing with the ``true''
invariant mass in the phase-space factors.
For the dicing of the offshell mass $m_a^\prime$ of a particle $a$ decaying in
a two-particle decay $a \to b c$, the emerging correction weight $w$ reads
\bea
    w_{\rm PS}
\;=\;\frac{\done\Phi^{(2)}(m_a^{\prime 2}, m_b^2, m_c^2)}{\done\Phi^{(2)}(m_a^2, m_b^2, m_c^2)}
\;=\;\frac{m_a^3}{m_a^{\prime 3}}\,\sqrt{\frac{\lambda(m_a^{\prime 2}, m_b^2, m_c^2)}{\lambda(m_a^2, m_b^2, m_c^2)}}
\eea
with $\lambda(a,b,c)=(a-b-c)^2-4bc$.
For n-particle decays, one can resort to an approximation using the mass of the
heaviest decay product for $m_b$ and the sum of the remaining decay-product
masses for $m_c$. In \Sherpa these phase-space correction weights for updated
masses are used, but no correction for the matrix element is applied.

\subsection{\tops{$\tau$}{tau}-lepton decays}
\label{sec:decays:taus}
\newcommand{\chpt}{\ensuremath{\rm \chi PT}}
\newcommand{\rcht}{\ensuremath{\rm R\chi T}}
\newcommand{\ks}{\ensuremath{\rm KS}}
\newcommand{\KS}{K\"uhn-San\-ta\-mar\'ia}

The $\tau$-lepton is the only lepton being heavy enough to decay into 
lighter hadrons.   Consequently, it provides an excellent laboratory 
for measuring hadronic currents and, additionally, to test the Fermi 
point-like interaction assumption and the 
CVC (PCAC) hypothesis\cite{GellMann:1960np}.

The mass of the $\tau$-lepton is roughly 1.77~GeV, which does not suffice
to produce a charmed hadron while decaying. Therefore,  its decay
products are either leptons or light hadrons consisting merely of up, 
down, and strange quarks.  For both, the matrix element of the decay
\begin{equation} 
	\tau^- (P) \to \nu_\tau (k) + \mbox{products}
\end{equation} 
is given by
\begin{equation} 
  \mathcal{M} = \frac{G_F}{\sqrt{2}}\,\bar{u}(k,\lambda_{\nu_\tau}) 
		(\gamma^\mu-\gamma^\mu\gamma_5) 
		u(P,\lambda_\tau) \;J_\mu
\end{equation} 
where the current $J_\mu$ depends on the chosen decay channel.

The branching ratio of the leptonic decay channels is about 35\%. This is a 
pure weak process, which can be calculated analytically and serves as a suitable
testbed for the left-handed leptonic currents.

The more involved hadronic channels with more than one hadron have
complicated resonance structures owing to intermediate particles 
with a short life time such as the $\rho$ meson.  The 
corresponding decays are described by form factors, which in \Sherpa
can be parametrised according to the following models:
\begin{itemize} 
	\item \KS (KS) \cite{Kuhn:1990ad} parametrisation,
	\item Resonance Chiral Theory (\rcht) \cite{Weinberg:1978kz,*Gasser:1983yg,*Gasser:1984gg,*Ecker:1988te}.
\end{itemize} 
The KS model is a phenomenological approach describing resonances through
their Breit--Wigner form.  They are introduced 
``by hand'' into the matrix element.  \rcht\/ on 
the other hand is an effective field theory: the extrapolation of Chiral 
Perturbation Theory (\chpt) to higher energies.  Using this approach, the 
decay matrix elements can be calculated analytically including all occuring
resonances.

\begin{figure}[t]
  \begin{minipage}{0.48\textwidth}
    \includegraphics[width=\linewidth]{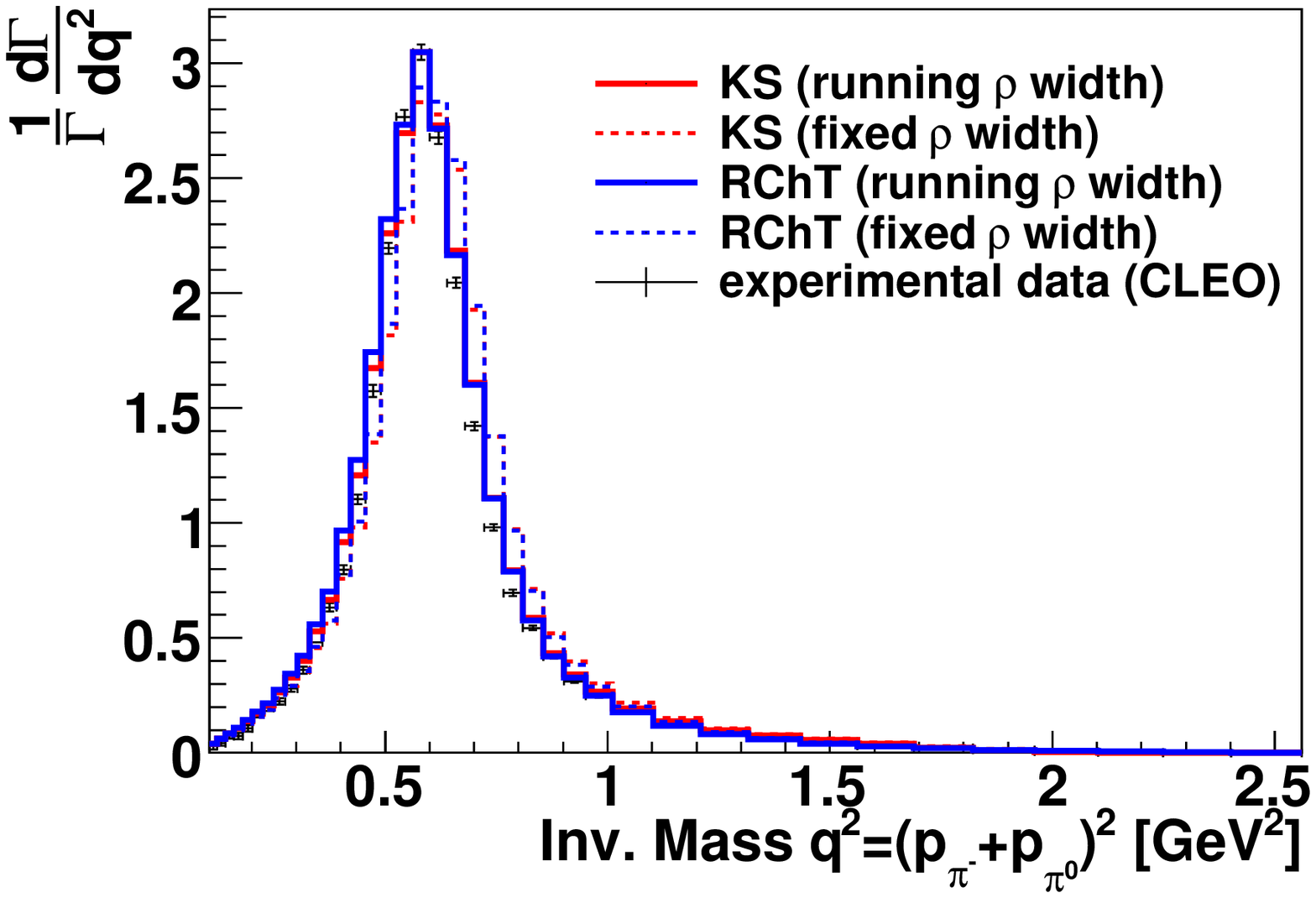}\\
    \myfigcaption{\textwidth}
{Invariant mass distribution of the two outgoing pions for
  $\tau^- \to \nu_\tau \pi^- \pi^0$.
\label{fig:decays:q2pipi}
}
  \end{minipage}
  \begin{minipage}{0.48\textwidth}
    \includegraphics[width=\linewidth]{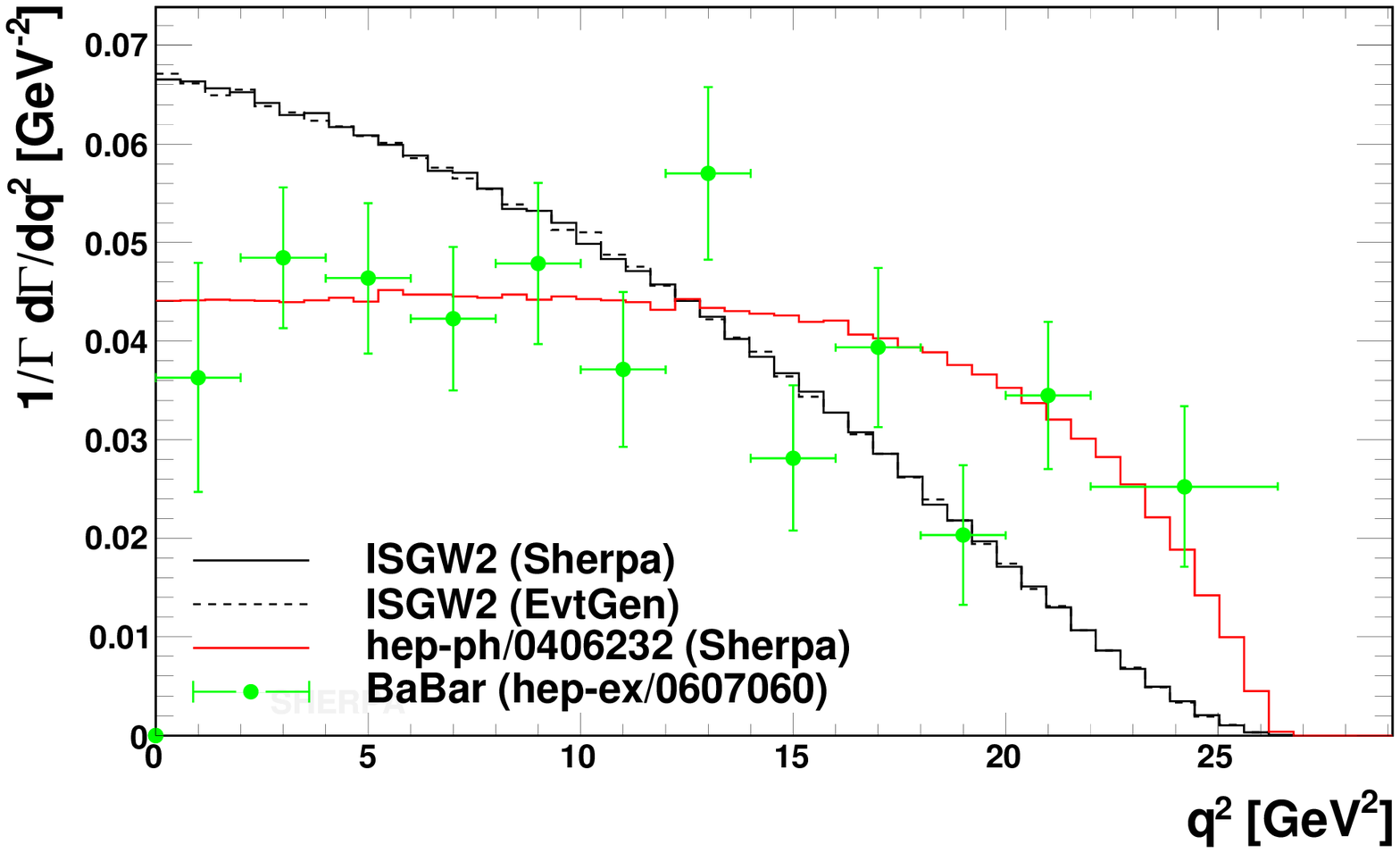}\\
    \myfigcaption{\textwidth}
{Invariant mass of the lepton-neutrino pair for the decay $B \to \pi \nu_e e$.
\label{fig:decays:q2Bpi}
}
  \end{minipage}
\end{figure}

Such form factors have been implemented for decays into up to three
pseudoscalars of the type $\pi^\pm$, $\pi^0$, $K^\pm$, $K^0$ and the two modes
producing four pions. As an example, Fig.~\ref{fig:decays:q2pipi}
shows a comparison of four different form-factor models with CLEO
data~\cite{Anderson:1999ui}.

\subsection{Hadron decays}

With approximately 200 decay tables consisting of 2500 decay channels the
majority of the decaying particles are hadrons. For some of them, form
factors have been very accurately measured, while for others not even
the branching ratios are well known.
In the following the features of hadron-decay simulations in \Sherpa will be
described thereby focussing on heavy mesons.
Details about the complete package can be found in~\cite{hadrons}.

\subsubsection{Heavy mesons}
\label{sec:decays:heavymesons}

Mesons containing one heavy quark ($m_q \gg \Lambda_{\rm QCD}$) as a
constituent can appear as 
\begin{itemize}
  \item $D^+$ ($c \bar d$) and $D^-$ ($\bar c d$),
  \item $D^0$ ($c \bar u$) and $\overline{D^0}$ ($\bar c u$),
  \item $D_s^+$ ($c \bar s$) and $D_s^-$ ($\bar c s$),
  \item $B^+$ ($\bar b u$) and $B^-$ ($b \bar u$),
  \item $B^0$ ($\bar b d$) and $\overline{B^0}$ ($b \bar d$),
  \item $B_s^0$ ($\bar b s$) and $\overline{B_s^0}$ ($b \bar s$),
  \item $B_c^+$ ($\bar b c$) and $B_c^-$ ($b \bar c$).
\end{itemize}
In most weak decays, only the heavy quark undergoes the flavour change and
enters into the $V-A$ current. The second quark only serves as
``spectator'' being exchangeable among the light quarks.

For mesons with heavy spectator quarks, additional decay modes may appear,
because the spectator quark can decay weakly as well. Differences may
also appear in the form factors, such that separate ones are
implemented for decays of the $B_c$~meson.

Excited mesons, like the $B^{*}$, do not feature very
diversified decay channels, but decay to their appropriate ground states
emitting an additional pion or photon. No form factors have been
implemented for these two-body decays, but they are treated according
to the respective generic matrix element.

The main decay modes of the pseudoscalar ground states fall into
three categories: leptonic decays, semileptonic decays and purely hadronic
decays.
Leptonic decay channels are easily parametrised using decay constants and
generic matrix elements.

For semileptonic decays, just like for $\tau$ decays, it is useful to
simply assume a factorised form of the decay 
matrix elements: in this case the heavy-to-heavy or heavy-to-light 
transitions can be modelled with form factors, which are determined
by effective field-theory approaches such as Heavy Quark Effective Theory
(HQET)~\cite{Neubert:1993mb,*Caprini:1997mu,*Richman:1995wm}, quark-model
predictions~\cite{Isgur:1988gb,*Scora:1995ty,*Goity:1994xn} or
QCD sum rules\cite{Ball:2004ye,*Ball:2004rg,*Ball:2007hb,*Aliev:2007uu}.
Fig.~\ref{fig:decays:q2Bpi} exemplifies the use of different form
factor models in the decay $B \to \pi \nu_e e$ by comparing to
predictions by the EvtGen simulation package~\cite{Lange:2001uf} and
data from BABAR~\cite{Aubert:2006px}.
For hadronic decays, again a factorisation between the two hadronic currents is
assumed, using the heavy currents from the semileptonic decays and combining
them with a hadronic current taken over from $\tau$-decays.
In fact, for a certain class of heavy-to-light transitions, this 
factorisation ansatz has been proven to yield results correct 
up to higher orders \cite{Beneke:2000ry},
and it will be used for these and many other decays in \Sherpa.

\subsubsection{Meson oscillations and CP violation}
\label{sec:decays:osci}

In addition to spin correlations and finite-width effects, more improvements
beyond the decay cascade are related to the simulation of mixing phenomena in
systems of neutral mesons, most notably $B\bar B$-mixing, and the
modelling of CP-violating effects.

\newcommand{\ket}[1]{|\,#1\rangle}
In the following the relationship between mass eigenstates $M_{L/H}$ and flavour
eigenstates $M^0/\overline{M^0}$ of a neutral meson $M$ is defined as
\begin{align}
  \ket{M_L} \;=\; p \ket{M^0} + q \ket{\overline{M^0}}
  \hspace{5mm} \mbox{and} \hspace{5mm}
  \ket{M_H} \;=\; p \ket{M^0} - q \ket{\overline{M^0}}\,.
\end{align}
This leads to a non-trivial time evolution $M_{\mathrm{phys}}^0(t)$ for flavour
eigenstates produced at $t=0$ in fragmentation and decay processes.
Four related effects are implemented in \Sherpa:

\begin{enumerate}
\item Explicit mixing in the event record can be accomplished by setting the
parameters $\Delta m = m_{H}-m_{L}$ and
$\Delta \Gamma=\Gamma_{H}-\Gamma_{L}$ of the neutral meson. According to
the time evolution of the flavour eigenstates, the mesons will then decay as
the appropriate flavour.

\item CP violation ``in mixing'' can be simulated by setting the value of
$|\tfrac{q}{p}|^2 \neq 1$.

\item Direct CP violation ``in decays'' can be accounted for by specifying different
decay tables for particles and their antiparticles.

\item CP violation ``in the interference of mixing and decays'' appears in
the decay to final states common to meson and anti-meson, because of
interference terms between the amplitude with a mixed meson and the unmixed
amplitude. It leads to a time dependent rate asymmetry
\newcommand{\Bz}{M^0}
\newcommand{\Bzb}{\bar{M}^0}
\newcommand{\Bzp}{M^0_{\mathrm{phys}}}
\newcommand{\Bzbp}{\bar{M}^0_{\mathrm{phys}}}
\begin{equation}
  A_{CP}(t) = 
	\frac{\Gamma(\Bzbp(t) \to f) \;-\; \Gamma(\Bzp(t) \to f)}
	     {\Gamma(\Bzbp(t) \to f) \;+\; \Gamma(\Bzp(t) \to f)}.
\end{equation}
Defining the quantity
$\lambda_f = \frac{q}{p} \frac{\bar{\mathcal{M}}_f}{\mathcal{M}_f}$ for a
decay to the common final state $f$ through the matrix elements $\mathcal{M}$,
the asymmetry can be expressed as
\bea \label{eqn:decays:acp}
  A_{CP}(t) \;=\; 
	\frac{S \cdot \sin(\Delta m t) - C \cdot \cos(\Delta m t)}
                   {\cosh(\Delta\Gamma \frac{t}{2}) -
 	\frac{2 \mathrm{Re} \lambda_f}{1+|\lambda_f|^2}
	\sinh(\Delta\Gamma\frac{t}{2})}
        \;\stackrel{\Delta\Gamma=0}{=}\;
	S \cdot \sin(\Delta m t) - C \cdot \cos(\Delta m t)\,,
\eea
where $S=\frac{2 \mathrm{Im} \lambda_f}{1+|\lambda_f|^2}$ and
$C=\frac{1-|\lambda_f|^2}{1+|\lambda_f|^2}$ denote the mixing-induced and 
direct contributions to the CP asymmetry, respectively.

The factors $S$ and $C$ have been measured or calculated for many decay channels
of the $B$~meson, cf.\ \cite{Amsler:2008zz} for a listing of measured values.
If
$\Delta\Gamma \neq 0$ is chosen, \Sherpa calculates $\lambda_f$ from the given
values of $S$ and $C$, 
\begin{align}
  \left|\lambda_f\right|^2 \;=\;& \frac{1-C}{1+C} \,,\\
  \mathrm{Im}(\lambda_f) \;=\;& \frac{S}{1+C} \,,\\
  -\frac{2\mathrm{Re}\lambda_f}{1+|\lambda_f|^2} \;=\;& -\sqrt{1-C^2-S^2}\,,
\end{align}
and simulates according to the non-simplified expression for
$A_{CP}(t)$ given in Eq.\ (\ref{eqn:decays:acp}).
\myfigure{t}
{\includegraphics[width=0.7\linewidth]{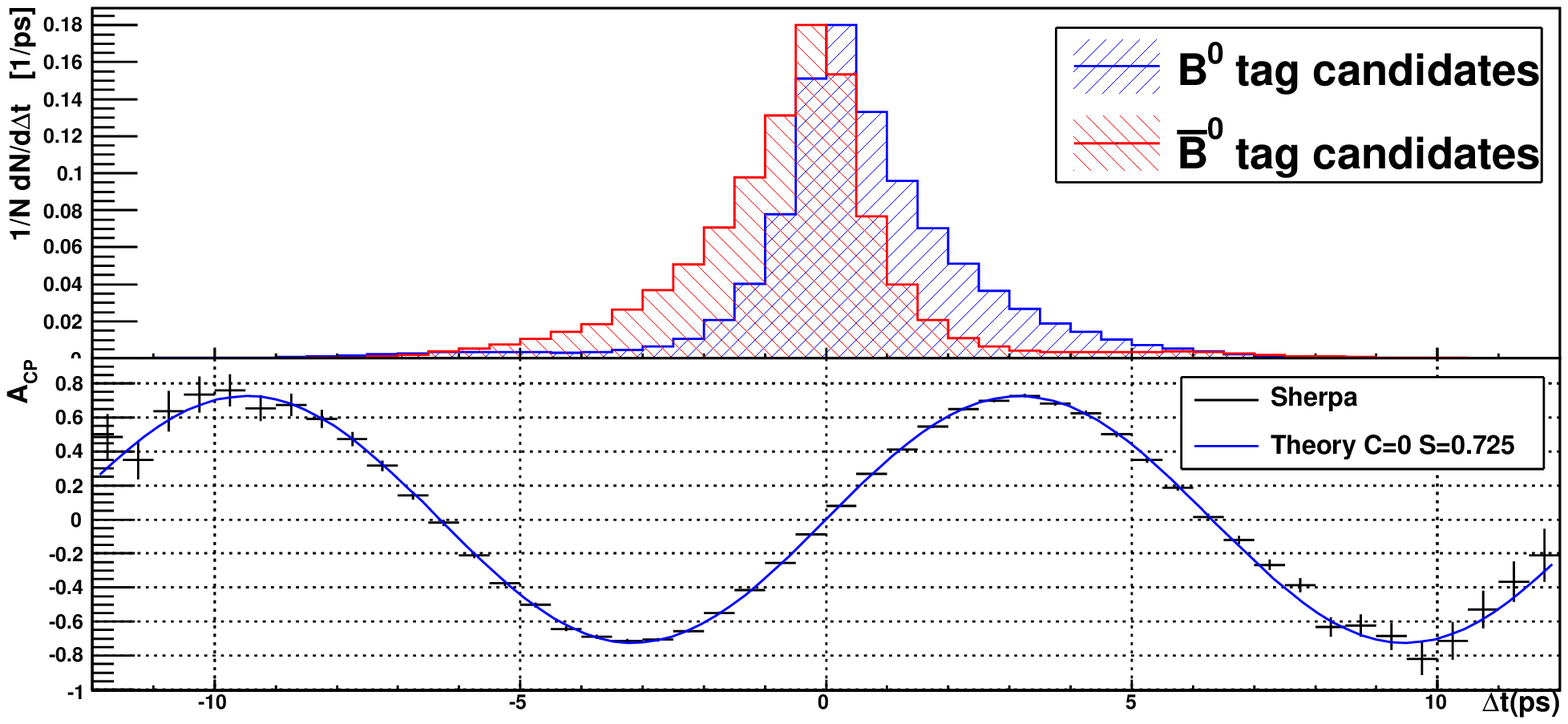}}
{CP violation in the interference between $B^0 \to J/\Psi K_S$ and
$\bar{B}^0 \to J/\Psi K_S$ with $\Delta\Gamma=0$, $S=0.725$ and $C=0$.
\label{fig:decays:mixing}
}
\end{enumerate}

A simulation of CP violation owing to interference effects in the 
decays of $B^0 \bar{B}^0$ pairs created from $\Upsilon(4S)$ decays
is presented in Fig.~\ref{fig:decays:mixing}. Here, one of the $B$
mesons is randomly selected to decay semileptonically at time $t_0$.
This provides a flavour tag at $t_0$ for the other $B$, decaying at
$t_1$ into $J/\Psi K_S$. Since both $B$ mesons have been produced 
coherently in the $\Upsilon(4S)$ decay, the reference time for the 
evolution of both flavour eigenstates is set by the semileptonic decay. 
Thus the rate asymmetry is governed by the time difference 
$\Delta t=t_1-t_0$, which can be positive or negative.

\section{Multiple emission of photons}

In the previous sections nearly all possible aspects of event 
generation were adressed. However, higher order QED corrections
in the form of soft-photon radiation were neglected. Nonetheless,
soft-photon radiation occurs at any stage of event generation whenever
charged particles are involved. They affect, e.g., production cross 
sections, (partial) decay widths of hadrons and $\tau$-leptons, and 
broaden momentum and invariant-mass distributions of final-state 
particles. Therefore these corrections have to be taken into account
in a realistic simulation. In \Sherpa this is accomplished by the 
\Photonspp module, which is based on the approach of Yennie, Frautschi 
and Suura (YFS) \cite{Yennie:1961ad}.

The corresponding algorithm relies on the idea of resumming leading 
soft logarithms to all orders, rather than focusing on leading collinear 
terms, which are taken into account in conventional parton showers 
(cf.~Sec.~\ref{sec:shower}). Soft logarithms are largely independent of 
the inner process characteristics and can be calculated from the charges 
of the external particles and their four-momenta only. The YFS formalism 
allows a systematic improvement of the eikonal approximation 
order-by-order in the QED coupling constant -- which has rendered it the 
standard approach for precision-calculations of QED radiation effects 
\cite{Jadach:1999vf,*Jadach:2000kw,*Jadach:2001mp,*Hamilton:2006xz}.  

At present the \Photonspp module is only capable to handle 
single-particle initial states and, hence, can only effect these 
corrections for particle decays. For the hard process, collinearly
enhanced photon emission off charged leptons and quarks is accounted
for during the parton-shower evolution, cf.\ Sec.~\ref{sec:shower}.

\subsection{The formalism}

The YFS formalism shall be reviewed very briefly to illustrate the
mechanisms of the resummed soft-photon contributions and the
order-by-order improvement for the hard emission region.

Consider a process at zeroth order in the electromagnetic coupling $\alpha$,
described by the matrix element $\mathcal{M}_0^0$. The sub- and 
superscripts denote the number of real photons involved and the order 
of $\alpha$, respectively. The four-momenta of the final-state particles 
are labelled $p_f$, while the incoming momenta are denoted by $p_i$. The fully
inclusive cross section for having {\it any} number of additional soft photons 
with momenta $k$ can then be written as
\bea
\!\sigma & \sim & \!\!\sum_{n_R = 0}^\infty\!\!\frac{1}{n_R !}\!
\int\!\done\Phi_p\,\done\Phi_k\,
(2\pi)^4\delta^4\!\left(\sum p_i-\!\sum p_f-\!\sum k\right)
\left|\sum_{n_V=0}^\infty\mathcal{M}_{n_R}^{n_V+\frac{1}{2}n_R}\right|^2
\label{eq:all_order_sigma}
\eea
where $n_V$ and $n_R$ respectively count the numbers of virtual and
real photons involved.

The starting point of the YFS algorithm is to approximate the dressed 
matrix elements by the zeroth-order expression times eikonal factors
that depend on the external momenta only. The correct result can then be 
restored order-by-order in perturbation theory by supplementing the 
non-leading, process-dependent pieces. In the case of adding just one 
virtual photon this can be formalised as
\bea\label{Eq:virtual_factorisation}
 \mathcal{M}_0^1 & = & \alpha B \mathcal{M}_0^0 + M_0^1\,,
\eea
with $M_0^1$ being the infrared-subtracted matrix element including one 
extra virtual photon. Note that accordingly $M_0^1$ is finite in the
limit of vanishing photon momentum, i.e.\ $k\to 0$. All soft
divergences owing to the virtual-photon insertion are contained in the
integrated eikonal $B$, which is process-independent and
universal\footnote{The universal integrated virtual eikonal is a sum
  over the eikonals of every QED subdipole, i.e.\
	\bea
	B 
	& = & \sum_{i<j}B_{ij} \,.
	\eea
	Details can be found in \cite{Schoenherr:2008av}.
                               }. 
When summing over all subsequent insertions of further virtual photons,
thereby iteratively applying Eq.~(\ref{Eq:virtual_factorisation}), the very 
appealing all-order expression
\bea
\sum_{n_V=0}^\infty\mathcal{M}_0^{n_V} 
\; = \; \sum_{n_V=0}^\infty\sum_{r=0}^{n_V}M_0^{n_V-r}\frac{(\alpha B)^r}{r!}
\; = \; \exp(\alpha B)\sum_{n_V=0}^\infty M_0^{n_V}\,,
\eea
can be derived. This can be generalised to any number of real photons
by exploiting the unique characteristics of abelian QED and the fact
that virtual photons inserted in closed charged loops do not produce
any additional infrared singularity. Hence, all
$M_{n_R}^{n_V+\frac{1}{2}n_R}$ are free of soft singularities from
virtual-photon exchange but may still exhibit such due to real-photon
emission.

However, YFS were able to show that real-photon emission processes can also be
factorised -- on the level of the squared matrix elements rather than the 
amplitudes. For a single radiated photon, one obtains
\bea
\frac{1}{2(2\pi)^3}\left|\sum_{n_V=0}^\infty 
              M_1^{n_V+\frac{1}{2}}\right|^2
& = & 
\tilde{S}(k)\left|\sum_{n_V=0}^\infty M_0^{n_V}\right|^2 + 
\sum_{n_V=0}^\infty\tilde{\beta}_{1}^{n_V+1}(k)\,.\label{eq:real_factorisation} 
\eea
Here, $\tilde{S}(k)$ is an eikonal factor containing the soft divergence
related to real-photon emission\footnote{
	The universal real eikonal is defined as a sum over subdipoles
	\bea
	\tilde{S}(k)
	& = & \sum_{i<j}\tilde{S}_{ij}(k) \,.
	\eea
	For details see \cite{Schoenherr:2008av}.
                    }. 
$\tilde{\beta}_{n_R}^{n_V+n_R}$ denotes the complete IR-finite 
(subtracted) squared matrix element for the basic process dressed with 
$n_R$ real and $n_V$ virtual photons. 
With Eq.~(\ref{eq:real_factorisation}) at hand the perturbative series 
can be reordered such that a complete partition of infrared finite 
and divergent terms is achieved. Moreover, the singular terms can be 
exponentiated by separating the real-emission phase space into a
region $\Omega$ containing the infrared divergence such that
$(1-\Omega)$ is completely IR finite. Introducing the YFS form factor\footnote{
	The integrated real-emission eikonal -- in analogy to the universal 
	virtual eikonal $B$ -- is defined as
	\bea
	2\alpha\tilde{B}(\Omega)
	& = & \int\frac{\dthree k}{k}\,\tilde{S}(k)\left(1-\Theta(k,\Omega)\right) \,,
	\eea
	with $\Theta(k,\Omega)=1$ if $k\notin\Omega$ and zero otherwise.
}
\bea
 Y(\Omega)
 & = & 2\alpha\left(B+\tilde{B}(\Omega)\right)
\eea
the cancellation of real and virtual infrared divergences
can be made explicit. The rearranged form of the exact cross 
section given in Eq.~(\ref{eq:all_order_sigma}) reads
\bea\label{Eq:exact_distribution}
\!\sigma 
& \sim & \!\!\sum_{n_R = 0}^\infty\!\!\frac{1}{n_R !}\!
	\int\!\done\Phi_p\,\done\Phi_k^\prime\,
	(2\pi)^4\delta^4\!\left(\sum p_i-\!\sum p_f-\!\sum k\right)
	e^{Y(\Omega)}\prod_{i=1}^{n_R}\tilde{S}(k_i)\Theta(k_i,\Omega) \nnb\\
&& \hspace{79mm}\times\,
	\left(\tilde{\beta}_0^0+\tilde{\beta}_0^1+
	\sum_{i=1}^{n_R}\frac{\tilde{\beta}_1^1(k_i)}{\tilde{S}(k_i)}
	+\mathcal{O}(\alpha^2)\right)\,.
\eea
The original perturbative series, organised in powers of the electromagnetic 
coupling constant $e$ and based on amplitudes, has been rearranged as a 
perturbative series in $\alpha$ based on squared matrix elements, whose 
infrared singularities have been extracted. While the
squared matrix elements encode all the process-specific information, 
e.g. spin correlations, interferences, and hard-emission properties, the 
YFS form factor describes the resummed universal soft limit.

In this apparent form of the all-orders cross section the leading-order 
process $\tilde\beta_0^0$ can be factored out enabling the
construction of a Monte Carlo algorithm to correct (arbitrary) cross
sections or decay rates for real and virtual QED radiation.

\subsection{The algorithm}

In Ref.~\cite{Schoenherr:2008av} a concrete version of such an algorithm is presented.
Two basic aspects of this ansatz shall briefly be discussed here,
the definition of the soft region $\Omega$ and the reconstruction algorithm 
for the leading-order particles' momenta.

To simplify calculations the singular phase-space region $\Omega$ has been 
chosen to be bounded by an isotropic hypersurface in the rest frame of 
the multipole, i.e.\ the ensemble of charged particles involved 
in the considered process. It is specified by an energy cut-off $\omega$. 
Photons outside $\Omega$ are generated exclusively. Their number, energy 
and angular distribution are determined by 
$\int\frac{\dthree k}{k}\tilde{S}(k)\Theta(k,\Omega)$, the soft
real-emission eikonal. The higher-order hard-emission corrections are then 
accomodated by introducing them as a relative weight.

This can be achieved by approximating the real-emission matrix elements 
in the quasi-collinear limit using the QED variant of the spin-dependent 
dipole splitting functions $\mbox{\sl g}_{ij}$ of 
\cite{Dittmaier:1999mb,*Catani:2002hc}. However, the terms describing the 
soft limit need to be subtracted as they are already included in the
YFS form factor. This yields new, soft-subtracted, dipole splitting
functions denoted by $\bar{\mbox{\sl g}}_{ij}$. For the collinearly
approximated hard-emission matrix elements one then obtains
\bea
 \tilde{\beta}_1^1 
& = & -\frac{\alpha}{4\pi^2}\sum_{i<j}Z_iZ_j\theta_i\theta_j
	\left(\bar{\mbox{\sl g}}_{ij}
		+\bar{\mbox{\sl g}}_{ji}\right)\tilde{\beta}_0^0 \,.
\eea
However, the $\tilde{\beta}_1^1$ lack any interference terms between 
the different amplitudes entering. These can only be incorporated using 
the exact real-emission matrix element. This, however, is process specific 
and, hence, is only available for some special cases (a list of available
matrix elements is given in Tab.~\ref{Tab:YFS_ME}). Furthermore, to maintain 
a process-independent formulation as far as possible, extended, composite 
objects, e.g.\ hadrons, need to be treated as point-like ones. To achieve a 
higher accuracy in selected channels form-factor models can be used. 

\mytable{htbp}
{\begin{tabular}{|c|c|c|}
  \hline
  matrix element & real $\order(\alpha_{\QED})$ & virtual $\order(\alpha_{\QED})$ \\
  \hline
  $V^0\to F^+F^-$ & $\surd$ & $\surd$ \\
  $V^0\to S^+S^-$ & $\surd$ & $\surd$ \\
  $S^0\to F^+F^-$ & $\surd$ & $\surd$ \\
  $S^0\to S^+S^-$ & $\surd$ & $\surd$ \\
  $W^\pm\to\ell^\pm\nu_\ell$ & $\surd$ & $\surd$ \\
  $\tau^\pm\to\ell^\pm\nu_\ell\nu_\tau$ & $\surd$ & - \\
  $S^0\to S^\mp\ell^\pm\nu_\ell$ & under construction & -\\
  $S^0\to V^\mp\ell^\pm\nu_\ell$ & under construction & -\\
  \hline
 \end{tabular}
 \vspace{3pt}
}
{List of available generic and specific infrared 
subtracted squared real-emission ($\tilde\beta_1^1$) 
and virtual-correction ($\tilde\beta_0^1$) matrix 
elements ($V$ - vector, $F$ - spin-$\tfrac{1}{2}$ fermion, 
$S$ - scalar).\label{Tab:YFS_ME} 
}

A comparison of the impact of the various possibilities to incorporate 
hard-emission effects can be found in Fig.~\ref{Fig:photons:Jpsi}, 
which depicts the total amount of energy radiated and the angular 
radiation spectrum in leptonic
$J/\psi$-decays, respectively. Both distributions highlight the importance 
of properly accounting for hard radiation and reveal the shortcomings 
of the approximated matrix elements when compared to the exact result. 
The approximation overestimates hard radiation at large angles, thus
enhances the total fraction of events with hard photons close to 
the kinematic limit at half the mass of the decaying particle. This 
effect originates from the missing interference terms and is the more 
pronounced the higher the mass of the emitting particle. The 
soft limit of course is the same in all approaches, since it is 
incorporated in the YFS form factor $Y(\Omega)$ to all orders in 
$\alpha$.

\mywidefigure{htbp}{
  \includegraphics[width=0.47\textwidth]{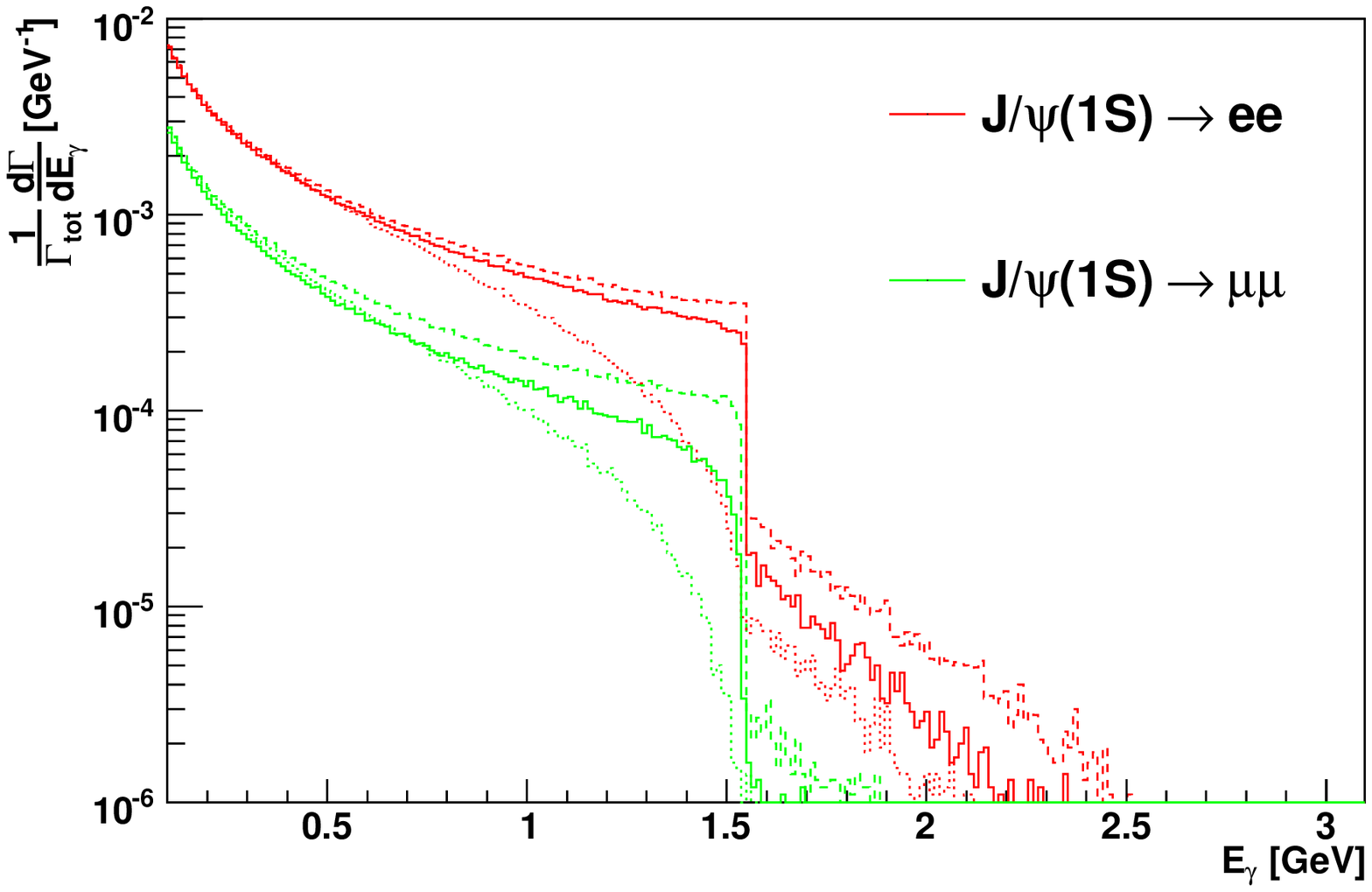}\hspace*{5mm}
  \includegraphics[width=0.47\textwidth]{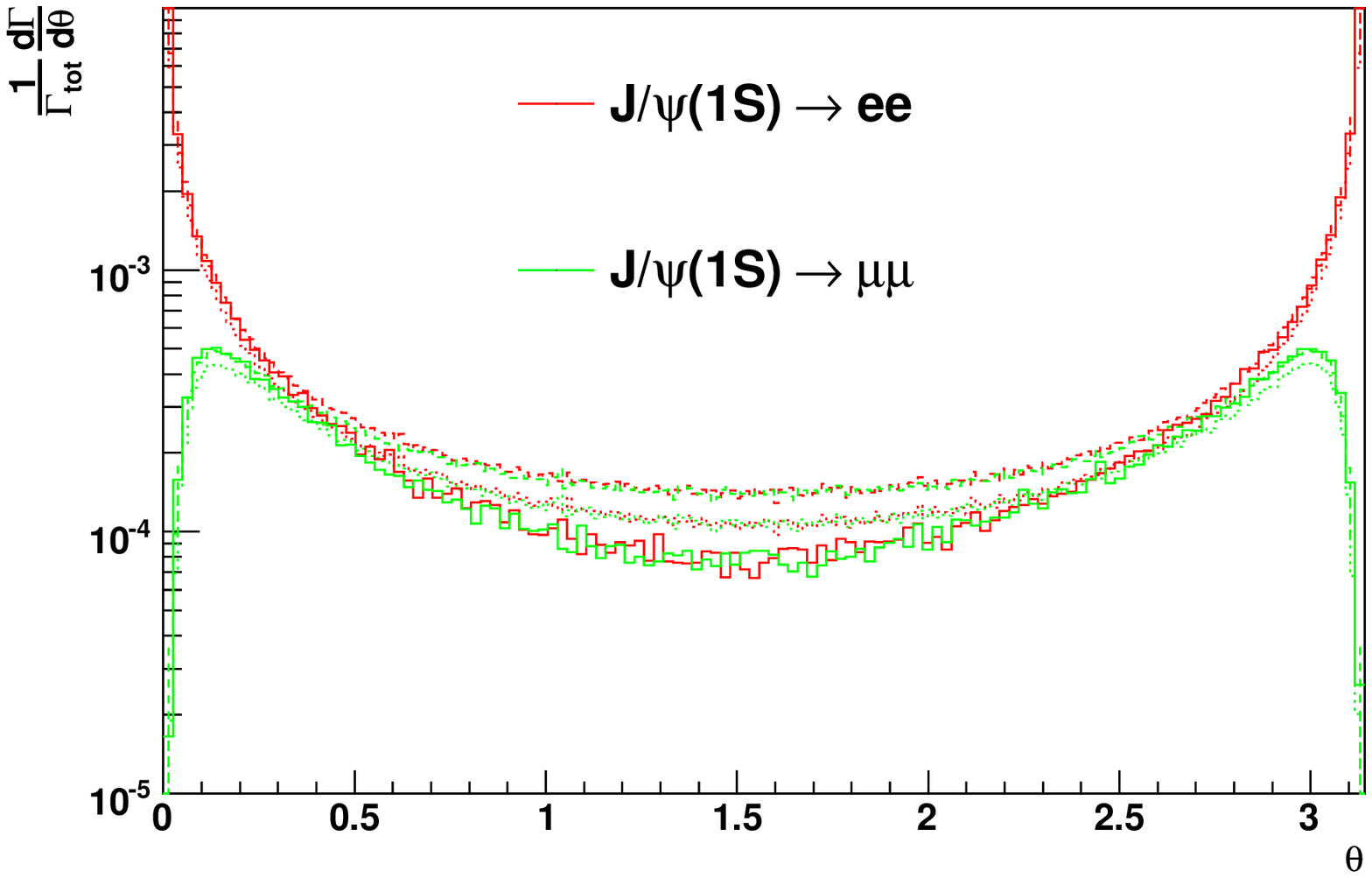}
}{Left panel: Total photon energy radiated for $J/\psi(1S)\to\ell^+\ell^-$ in
the $J/\psi$ rest frame. Right panel: Angular distribution of photon radiation 
for $J/\psi(1S)\to\ell^+\ell^-$ in the $\ell^+$-$\ell^-$ rest frame. The effects 
of different types of higher-order hard-emission corrections are shown: no 
corrections (dotted), approximated real-emission matrix-element correction 
(dashed) and exact real-emission matrix-element correction (solid).
\label{Fig:photons:Jpsi}
}

The original particles' momenta need to be adapted to satisfy momentum
conservation after the higher-order QED corrections have been included.
This is achieved by distributing the additional photons' momenta uniformly 
among all, charged and neutral, particles of the process, supplemented
by a common rescaling of all final-state three-momenta in the 
multipole's rest frame. This prescription necessitates a change of the 
initial-state momenta as well. However, since these were already fixed 
when calculating the leading-order process, this change is interpreted 
as a shift of the multipole's rest frame. Additional complications arise 
for initial states involving more than one particle, however, at 
present \Photonspp is limited to particle decays.

On the other hand, photons emitted inside $\Omega$ are treated 
inclusively and although their number is infinite, the sum of their
momenta is assumed to have a negligible effect on the overall momentum
distribution.

\section{Summary}

In this article the original publication on the proof-of concept 
version of \Sherpa \cite{Gleisberg:2003xi}, dating back to 2003, has 
been updated.  Since then, \Sherpa has matured considerably, especially
in terms of physics capabilities, reliability and user support.  At the
same time, however, the original strategy of a separation of physics
modules and structures for event generation has been maintained.  In 
fact, the underlying idea of the physics being confined to individual
modules, steered by corresponding handler classes, which 
in  turn are employed by appropriate event phase classes during event 
generation, has proven to be extremely flexible, robust and versatile.  
Eventually, this construction paradigm has allowed a comparably quick 
and painless incorporation of many new physics aspects and additional
user interfaces.  This has helped to
augment \Sherpa with many hitherto missing physics aspects, for example
hadronisation, hadron decays and the radiation of soft photons, thus replacing 
interfaces to already existing Fortran routines from other codes\footnote{
  For the sake of backward-compatibility, these interfaces 
  are kept available.}.  
These additions render the program a truly self-contained event generator for 
lepton--lepton and hadron--hadron collisions.  In addition, many existing 
physics modules have been upgraded or improved.  This ranges from the 
implementation of some models for new physics over a significantly enhanced 
and validated merging prescription for matrix elements and parton showers 
to the incorporation of a number of interfaces to, e.g., the Les Houches 
accord files for transmitting MSSM parameters or the LHAPDF library of PDFs. 
In addition, the user-interface has significantly improved, now including,
among others, various event formats, a large variety of phase-space cuts and the 
greatly enhanced \Sherpa-internal analysis tool.  It is worth stressing 
that all these improvements certainly enhance both the reliability and the 
physics abilities of \Sherpa.

In addition to the published features of \Sherpa described in this 
article, many other features have established, but are not
yet part of a publicly available version.  These features include two new 
QCD shower modules, one based on Catani--Seymour dipole 
factorisation~\cite{Schumann:2007mg} and a dipole shower~\cite{Winter:2007ye}. 
In the matrix-element sector, an automated Catani--Seymour dipole subtraction 
algorithm has been implemented within \Amegic~\cite{Gleisberg:2007md}, and 
a new matrix-element generator for very large multiplicities,
\Comix \cite{Gleisberg:2008fv}, has been constructed.  

These new modules, have not been released yet as part of \Sherpa, but
are planned to be included in a version 1.2, which will be
made available in 2009.  Their incorporation can be seen as a first step
of the package being upgraded to next-to-leading order accuracy in the 
matrix-element sector.
In this context, the initial construction paradigm of \Sherpa again
proves to yield a very flexible code, which can easily support
the newly emerging structures.

\section*{Acknowledgements}
T.G.'s research was 
supported by the US Department of Energy, contract DE-AC02-76SF00515.  S.H.\ 
acknowledges funding by the HEPTOOLS Marie Curie Research Training Network
(contract number MRTN-CT-2006-035505) and the Swiss National Science 
Foundation (SNF, contract number 200020-117602).  F.S.'s research 
was funded by MCnet (contract number MRTN-CT-2006-035606). 
S.S.\ acknowledges funding by the UK Sience and Technology Facilities Council 
(STFC). J.W.\ thanks O.~Gonz\'alez for helpful discussions.
Fermilab is operated by Fermi Research Alliance, LLC, under contract
DE-AC02-07CH11359 with the United States Department of Energy. Financial
support from BMBF is gratefully acknowledged.
\bibliographystyle{amsunsrt_mod}
\bibliography{journal}
\end{document}